\documentclass[11pt,a4paper]{article}
\usepackage{jheppub,amsmath, amsthm, amssymb,slashed,url}
\usepackage{graphicx}
\usepackage{epstopdf}
\def\t{{ \sf t}} 
\def\ct{{\cmmib t}}

\def\be{\begin{equation}}
\def\ee{\end{equation}}
\def\bt{{{t}}}
\def\hat{\widehat}
\def\tilde{\widetilde}
\def\frak{\mathfrak}
\def\D{{\mathcal D}}
\def\MM{M_4}
\def\k{{\cmmib k}}
\def\Bbb{\mathbb}
\def\Mthree{M_3}
\def\A{{\mathcal A}}
\def\d{{\mathrm d}}

\def\R{{\mathbb R}}
\def\C{{\mathbb C}}
\def\U{{\mathcal U}}
\def\CP{{\mathbb{CP}}}
\def\N{{\mathcal N}}
\def\T{{\mathcal T}}
\def\F{{\mathcal F}}
\def\Tr{{\mathrm {Tr}}}
\def\Z{{\mathbb Z}}
\def\Q{{\mathbb Q}}
\def\CC{{\mathcal C}}
\def\ad{{\mathrm{ad}}}
\def\Tr{{\mathrm{Tr}}}
\def\Btriv{{\mathcal B}_{\mathrm{triv}}}
\def\L{{ L}}
\def\Bcc{{\mathcal B_{\mathrm{cc}}}}
\def\K{{\mathcal K}}
\def\V{{\mathcal V}}
\def\J{{\mathcal J}}
\def\I{{\mathcal I}}
\def\P{{\mathcal P}}
\def\B{{\mathcal B}}
\def\Boper{{\mathcal B_{\mathrm{oper}}}}
\def\NS5{\mathrm{NS5}}
\def\uD5{\mathrm{D5}}
\def\M{{\mathcal M}}
\def\W{{\mathcal W}}
\def\P{{\mathcal P}}
\def\H{{\mathcal H}}
\def\ZZ{\eusm B}
\def\cZ{\mathcal Z}
\def\sW{{ W}}
\def\sY{{ Y}}
\def\nY{{\mathcal Y}}
\def\ca{{\cmmib a}}
\def\ss{{d}}
\def\m{\cmmib m}
\def\tilde{\widetilde}
\def\bar{\overline}
\font\teneurm=eurm10 \font\seveneurm=eurm7 \font\fiveeurm=eurm5
\newfam\eurmfam
\textfont\eurmfam=\teneurm \scriptfont\eurmfam=\seveneurm
\scriptscriptfont\eurmfam=\fiveeurm

\font\teneusm=eusm10 \font\seveneusm=eusm7 \font\fiveeusm=eusm5
\newfam\eusmfam
\textfont\eusmfam=\teneusm \scriptfont\eusmfam=\seveneusm
\scriptscriptfont\eusmfam=\fiveeusm
\def\eusm#1{{\fam\eusmfam\relax#1}}
\font\tencmmib=cmmib10 \skewchar\tencmmib='177
\font\sevencmmib=cmmib7 \skewchar\sevencmmib='177
\font\fivecmmib=cmmib5 \skewchar\fivecmmib='177
\newfam\cmmibfam
\textfont\cmmibfam=\tencmmib \scriptfont\cmmibfam=\sevencmmib
\scriptscriptfont\cmmibfam=\fivecmmib
\def\cmmib#1{{\fam\cmmibfam\relax#1}}

\title{Knot Invariants from Four-Dimensional Gauge Theory}
\author{Davide Gaiotto$^1$}
 \author{and Edward Witten$^{1,2}$}
\affiliation{$^1$School of Natural Sciences, Institute for Advanced Study,\\ 1 Einstein Drive, Princeton, NJ 08540 USA\\ \vskip.1cm $^2$Department of Physics, Stanford University,
Palo Alto CA 94305}
\abstract{It has been argued based on electric-magnetic duality and other ingredients that
the Jones polynomial of a knot in three dimensions can be computed by counting the solutions of certain
gauge theory equations in four dimensions.  Here, we attempt to verify this  directly by analyzing
the equations and counting their solutions, without reference to any quantum dualities.  
After suitably perturbing the equations to make their behavior more generic, we are able
to get a fairly clear understanding of how the Jones polynomial emerges.  The main ingredient
in the argument is a link between the four-dimensional gauge theory equations in question and
conformal blocks for degenerate representations of the Virasoro algebra in two dimensions.
Along the way we get a better understanding of how our subject is related to a variety of new
and old topics in mathematical physics, ranging from the Bethe ansatz for the 
Gaudin spin chain to the $M$-theory description of BPS monopoles and the 
relation between   Chern-Simons gauge theory and Virasoro conformal blocks. }
\begin{document} \maketitle

\section{Introduction}

The Jones polynomial \cite{Jones}  is an invariant of knots that has multiple  relations 
to many aspects
of mathematical physics, including integrable lattice statistical mechanics, two-dimensional
conformal field theory and associated representations of braid groups, and three-dimensional
Chern-Simons gauge theory.  Khovanov homology \cite{Khovanov} is a more recent 
topological theory in
four dimensions; in this theory, a knot is viewed as an object in three-dimensional space 
and the invariant associated to a knot is
a vector space (of physical states) rather than a number. The relation between the two theories
is that the four-dimensional theory associated to Khovanov homology, when compactified 
on a circle,
reduces to the three-dimensional theory that yields the Jones polynomial.

 Khovanov homology has been interpreted physically \cite{GSV} in terms of 
 topological strings, building on earlier work on BPS states of open
strings \cite{OV}. See \nocite{LMV,NV,Ma,DGR,DVV,AY,CNV,ACDKV,AgS} \cite{LMV}-\cite{AgS} for a sampling of
additional developments.  An alternative but closely related 
physical interpretation of Khovanov homology has
been given in \cite{fiveknots}, where more detailed references can be found concerning 
the Jones polynomial, Khovanov homology, and their relations to mathematical physics.

According to this more recent proposal,
the Jones polynomial can be computed by counting the solutions of
certain elliptic partial differential equations in 4 dimensions, and Khovanov 
homology can then be constructed by counting the solutions of related  
equations in 4+1 dimensions.  The reasoning that led to this proposal relied on 
electric-magnetic duality of $\N=4$ super Yang-Mills theory in four dimensions to 
transform one description that is rather ``quantum'' in nature (being closely related to 
Chern-Simons gauge theory on a bounding three-manifold) to another that is 
``semiclassical'' in the sense that the partition function can be computed just by 
suitably counting the classical solutions of certain differential equations.

Instead of relying on electric-magnetic duality to predict this perhaps mysterious result, 
can we understand it by a direct study of the equations?  This is the goal of the present paper.  
We will gain a reasonable degree of understanding of the Jones polynomial and a good
foundation for understanding Khovanov homology.

\subsection{A Brief Review}\label{review}

\begin{figure}
 \begin{center}
   \includegraphics[width=3.5in]{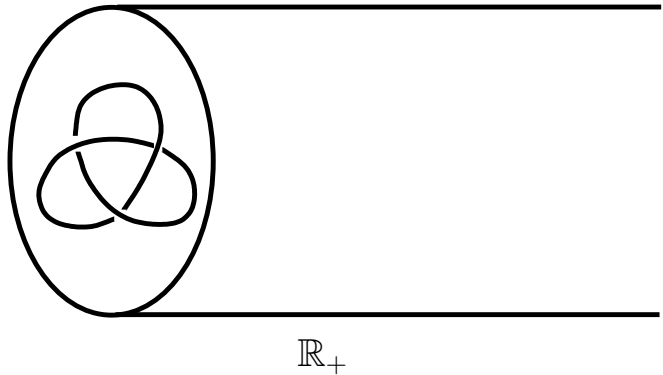}
 \end{center}
\caption{\small  A knot has been placed at the boundary of the four-manifold $\MM=W\times\R_+$. }
 \label{knotboundary}
\end{figure}

The four-dimensional equations in question can be described as follows.  The gauge group is a compact
Lie group\footnote{We reverse notation from \cite{fiveknots}, writing $G^\vee$ for the gauge group in
the Chern-Simons description and $G$ for the gauge group in the dual ``magnetic'' description, on which we focus in this paper.}   $G$.  The fields in the equations are a gauge field $A$ which is a connection
on a $G$-bundle $E\to \MM$, with $\MM$ an oriented Riemannian four-manifold, and another field $\phi$ that is a one-form valued in the adjoint representation of $G$.  The equations, which were first studied in relation to the geometric
Langlands correspondence \cite{KW}, read
\begin{align}\label{bpseqns} (F-\phi\wedge \phi + \t \,\d_A\phi)^+&=0   \cr
                             (F-\phi\wedge\phi-\t^{-1}\d_A\phi  )^- &=0 \cr
                               \d_A\star\phi&=0,                     \end{align}
where  the selfdual and anti-selfdual projections of a two-form
$b$ are denoted $b^\pm$; $\d_A=\d+[A,\,\cdot\,]$ is the gauge-covariant
exterior derivative; $F=\d A+A\wedge A$ is the Yang-Mills field strength;
$\star$ is the Hodge star operator; and $\t$ is a real parameter.
(Actually, $\t$ takes values in $\mathbb{RP}^1=\mathbb{R}\cup\infty$;
for $\t\to 0$ or $\t\to\infty$, one multiples the second equation by $\t$ or
the first by $\t^{-1}$.)  To study knot invariants, one specializes to
$\MM=W\times \R_+$, where $W$ is a three-manifold and $\R_+$ is the half-line $y\geq 0$ (fig. \ref{knotboundary}).

The boundary condition at $y=0$ is
slightly subtle but can be easily described in the absence of knots.  Suppose first that $\t=1$.
Since the boundary condition is local, we can specialize to $W=\R^3$ in describing it.  (In any case,
that is the main example for the present paper.)  Consider a classical solution that is invariant
under translations along $\R^3$ and such that $A$ and the part of $\phi$
normal to the boundary vanish.  The equations then reduce to Nahm's
equations for $\vec\phi$, the part of $\phi$ tangent to the boundary:
\begin{equation}\label{nahm}\frac{\d \vec\phi}{\d y}+\vec\phi\times \vec\phi = 0. \end{equation}
These equations have a singular solution, first introduced by Nahm in his work on monopoles.  Pick
an embedding  $\rho:\frak{su}(2)\to \frak g$ (where $\frak{su}(2)$ and $\frak g$ are the Lie algebras
of $SU(2)$ and $G$, respectively), given by a triple of elements $\vec \ct\in \frak g$ obeying
$[\ct_1,\ct_2]=\ct_3,$ and
cyclic permutations.  Then the solution is
\begin{equation}\label{nahmpole}\vec\phi=\frac{\vec \ct}{y}. \end{equation}
Though any $\rho $ gives a solution, the case we want is that $\rho$ is a principal embedding. For $G=SU(N)$, this means that $\rho$ is
an irreducible embedding of $SU(2)$ in $G$; for any $G$, it means that the raising operator
$\ct_+=\ct_1+i\ct_2$ is a ``regular'' element of the complexified Lie algebra
$\frak g_\C$ (this means that the subalgebra of $\frak g_\C$ that commutes with $\ct_+$ has the minimum possible
dimension).  Then one can define a boundary condition
by allowing precisely those solutions of (\ref{bpseqns}) that can
be approximated for $y\to 0$ by the model solution (\ref{nahmpole})
with the regular Nahm pole. This has
an analog for any $\t$; the starting point, as explained in an appendix, is to set the tangential part
$\vec A$ of the gauge field to be a specific multiple
of $\vec\phi$, so that the equations reduce again to Nahm's equations.

When a link $L\subset W$ is included, this boundary condition is modified along $L$.
A link is simply the union $L=\cup_iK_i$ of disjoint embedded circles $K_i$.
The $K_i$ are labeled by representations $R^\vee_i$ of the
Langlands or GNO dual group $ G^\vee$ to $G$, and in this description
the knots enter the formalism only via the way they enter the boundary
conditions.  Roughly speaking, the modification is made by requiring the
presence of singular BPS monopoles supported along the $K_i$ with magnetic
charges given by the $R^\vee_i$.

The $G$-bundle $E\to \MM$ has an instanton number $P$
defined in the usual way as a multiple of $\int_{\MM}\Tr\,
F\wedge F$.  (The definition of $P$ as a topological
 invariant involves some subtleties that are described in
 \cite{fiveknots}; roughly speaking, the boundary conditions
 at the finite and infinite ends of $\R_+$ give suitable trivializations
 of $E$, enabling one to define the instanton number.\footnote{For
 general $W$ and a general choice of the boundary condition at $y=\infty$,
 $P$ takes values not in $\Z$ but in a certain coset of $\Z$ in $\Q$.})
For each value $n$ of the instanton number, one defines
an integer $a_n$ by ``counting'' (with signs
that are determined by the sign of the fermion
determinant of $\N=4$ super Yang-Mills theory) the number
of solutions of the supersymmetric equations (\ref{bpseqns}) with instanton number $n$.  Then
the partition function of a certain version of twisted $\N=4$ super Yang-Mills theory on $\MM$ is
\begin{equation}\label{partfn} Z(q)=\sum_n a_nq^n, \end{equation}
where the definition of $q$ in terms of parameters of $\N=4$
super Yang-Mills theory was explained in
\cite{fiveknots}.

To get the Jones polynomial and its analogs for
other groups and representations, one specializes to $W=\R^3$ and
takes the boundary condition at $y=\infty$ to be simply $A,\phi\to 0$.  Then for example
for $G^\vee=SU(2)$ and $R^\vee$ the
two-dimensional representation of $SU(2)$, $Z(q)$ is supposed
to become the Jones polynomial.  Since $W=\R^3$ is the case relevant to the Jones
polynomial, it will be the main example in the present paper.  However, many of our
considerations 
apply also  for $W=\R\times C$ where $C$ is a Riemann surface, so we will  consider this case
as well.

A slight generalization of the above-described procedure is to modify the boundary condition at infinity
  so that $A$ and $\phi_y$ vanish but $\vec\phi$ approaches, up to a gauge transformation,
  a specified triple $\vec \ca$ of elements of $\frak t$, the Lie algebra of a maximal torus $T$ of
  $G$. Physically, this means that one takes the vacuum at infinity to be specified by a given point
  on the Coulomb branch. (In the presence of the Nahm pole boundary condition, turning on
  $\phi_y$ or the other two scalars of $\N=4$ super Yang-Mills theory -- called $\sigma,\bar\sigma$
  in \cite{fiveknots} --  would break supersymmetry; so $\vec \ca$ are the only useful Coulomb branch parameters.)   Continuously turning on Coulomb branch parameters should not affect
  the counting of solutions of an elliptic equation, so this procedure should give
  a slightly more general way to compute the Jones polynomial.  To describe the basic solution of
   the equations (\ref{bpseqns}) with a specified choice of $\vec \ca$ at infinity,
  one looks for a solution that still has $A=\phi_y=0$
and is still invariant under translations along $\R^3$, but now obeys $\lim_{y\to\infty}\vec\phi = g \vec \ca g^{-1}$, for some $g\in G$.  The equations still reduce to Nahm's equations (\ref{nahmpole}).
A general theorem \cite{Krontwo} says that for any simple Lie group $G$,
and any specified choice of $\vec \ca$, there is a unique solution of Nahm's equations
with a regular Nahm pole at $y=0$ and the required behavior for
$y\to \infty$.  This solution describes the ground state at the given point on the Coulomb branch
in the absence of any 't Hooft operators on the boundary.

\subsubsection{Lift To Khovanov Homology}\label{lift}

Though our main focus will be to recover the Jones polynomial from this framework, we will also
briefly sketch how Khovanov homology is supposed to arise.
A primary purpose of this is to explain the extent to which the particular
values $\t=\pm 1$ are or are not special, since this will be important
later.

To get Khovanov  homology instead of the Jones polynomial, we are supposed to ``categorify'' the above-described situation, which is just a fancy way to say that we must obtain everything that has been described so far from a theory in one dimension higher.  For this, let $x^1,x^2,x^3$ be local coordinates on $W$ and
decompose $\phi$ as $\phi =\vec\phi\cdot \d\vec x+\phi_y \d y$, where $\phi_y$ is the component of $\phi$ in the $y$ direction.  Categorification is accomplished by introducing a new time coordinate
$x^0$ and replacing $\phi_y$ by the covariant derivative $D/D x^0$.  This replacement makes sense
in that, since $\phi_y$ only appears in (\ref{bpseqns}) inside commutators and covariant derivatives,
the replacement does give a differential equation (rather than a differential operator), now on the five-manifold
$\R\times W\times \R_+$.  Moreover this differential equation, whose details are described in section
5 of \cite{fiveknots}, is elliptic so problems of counting its solutions make sense.\footnote{This equation has also been formulated and some basic properties described in \cite{Haydys}.}

This five-dimensional lift of the four-dimensional equations (\ref{bpseqns}) also has a surprising
four-dimensional symmetry, provided we set $\t=\pm 1$.  The original four-dimensional symmetry
relating the different directions in $\MM=W\times\R_+$ has been
spoiled by the replacement $\phi_y\to D/D x^0$.  But at $\t=\pm 1$,
the five-dimensional equations acquire a new four-dimensional symmetry:
one can replace $\R\times W$ by a general
oriented Riemannian four-manifold $M$, without additional
structure, and formulate these equations on $M\times \R_+$.  For
studying the Jones polynomial, the values $\t=\pm 1$ are not
particularly distinguished; the
counting of solutions of the elliptic equations (\ref{bpseqns}) is independent of $\t$.
Moreover, categorification -- the substitution $\phi_y\to D/D x^0$ -- is not limited to $\t=\pm 1$.
What is special about $\t=\pm 1$ is the four-dimensional symmetry of the categorified
theory, which is likely to have important implications for Khovanov homology and its
analogs on other manifolds.

From a physical point of view, the five-dimensional lift of
the equations (\ref{bpseqns}) are BPS conditions of a certain twisted
version of five-dimensional super Yang-Mills theory, formulated on $\R\times W\times \R_+$; they
describe configurations that are invariant under one of the supercharges, which we will call $Q$.
This operator obeys $Q^2=0$, and the space of supersymmetric
ground states is the same as the cohomology of $Q$.  This is the
candidate for Khovanov homology.  Mathematically,
the five-dimensional equations can be interpreted as Morse theory flow equations,
and the space of
supersymmetric ground states is the analog of Floer homology for this
situation.  Physically, to construct the space of supersymmetric ground states,
one starts with time-independent solutions of the five-dimensional equations --
these are simply the solutions of the  original uncategorified equations (\ref{bpseqns}) in four
dimensions.  Expanding around any one of these solutions, one can construct
an approximate supersymmetric state, and these
furnish a basis for the space of supersymmetric states in the classical
approximation. Then one computes quantum corrections by taking account of
tunneling between classical
vacua; the tunneling events are solutions of the full five-dimensional equations.

From this point of view, the link between Khovanov homology and
the Jones polynomial comes from the fact that the classical solutions that
give a basis for the classical approximation to Khovanov homology are the
same ones that must be counted to compute the Jones polynomial.

\subsection{Methods Used In This Paper}\label{methods}

{\it A priori}, to count the solutions of the nonlinear partial differential equations
(\ref{bpseqns}) is a daunting problem.
Our attempts to simplify this problem are based on three ideas.

\begin{figure}
 \begin{center}
   \includegraphics[width=3.5in]{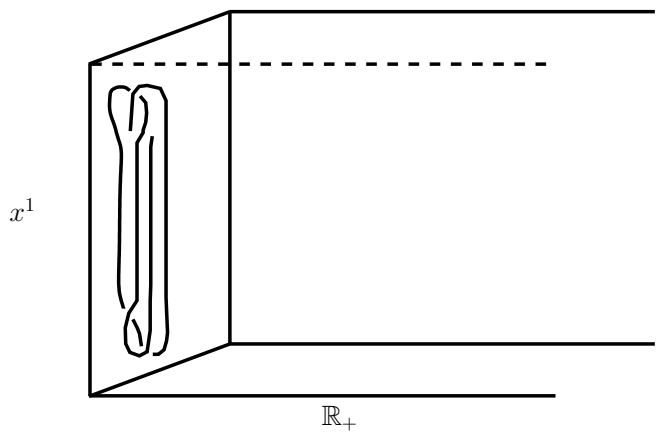}
 \end{center}
\caption{\small   Stretching a knot in one direction -- here taken to be the $x^1$ direction --  to reduce to a situation that
almost everywhere is nearly independent of one coordinate.  After much stretching, the
knot is everywhere nearly independent of $x^1$, except near the finite set of critical values of $x^1$ at which a pair of strands appears or disappears.  (In the figure, these occur only
at the top and bottom.)}
 \label{stretching}
\end{figure}

The first is a standard idea in topological field theory.  We consider knots in $W=\R\times C$, where $C$ (which may be simply $\R^2$) is a two-manifold, and we parametrize $\R$ by $x^1$.  We stretch our knots in the $x^1$ direction, so that except at a few exceptional values of $x^1$ where the number of strands changes, the boundary conditions
are nearly independent of $x^1$ (fig. \ref{stretching}).
We hope that, away from the exceptional values of $x^1$, the solutions
can be approximated by solutions that are independent of $x^1$.  Once one drops $x^1$, the equations
reduce to equations in three dimensions.  The reduced equations preserve more supersymmetry and one may hope to understand their solutions.

\begin{figure}
 \begin{center}
   \includegraphics[width=3in]{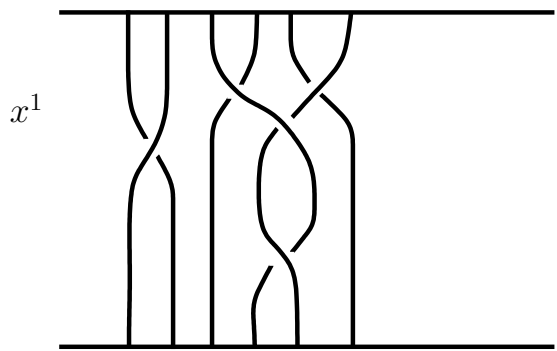}
 \end{center}
\caption{\small  A braid in $I\times C$; by gluing together the top and bottom, one can make
a closed braid in $S^1\times C$.  After much stretching, a braid can be described by adiabatic
evolution in $x^1$, with no exceptional values where this description breaks down.}
 \label{braid}
\end{figure}

After finding the three-dimensional solutions, to recover a four-dimensional picture, we 
have to take
into account an adiabatic variation of the parameters in the three-dimensional equations.
This is because our
knot, even after stretching, is not quite independent of $x^1$.
We also have to consider the jumping that occurs
when the number of strands changes.  Actually, there is an
important special case in which one only has to consider the adiabatic variation of
parameters.  This is the case (fig. \ref{braid}) that $\R\times C$
is replaced by  $S^1\times C$ (or $I\times C$ where $I$ is a closed interval,
though this introduces
questions about boundary conditions) and the link is replaced by a braid.  In this situation,
 one would study not the Jones polynomial but its associated braid group representations,
 which are also of great interest.

\begin{figure}
 \begin{center}
   \includegraphics[width=3.5in]{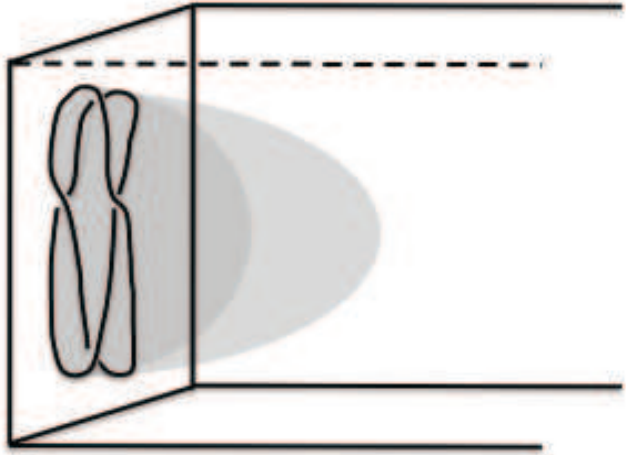}
 \end{center}
\caption{\small  As a knot is stretched  along the boundary, a solution
of the supersymmetric
equations might become delocalized in the $y$ direction, normal to the boundary.  This
is schematically indicated here; the shaded region indicates the spatial extent of a solution
-- that is, of the region over which the chosen solution  deviates significantly from the one that
describes the vacuum in the absence of knots -- and its thickness is proportional to the amount
that the knot has been stretched.}
 \label{failure}
\end{figure}

One important thing to mention about this program is that it is not guaranteed
to work.  As one stretches a knot
in the $x^1$ direction, the solution might simultaneously ``spread'' in the $y$
direction (fig. \ref{failure}) so that even after stretching, the solution might not approach
an $x^1$-independent limit.  In fact, we will find  that this happens under some conditions.
One of our main tasks will be to understand conditions under which the sort of behavior
suggested in fig. \ref{failure} does not occur.

In carrying out the program that we have
just described, we start in section \ref{teqone} at $\t=\pm 1$ because these
are special values for Khovanov homology (as we recalled in section \ref{lift} above)
and also because some simplifications in the three-dimensional equations at $\t=\pm 1$
were already found in section 3.6 of \cite{fiveknots}.

Because we encounter some puzzling phenomena (which we will ultimately understand along
the lines of fig. \ref{failure}), we look for some additional simplifications.  In doing
so, we primarily exploit two ideas.

The first idea is to modify the
boundary conditions to incorporate gauge symmetry breaking.  The basic
idea was already explained at the end of section \ref{review}:
instead of asking for $\vec\phi$ to vanish at infinity,
we ask for $\vec\phi\to g\vec \ca g^{-1}$, where $g\in G$ and
the three components
of $\vec \ca=(\ca_1,\ca_2,\ca_3)$ take values in a Cartan subalgebra of the Lie algebra
$\frak g$ of $G$. (In the more general
case $W=\R\times C$, we would similarly modify the boundary
condition to require that the component of $\phi$ in the $\R$ direction is in a specified conjugacy
class at infinity.)
Continuously changing the boundary conditions in this way should not change the counting
of solutions that leads to the Jones polynomial.  On the other hand, in such counting problems
one often finds that perturbing to a more generic situation can make things easier.  Moreover,
in the present case, taking $\vec \ca$ to be generic reduces the nonabelian gauge theory that
we are studying to an abelian theory at low energies.  If we scale up our knots so that all
relevant directions of the $K_i$ are large compared to $1/|\vec \ca|$, then we can reasonably
hope to find some
sort of effective abelian description of the relevant phenomena.

The second idea that we exploit is perhaps even more obvious. Since the
equations that arise at $\t=1$ with the Nahm pole boundary conditions described above are
rather special, we perturb the value of $\t$ and/or the Nahm pole boundary conditions
to something more
 generic.  This proves to be very fruitful, especially when combined with gauge symmetry breaking.

\subsection{Outline And Results}\label{results}

In section \ref{teqone}, we analyze the three-dimensional reduction of  equations (\ref{bpseqns})
at $\t=1$.
We get an interesting description in terms of Higgs bundles with some additional structure,
but it becomes clear that the program suggested in fig. \ref{stretching} will encounter some
difficulties at $\t=1$.    In section \ref{analog}, we perturb the equations to $\t\not=1$ and
find that this offers a much more promising framework for understanding the Jones polynomial.
The equations for generic $\t$ have surprising and useful relations to a variety of topics
in mathematical physics, including the Bethe equations for an integrable spin system known
as the Gaudin model, and certain special ``degenerate'' conformal blocks of the Virasoro algebra;
the rest  of the paper is based on these relations.
In section \ref{fourd}, we discuss the general framework for constructing braid 
group representations from adiabatic evolution of the parameters governing time-independent 
solutions.   The general framework is a little abstract, but in section \ref{fthimconformal},
we show that in our particular problem, it can be made very concrete using
the free field representation of certain Virasoro conformal blocks.
  In section \ref{brrep}, we implement that idea
in detail.  This finally enables us to understand how the Jones polynomial and the braid
group representations associated to it can be recovered by counting solutions of the 
four-dimensional BPS equations (\ref{bpseqns}).    Section \ref{supermon} is devoted to
describing an effective superpotential for BPS monopoles that can be used to understand some
of the subtle results of sections \ref{teqone} and \ref{analog}.  In section \ref{opbranes}, 
we place some structures encountered in this paper in a wider context of 
mathematical physics.  Three appendices fill in details of
the derivations.

\section{Analysis At \t=1}\label{teqone}

\subsection{Some Preliminaries}\label{preliminaries}

\begin{figure}
 \begin{center}
   \includegraphics[width=3.5in]{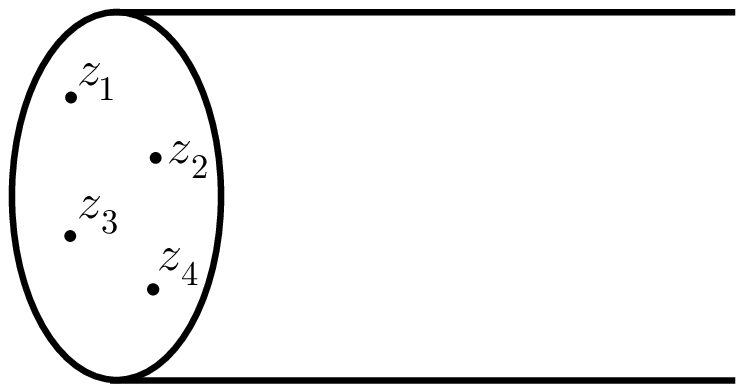}
 \end{center}
\caption{\small  In a time-independent situation, we look for solutions on a three-manifold $M_3=C\times \R_+$,
where $C$ (taken here to be a two-sphere) is a Riemann surface.  Knots are placed at points on the boundary of $M_3$,
labeled here as $z_1,\dots,z_4$.}
 \label{redthree}
\end{figure}

As explained in the introduction,
after stretching a knot along the first factor of $\MM=\R\times C\times\R_+$, we want
to find the solutions that are independent of the first coordinate (which we call $x^1$),
that is the solutions that obey reduced
equations on the three-manifold\footnote{\label{confusion}
To minimize confusion, we note the following.   In this paper, we use two different decompositions of $\MM=\R\times C\times\R_+$  as the product of a three-manifold and a one-manifold: we write
$\MM=W\times \R_+$ with $W=\R\times C$, but also $\MM=\R\times \Mthree$ with
$\Mthree=C\times\R_+$.} $\Mthree=C\times \R_+$.
Here  $C$ is a Riemann surface (which may be simply $\R^2$), and $\R_+$ is the half-line $y\geq 0$. If knots are present, we take their support
to be of the time-independent form  $\R\times z_i\times\{0\}$,  where the $z_i$ are points in $C$, and
$\{0\}$ is the endpoint $y=0$ of $\R_+$.  The picture is sketched in fig. \ref{redthree}.

Solutions that can be derived from three dimensions preserve   more than
the generic amount of supersymmetry -- they preserve four supercharges, to be precise --
and accordingly, as described in section
3.6 of \cite{fiveknots}, their
structure simplifies.  In all cases that we will encounter in the present paper,
the reduced equations
can be usefully described in terms of three differential operators $\D_i$, $i=1,2,3$.
The equations say that the $\D_i$ commute
\begin{equation}\label{commuting}[\D_i,\D_j]=0,\end{equation}
and obey a moment map condition
\begin{equation}\label{momentmap}\sum_{i=1}^3 [\D_i,\D_i^\dagger]=0,\end{equation}
where $\D_i^\dagger $ is the adjoint of $\D_i$ in a natural sense.  The commutativity
constraint (\ref{commuting}) is invariant under complex-valued gauge transformations
$\D_i\to g\D_i g^{-1}$,
where $g$ is a $G_\C$-valued gauge transformation ($G_\C$ is the complexification of
$G$), while the moment map condition (\ref{momentmap})
is only invariant under $G$-valued gauge transformations.  What will make our problem
tractable is that solutions of the combined system of equations modulo $G$-valued gauge
transformations are equivalent to solutions of just the commutativity constraint
(\ref{commuting}) modulo
$G_\C$-valued gauge transformations.  But solutions of the commutativity
constraint modulo complex gauge transformations can be described in terms of
holomorphic quantities, so it is possible to
understand them.

Different instances of this structure vary by the construction of the $\D_i$, which
depends on the choice of $\t$, and also on the boundary conditions that we assume at
$y=0$ and at $y=\infty$.  The most basic case considered in
\cite{fiveknots}
is that $\t=1$  (or $-1$) and the boundary condition is given by a Nahm pole in the
part of $\phi$ tangential to the
boundary.  That boundary condition sets to zero $\phi_y$, the normal part of $\phi$, at $y=0$.
If we also require $\phi_y$ to vanish for $y\to\infty$, then
a simple vanishing  argument shows that in a solution that is independent of $x^1$, $\phi_y$ is identically zero; similarly,
$A_1$, the component of $A$ in the $x^1$ direction, vanishes in a three-dimensional
solution.  Once $\phi_y$ and $A_1$
are set to zero, the equations can be put in the above-described form, as shown in detail
in \cite{fiveknots}, section 3.6,\footnote{Our notation differs from the notation used there by
a relabeling of the coordinates $x^i\to x^{i+1}$ (whose purpose is to make ``room'' for a new
time coordinate $x^0$ upon categorification).  Also, for later convenience we permute the
$\D_i$ in an obvious way.} with
\begin{align}\label{eqform} \D_1 & = \frac{D}{Dx^2}+i\frac{D}{Dx^3}    \cr
                                   \D_2 & = [\phi_2-i\phi_3,\,\cdot\,] \cr
                             \D_3 & = \frac{D}{Dy}-i[\phi_1,\,\cdot\,]  \end{align}
and the moment map condition
\begin{equation}\label{mmtmap}0=
\sum_{i=1}^3[\D_i,\D_i{}^\dagger]=F_{23}-[\phi_2,\phi_3]-D_y\phi_1.\end{equation}
 In writing these formulas, we have simply taken $C=\R^2$
with coordinates $x^2,x^3$.  However, it is helpful to introduce a complex coordinate
$z=x^2+ix^3$, and to write $\phi_2\d x^2+\phi_3\d x^3=\varphi \d z+\bar\varphi \d\bar z$;
also we introduce a complex connection $\A_y=A_y-i\phi_1$ for parallel transport in the $y$
direction and write $\D_y=\d_y+[\A_y,\,\cdot\,]$.  Then we can write
\begin{align}\label{neqform} \D_1 & =2\frac{D}{D\bar z} \cr
                                                              \D_2 & = 2[\varphi,\,\cdot\,]\cr
                                                               \D_3& = \frac{\D}{\D y}. \end{align}
With this way of writing the $\D_i$, they make sense on an arbitrary Riemann surface
$C$, with $D/D\bar z$ understood as the $\bar\partial $ operator and $\varphi$ as
a $(1,0)$-form on $C$.

As one would expect in a geometry that preserves four supercharges, the commutativity constraint
$[\D_i,\D_j]=0$ can be derived from a superpotential, namely
\begin{equation}\label{superone}\W= \frac{1}{4\pi i}\int_{C\times \R_+}\Tr\,\varphi \F_{y\bar z},\end{equation}
with $\F_{y\bar z}=[\D_y,D_{\bar z}]$.  To be more exact, in varying $\W$
to derive the conditions $[\D_i,\D_j]=0$, one
should require the variation of $A$ at $y=0$ to vanish.  Otherwise, the variation of $\W$ contains
additional delta function terms at $y=0$.

The equations $[\D_i,\D_j]=\sum_i[\D_i,\D_i{}^\dagger]=0$ have been called the extended
Bogomolny equations in \cite{KW}; actually,
these equations are a hybrid of the equations of Nahm, Hitchin,
and Bogomolny.  They reduce to Nahm's equations if we drop the dependence on $z$,
to Hitchin's equations if we drop the dependence on $y$, and to the Bogomolny equations if
we set $\varphi=0$. This is not just an analogy: we can borrow standard strategies
from the theory of moduli spaces of Nahm, Hitchin, or Bogomolny equations.

The equation $[\D_1,\D_2]=0$, taken for fixed $y$, defines a Higgs bundle $(E,\varphi)$ in the sense of Hitchin.
The fact that $\D_1$ and $\D_2$ commute with $\D_3$ simply means that the Higgs bundle is independent of $y$, up to a complex-valued gauge transformation.
When we specialize to $C=\R^2\cong \C$, we get a Higgs bundle on $\C$ that can be understood as a Higgs
bundle on $\CP^1=\C\cup \infty$, possibly with a singularity at infinity.  This case will be considered
in section \ref{simvar}.  There is another natural way for singularities of the Higgs bundle to arise.
As in section 6 of \cite{fiveknots}, one may include surface defects supported on
codimension two submanifolds in $\MM$.  Taking these to be  of the form $\R\times q_i \times \R_+$, where the $q_i$ are points in $C$, the construction summarized above still applies and  the Higgs bundles acquire singularities at the points $q_i$.  We consider this situation in section
\ref{mondr}.

Additional structure arises from boundary conditions at $y=0$ and $y=\infty$.
 We will discuss the consequences of the Nahm pole at $y=0$ in section \ref{bc}.
The analogy to Nahm's equations will be useful: we can extract some holomorphic data from the commuting pair $\D_2$,$\D_3$ for any given point in $C$.
 As $\D_2$,$\D_3$ commute with $\D_1$, this data varies holomorphically on $C$, or possibly meromorphically in the presence of singularities.

Finally, an analogy with the Bogomolny equations will help us understand the physical content of our solutions, especially
when we turn on gauge symmetry breaking for $y\to\infty$.
Indeed, the holomorphic data in the commuting pair $\D_1$,$\D_3$ is analogous to
the data which specifies the position of BPS monopoles
in a solution of the Bogomolny equations.

\subsection{The Boundary Condition}\label{bc}
Our next task is to analyze the boundary conditions, first in the absence of singular monopoles.
The model solution (\ref{nahmpole}) at $\t=1$ has a singularity in
$\D_2$ and $\D_3$, but not in $\D_1$, simply because $\D_1$ does not contain
the scalar fields. For $ \frak g = \frak{su}(2)$, with a standard choice of the Lie algebra
elements $\ct$, this solution is explicitly
\begin{equation}\label{burg}
\varphi = \frac{1}{y} \begin{pmatrix} 0 & 1 \cr 0 & 0\end{pmatrix},
\qquad A_{\bar z}=0, \qquad \A_y = \frac{1}{2y} \begin{pmatrix} 1 & 0 \cr 0 & -1\end{pmatrix}.
\end{equation}
We consider the matrices here to act on the fiber of a trivial rank two complex vector bundle
$E\to \R^2\times\R_+$.  For $G=SO(3)$, on a general Riemann surface,
 there could be a global obstruction to defining  $E$ as a rank two bundle and
 one would then consider instead the corresponding adjoint
bundle $\ad(E)$.  Because our considerations will be local along the Riemann surface $C$, a
reformulation in terms of the adjoint representation
does not change much, and we will omit this.

The  solution (\ref{burg}) can be written
\begin{equation}\label{nolf} \varphi= g\varphi_1 g^{-1},
~~ \D_y=g\frac{\d}{\d y}g^{-1} \end{equation}
with
\begin{equation}\label{zolf} \varphi_1=\begin{pmatrix} 0 & 1\cr 0 & 0 \end{pmatrix}\end{equation}
and $g$ a  $G_\C$-valued gauge transformation that is singular at $y=0$
\begin{equation}\label{singgauge}
g =\begin{pmatrix} y^{-1/2} & 0 \cr 0 & y^{1/2} \end{pmatrix}.
\end{equation}
In other words, the solution is obtained by the complex-valued gauge transformation $g$
from a trivial solution $\varphi=\varphi_1,$ $A_{\bar z}=\A_y=0$.

We want to consider solutions which look like the model solution
near $y=0$, up to a gauge transformation.
We would like to express this constraint in terms of the Higgs bundle data
$(E,\varphi)$ away from the boundary.
For that purpose, it is useful to consider the behavior of a local holomorphic section $s$ of the
gauge bundle $E$ that is invariant under parallel transport in the $y$ direction. We will first
do the calculation very explicitly for the model solution, and then identify which
features are valid more generally.

Let $s$ be a section of the
gauge bundle $E$ that obeys
\begin{equation}\label{onurf}\frac{\D s}{\D y}=0.\end{equation}
Since $\D/\D y= g (\d/\d y) g^{-1}$, the solutions of this equation are of the form
$s= gs_0$, where $s_0$ is independent of $y$.  Thus a general solution takes the form
\begin{equation}\label{wurf} s=\begin{pmatrix} a y^{-1/2}\cr b y^{1/2}\end{pmatrix},\end{equation}
with constants $a,b$.
In particular, a generic vector
in $E$, when parallel transported in the $y$ direction to
$y=0$, will blow up as $y^{-1/2}$.  There is a one-dimensional
subspace consisting of solutions of $\D s/\D y=0$ that actually
vanish as $y^{1/2}$ for $y\to 0$.  This subspace is simply characterized
by the condition $a=0$.

We write $E_y$ for the restriction of $E\to C\times \R_+$ to $C\times \{y\}$ for any fixed
$y>0$.  Parallel transport using $\D_y$ gives a natural identification of the $E_y$
for all $y$, and we write simply $E$ for $E_y$, regarded as a bundle over $C$.
Similarly, the restriction of $\varphi$ to $C\times y$ is independent of $y$, up to parallel
transport by $\D_y$.  So by restriction to $C\times y$, we get a Higgs bundle
$(E,\varphi)\to C$.

The ``small'' sections of $E$ -- the solutions $s$ of $\D s/\D y=0$ that vanish for $y\to 0$
-- generate a rank one sub-bundle $\L\subset E$.
In the model solution, it is simply the sub-bundle of sections of $E$ of
the form $\begin{pmatrix}0\cr b\end{pmatrix}$.   $\L$ is a holomorphic sub-bundle of $E\to C$;
a section of $\L$ is holomorphic if it is annihilated by $\D_1=2\,D/D\bar z$.
Concretely, in the model solution, a section $\begin{pmatrix}0\cr b\end{pmatrix}$ of $\L$ is
holomorphic if $b$ is a holomorphic function of $z$.

The fundamental reason that $\L$ is
holomorphic is that, as $[\D_2,\D_3]=0$, we can ask for a small solution of $\D_3s=0$
to also be annihilated by $\D_2=2D/D\bar z$.   However, we cannot also ask for a small solution to
be annihilated by $\varphi=\D_1/2$.  This is clear from the above formulas; in the model
solution, $\varphi$
annihilates a vector if the bottom component vanishes, not if the top component vanishes.

To measure the failure of $\L$ to be $\varphi$-invariant, we can proceed as follows.
If $\L$ is a trivial line bundle, which will be the case 
in our applications, then we can pick a section $s$ of $\L$ that is everywhere nonzero.
For our model solution, we just pick $s=\begin{pmatrix} 0 \cr 1\end{pmatrix}$.  Then we define
\begin{equation}\label{smallexp}  \kappa= s\wedge \varphi s.  \end{equation}
For the model solution, we see that $\kappa=1$, and in particular 
$\kappa$ is everywhere nonzero.
Nonvanishing of $\kappa$ means that $\varphi s $ is not a multiple of $s$, so $\L$ is not invariant
under multiplication by $\varphi$.
(If $\R^2$ is replaced by a general Riemann surface $C$, it might be impossible to pick an $s$
that is globally nonzero, but one can still pick a local section $s$ of $\L$ and measure the
failure of $\L$ to be
$\varphi$-invariant by computing $\kappa=s\wedge \varphi s$. Whether $\kappa$ vanishes
does not depend on the choice of $s$ as long as it is nonzero.)

Three basic properties of $s$ which held for the model solution  remain true for any solution
with a regular Nahm pole,
since they are unaffected by the subleading behavior of $\D_y$ and $\varphi$
as $y \to 0$:
\begin{itemize} \item A generic vector in $E$, when parallel transported in the $y$ direction to
$y=0$, will blow up as $y^{-1/2}$. \item  The sections $s$ that under parallel transport to
$y=0$ actually vanish span a rank 1 holomorphic sub-bundle  $\L\subset E$.  \item
Finally, we cannot also ask for a small section to
be annihilated by $\varphi=\D_1/2$. On the contrary, in a solution that can be approximated
by the model solution near $y=0$, $\varphi$ does not annihilate a small section at any point in 
$C$.\end{itemize}

Consequently, a solution of the full equations with a regular Nahm pole for $y\to 0$
gives not just a Higgs bundle $(E,\varphi)$.  Rather, there is some
additional structure: $E$ is endowed with a holomorphic line
 sub-bundle $\L$ which is nowhere stabilized by $\varphi$. On $\R^2$, $\L$ is inevitably trivial,
so we can
simply say
that an everywhere  non-zero
holomorphic section
$s$ of $E$ exists with $s \wedge \varphi s =1$.
Notice that this  condition is far from sufficient to determine $\varphi$.
If $s=\begin{pmatrix}0\cr 1\end{pmatrix}$,
then the condition $s\wedge \varphi s=1$ fixes the upper right matrix
element of $\varphi$ and puts
no condition on the others.  The reason that only one matrix element of $\varphi$ is fixed
in terms of $s$ is that  when we make a gauge transformation
that
behaves like (\ref{singgauge}) for
$y\to 0$, $\varphi$ acquires a singularity that only depends on its upper right matrix
element.  In an appropriate situation (on a Riemann surface $C$ of
higher genus, or on $\R^2$ in the presence of singular monopoles, as introduced shortly), there
can be a nontrivial moduli space of triples $(E,\varphi,\L)$ with $E$ and $\L$ fixed and only
$\varphi$ varying.  This is explained in section \ref{opbranes}.

For any triple $(E,\varphi,\L)$, we expect that the full system of equations can be solved
by a complex gauge transformation.   Suppose
we are given a $y$-independent Higgs bundle $(E,\varphi)$, and a holomorphic
sub-bundle $\L\subset E$
which is nowhere stabilized by $\varphi$.
In a basis of $E$ given by $\varphi s$ and $s$,
 we make the gauge transformation (\ref{singgauge}).
This will reproduce the Nahm pole singularity at $y=0$, but generically it will not give
a solution of the moment map condition (\ref{mmtmap}).   By further making a smooth
complex gauge transformation, one can hope to get a solution of the moment map condition.

\subsubsection{Adding Singular Monopoles}\label{addsing}

In order to add singular monopoles at the boundary, we need to replace the Nahm model
solution with a more general singular solution,
given in \cite{fiveknots}, section 3.6.
Let us consider the case of a single singular monopole, located at $y=z=0$.
The model solution has the same singularity as before for $y\to 0$ at $z\not=0$,
but has a more complicated form near $y=z=0$.
It can be obtained by a complex-valued gauge transformation, described explicitly in
\cite{fiveknots},  from a solution of the commutativity constraint with
\begin{equation}\label{solcom} \varphi=
\begin{pmatrix} 0 & z^k\cr 0 & 0 \end{pmatrix}, ~~A_z=\A_y=0.
\end{equation}
To express this in our present language, $\L$ is still spanned by
sections that in the gauge (\ref{solcom}) are multiples of
$s=\begin{pmatrix} 0 \cr 1 \end{pmatrix}$; this is because $g$ has
similar behavior as before for $y\to 0$.  We now have
\begin{equation} s\wedge\varphi s = z^k.\end{equation}
The zero of $s\wedge\varphi s$ is interpreted as the position of the singular
monopole, and its degree is the charge.  This interpretation suggests
immediately what a solution with several
singular monopoles should mean.  We consider a solution described by a Higgs
bundle $(E,\varphi)$ with
a sub-bundle $\L\subset E$ that is generically not $\varphi$-invariant.  If
\begin{equation}\label{olcomtwo} s\wedge \varphi s =\prod_{a=1}^s(z-z_a)^{k_a},\end{equation}
then we say that the solution has singular monopoles of charges $k_a$ at the locations $z_a$.
One hopes to be able to prove that given such data, there is a unique solution whose singularity
near each $z=z_a$ agrees with that of the singular model solution.

Though we mainly consider $\frak{su}(2)$ in the present paper, we can
readily generalize these statements to a more general Lie algebra. For simplicity, take
$G=SU(N)$ and view $E$ as a complex vector bundle of rank $N$.
Consider a Nahm pole based on the principal
embedding of $\frak{su}(2) \to \frak{su}(N)$. The eigenspaces of $\ct_3$
(in the fundamental $n$-dimensional representation of $\frak{su}(N)$) are one-dimensional,
and we have a line bundle $\L$ defined by sections which decrease
as fast as possible as $y \to 0$. The only constraint on $\L$ is that, away from the positions
of singular monopoles,
\begin{equation}
 \L \oplus \varphi \L \oplus  \cdots \oplus \varphi^{n -1} \L = E
\end{equation}
When we specialize to $C=\R^2\cong \C$, we can define the line sub-bundle $\L$ by a specific
section $s$ of $E$, defined up to rescaling, such that the sections
$s, \varphi s, \cdots,\varphi^{N-1}s$ are linearly independent.  This constraint is relaxed
at the location of singular monopoles, in a way which depends on their charges.  An equivalent
description in terms of zeroes of matrix elements of $\varphi$ is given in \cite{fiveknots}, eqn.
(3.59).

This can be extended to the case of a general Nahm pole at $y=0$, not
necessarily associated to a principal embedding of $\frak{su}(2)$.  (This extension will
not be studied in the present paper.)  In general,
for $y\to 0$, there are local sections growing as either integer or half-integer
powers of $y$.  
We can define a flag of holomorphic sub-bundles $E_n$ of $E$ by looking
at local sections which grow at most as a given power $y^{(n-N)/2}$
as $y \to 0$. Clearly, $E_{n+1} \subset E_n$; also  $E_0=E$, and $E_n=0$ for large enough $n$. Upon rescaling by $y^{-(n-N)/2}$, a
generic vector in $E_n$ has a finite limit as $y \to 0$,
and this limit is an eigenvector of $\ct_3$ with eigenvalue $(N-n)/2$. The kernel of this
map is $E_{n+1}$; hence the $y \to 0$ limit identifies the
quotient spaces $E_n/E_{n+1}$ with the eigenspaces of $\ct_3$.
Multiplication by $\varphi$ gives maps $E_n \to E_{n-2}$. As
$\varphi \sim \frac{\ct_+}{y}$, the holomorphic maps
$\phi_n : E_n/E_{n+1} \to E_{n-2}/E_{n-1}$ can be identified with
the action of $\ct_+$ on the eigenspaces of $\ct_3$.

\subsection{Solutions Without Symmetry Breaking}\label{solutions}

Now we want to use the ideas that have just been described to determine
some moduli spaces of solutions of the supersymmetric
equations (\ref{commuting}), (\ref{momentmap}). We will do this
for $G=SO(3)$, so that the dual group whose representations label
the singular monopoles is $G^\vee=SU(2)$.

First we work
at the origin of the Coulomb branch; this means that we consider solutions such that the scalar
fields $\vec\phi$ vanish for $y\to\infty$.  In particular, $\varphi$ must vanish at infinity.
Since the equation $\D_y\varphi=0$ means that the conjugacy class of
$\varphi$ is $y$-independent,
$\varphi$ can only vanish at infinity if it is everywhere nilpotent.  This means
that by a complex gauge
transformation, we can make $\varphi$ upper triangular:
\begin{equation}\label{uptri}\varphi=\begin{pmatrix} 0 & p(z) \cr 0 & 0 \end{pmatrix},\end{equation}
with some polynomial $p(z)$.  We cannot, however, put $s$ in a standard form at the same time.
So we simply take
\begin{equation}\label{nuptrix}s=\begin{pmatrix} P(z)\cr Q(z)\end{pmatrix},\end{equation}
with polynomials $P,$ $Q$.  Without changing $\varphi$, we can make
an upper triangular gauge transformation, shifting $P$ by a polynomial multiple of $Q$:
\begin{equation}\label{uptrix} P(z)\to P(z)+U(z) Q(z).\end{equation}
This is the only freedom, apart from a rescaling of $s$ by a complex constant.

Now let us ask how we can pick $p, $ $Q$, and $P$ to describe a configuration
with singular monopoles
of charges $k_a$ at the points $z_a$ in the boundary. Since $s\wedge \varphi s = pQ^2$,
the condition
that we need is
\begin{equation}\label{factors}
p(z) Q(z)^2 = \prod_a(z - z_a)^{k_a} := K(z).
\end{equation}
If the singular monopoles all have minimal charge, $k_a=1$, we can
only obey this with $p=K$ and $Q=\pm 1$.  Then we can set $P(z)=0$ by a
transformation (\ref{uptrix}), and changing the sign of $Q$
multiplies $s$ by an inessential constant. So at the origin of the Coulomb branch,
there is a unique solution with singular monopoles
of specified locations and minimal magnetic charge.

This fact is actually an obstruction to the program that was described in section \ref{methods} (see
fig. 2).
The case of singular monopoles of minimal charge is supposed to be dual to the Jones polynomial,
which is the invariant computed in Chern-Simons gauge theory for $G^\vee=SU(2)$ with all knots
labeled by the two-dimensional representation of $SU(2)$.  If there is only one solution, this means
that the physical Hilbert space associated to this problem is one-dimensional.  The Jones representations of the braid group would then be of rank 1.  This is certainly not the case.
We must be running into some version of the problem that was indicated in fig. 4.  We will
get a clearer picture of what is happening after including symmetry breaking in section
\ref{symbr}, and  after deforming to $\t\not=1$ in section \ref{analog}.  This will ultimately enable us to circumvent the obstacle just described.

What happens if some singular
monopoles have non-minimal charge?  In this case, we encounter moduli spaces of solutions.
In general we solve (\ref{factors}) by
\begin{equation}\label{formsum} p(z)=
\prod_a(z-z_a)^{m_a},~~ Q=\prod_a(z-z_a)^{r_a},\end{equation}
with
\begin{equation}\label{tonf}m_a+2r_a=k_a.\end{equation}
For $Q$ as in (\ref{formsum}), we cannot set $P$ to 1 by a transformation (\ref{uptrix});
rather, the values
of $P$ and its first $r_a-1$ derivatives are invariants at each point $z=z_a$.  So there are
$r_a$ moduli associated to each $z_a$, where the
possible values of $r_a$ are $\{0,1,\dots, [k_a/2]\}$.
We will argue beginning in section \ref{symbr} and in most detail in
section \ref{supermon} that the  moduli represent the positions in the
$y$ direction of some smooth BPS monopoles, together with some conjugate angles.

In implementing the program described in section \ref{methods},
if some knots are labeled by integers $k_a>1$ -- in other words, by representations
of $G^\vee=SU(2)$ of dimension $k_a+1>2$ --
one would have to  handle the evolution of  the four-dimensional
solution as a path in the
moduli space of three-dimensional solutions.  (This would be done by quantizing the
moduli space to get an appropriate space of physical states, in which the evolution
would take place.)   Instead of following that route,
we will perturb the equations that we have just analyzed to more
generic ones, with symmetry breaking at infinity or with $\t\not=1$.
This will eventually reduce all of our moduli spaces to finite collections of points.

\subsection{``Real'' Symmetry Breaking}\label{symbr}

Our first approach to getting a clearer understanding of the solutions -- and a more
useful reduction
to three dimensions -- will be to move on the Coulomb branch. The basic idea was
already explained
at the end of section \ref{review}.  We fix some constant, nonzero expectation values $\vec \ca$
for the tangential scalar fields $\vec \phi$, and consider only solutions of the supersymmetric equations
 (\ref{bpseqns}) such that $\lim_{y\to \infty}\vec\phi=g \vec \ca g^{-1}$, for some $g\in G$.

From our present point of view, we want to analyze the reduced three-dimensional equations in the context of symmetry breaking.  The reduction splits off the $x^1$ direction, so we write  $\vec\phi\cdot \vec \d x=\phi_1\,\d x^1+\varphi\, \d z +\bar \varphi\,\d \bar z$.  The effects of an expectation value for $\phi_1$
or for $\varphi$ will be quite different.
We first consider the case of turning on $\phi_1$ only.  As we will see, this has the effect of
making the solutions somewhat more physically transparent, without adding or removing solutions.

For $G=SO(3)$,   we can  pick a gauge where $\phi_1$ is constant at large $y$.
This means that  for large $y$, $\A_y=A_y-i\phi_1$ approaches a constant matrix at infinity,
say such that
\begin{equation}\label{gauga}
\A_y = \begin{pmatrix} \ca_1 & 0 \cr 0 & -\ca_1\end{pmatrix}.
\end{equation}
with $\ca_1>0$.
This expectation value breaks $SO(3)$ to $U(1)$.  We pick a generator of the unbroken
$U(1)$, normalized so that the off-diagonal components of an adjoint-valued field, which are
the fields of minimum electric charge for $G=SO(3)$, have
charges $\pm 1$:
\begin{equation}\mathcal Q=\begin{pmatrix}-1/2 & 0 \cr 0 & 1/2\end{pmatrix}\end{equation}
The choice of sign will be convenient.

The equation
$\D_y\varphi=0$ implies that the
lower-triangular component of $\varphi$ must
be zero, since otherwise it would grow exponentially fast at infinity.
The diagonal component must also be zero, since otherwise $\varphi$ would have a nonzero
limit at infinity.
Hence symmetry breaking
provides a natural frame in which $\varphi$ is strictly
upper-triangular for $y\to\infty$, just as the boundary
condition did for $y\to 0$.
  But $\varphi$  decays exponentially fast at large $y$.

In the low energy effective $U(1)$ theory, a field of electric charge 1 is a section of a line
bundle that we will call $\M$.  The first Chern class of $\M$, integrated over the plane
given by $y=y_0$, for some large constant $y_0$, is an integer that we will call the magnetic
charge $\m$.   The upper triangular matrix element of $\varphi$
\begin{equation}\label{tomigo}\varphi\sim \begin{pmatrix} 0 & \tilde p \cr 0 & 0 \end{pmatrix}
\end{equation}
is a section of $\M^{-1}$, and hence the magnetic charge, which is minus the first Chern class
of $\M^{-1}$, is minus the number of zeroes of $\tilde p$.  This is the same as minus the number of
zeroes of $p$ (defined in (\ref{uptri})), which is equivalent to $\tilde p$ by a complex gauge transformation.  So in
the notation of (\ref{tonf}), the magnetic charge is
\begin{equation}\label{magch} \m = -\sum_a m_a=\sum_a(-k_a+2r_a).\end{equation}
This means that for a given configuration of singular monopoles or 't Hooft operators at $y=0$,
the smallest possible value of $\m$ is $-\sum_ak_a$.  Every time that we add a zero to $Q$,
$\m$ increases by 2.

We propose that this fact can be interpreted in terms of smooth BPS monopoles.
If we simply set $\varphi=0$, the extended Bogomolny equations that we have been
studying reduce to the usual Bogomolny equations.  Far from the
boundary, the Bogomolny equations are a very good approximation, since $\varphi$ is so small.
The Bogomolny equations on $\R^3$ admit smooth monopole solutions of charge $2$.
(We measure magnetic charge in units such that the minimum magnetic charge allowed
by Dirac quantization is $\pm 1$.
This is also the magnetic charge of a minimum charge  singular monopole or 't Hooft
operator, but smooth monopoles have even charge.)
Our proposal is that if $Q$ has a zero of order $r_a$ at $z=z_a$, then there are $r_a$
smooth monopoles located at $z=z_a$.  This statement has a precise meaning only if
the monopoles are located at very large $y$, where the extended Bogomolny equations reduce
to the ordinary ones, and have smooth monopoles as solutions.  However, we have found
that if $Q$ has a zero of order $r_a$ at $z=z_a$, then there are precisely $r_a$ complex
moduli associated to the point $z=z_a$ in the complex $z$-plane.  We propose that these
moduli are the positions in the $y$ direction of $r_a$ smooth monopoles that are located
at $z=z_a$, along with conjugate angles.  Again, the precise meaning of this statement holds
when the $y$ positions in question are large.  We will make this interpretation quantitative in
section \ref{supermon}, but for now we consider qualitative arguments.

A first motivation for this proposal comes from $\m$; we interpret the formula (\ref{magch}) as the
sum of the magnetic charges of the singular monopoles at the boundary ($-\sum_ak_a$)
plus a contribution of 2 for every smooth monopole  (contributing $2\sum_ar_a$ in toto).
The fact that the charge of the singular monopoles is negative is worthy of note.  Prior to
symmetry breaking, the charge of the singular monopoles does not have a meaningful sign
-- it is dual to a representation of $G^\vee=SU(2)$ -- but after symmetry breaking, this sign
makes sense and with our normalization it is negative.

To learn more, we recall some facts about the Bogomolny equations.
For the Bogomolny equations on $\R^3$, the holomorphic data are the commuting
operators $\D_1$ and $\D_3$, defined exactly as in eqn. (\ref{neqform}).
Localized  monopole solutions of the Bogomolny equations manifest themselves in the
holomorphic data through
``bound states'' in the parallel transport by $\D_3$, i.e. through normalizable solutions to
\begin{equation}\label{zonurf}\frac{\D s}{\D y}=0.\end{equation}
Such solutions only appear at specific positions in the $z$-plane, which are interpreted
as the $z$ values of the monopole locations, because generically, if we pick $s$ to
decay exponentially for $y\to -\infty$, it will grow exponentially for $y\to +\infty$.

In our setup, we are limited to $y\geq 0$, and an analogous normalizable
solution exists precisely if
the ``small'' section $s$ that vanishes for $y\to 0$  also decays exponentially at large $y$.
If $\varphi$ also decays exponentially as well,  then $s\wedge \varphi s$ vanishes
for $y\to\infty$ and hence for all $y$.
So sections $s$ that vanish at both ends can arise only
 only at zeroes of $K(z)$.  Indeed, if
 \begin{equation}\label{sinfty}s=\begin{pmatrix} \tilde P\cr \tilde Q\end{pmatrix}\end{equation}
 near infinity, then given the form of  (\ref{gauga}), vanishing of $s$ for $y\to \infty$
is equivalent to $\tilde Q=0$.   Zeroes of $\tilde Q$ are the same as zeroes of $Q$.  
(Indeed,   $s\wedge \varphi s$ is
independent of $y$; its zeroes are the zeroes of $Q$ if $y$ is small or of $\tilde Q$ if
$y$ is large.)
So the relation
to the ordinary Bogomolny equations does indeed suggest that there are $r_a$ smooth
monopoles at each point $z=z_a$ at which $Q$ vanishes.
The position of the smooth monopoles in the $y$ direction
should be encoded in the values of $P$ and its first $r_a-1$ derivatives.
We will make this picture more precise in section \ref{supermon}; for now we simply observe that
the normalizable small section $s(z_a)$ behaves as $P(z_a) e^{ -m_1 y}$,
which suggests that increasing $|P(z_a)|$ moves the smooth monopoles
towards large $y$.

The limit $P(z_a) \to 0$ is somewhat singular, as $s$ is supposed
to be everywhere non-zero. The physical picture suggests that
the limit will correspond to a monopole bubbling situation, where
the smooth monopole is pushed to the boundary, and screens the
singular monopole's charge.  Monopole bubbling is a phenomenon (originally described in
\cite{Kronheimer} and rediscovered in section 10.2 of \cite{KW}) in which an 't Hooft operator absorbs
a smooth BPS monopole, lowering the magnitude of its magnetic charge.

\subsection{``Complex'' Symmetry Breaking}\label{simvar}

Now we consider the case of symmetry breaking in $\varphi$.  Suppose, for $G=SO(3)$, that the
eigenvalues of $\varphi$ for $y\to\infty$ are $\pm \ca$.  Then, as $\varphi$ commutes with
$\D_y$, its eigenvalues are $\pm \ca$ everywhere.  By a complex gauge transformation, we can reduce to the case that
$\varphi$ is a constant diagonal matrix:
\begin{equation}\label{constdiag}\varphi=\begin{pmatrix} \ca & 0 \cr 0 & -\ca \end{pmatrix}.\end{equation}
Writing  as usual $s=\begin{pmatrix}P \cr Q\end{pmatrix}$ for the small section,
we find $s\wedge \varphi s = 2\ca  PQ$.  After putting $\varphi$ in the form (\ref{constdiag}),
we can still make a gauge transformation by $\mathrm{diag}(\lambda,\lambda^{-1})$,
mapping
\begin{equation}\label{omog}P\to \lambda P,~~Q\to\lambda^{-1} Q.\end{equation}
In the presence of singular monopoles of charges $k_a$ located at $z=z_a$ (and $y=0$),
the condition we want to satisfy is $s\wedge\varphi s=\prod_{a=1}^\ss (z-z_a)^{k_a}:=K(z)$.
This becomes
\begin{equation}\label{counting}
2\ca P Q  = K(z).
\end{equation}
Solutions of these equations  are associated to
factorizations of $K(z)$ and (modulo a transformation (\ref{omog})) have no moduli.

In the case of $\ss$ boundary 't Hooft
operators that all have $k_a=1$, corresponding to minimum magnetic
charge, the number of solutions is precisely $2^\ss$.  The solutions
correspond simply to  the possible ways to distribute the
factors of $K(z)$ between $P$ and $Q$.   The number $2^\ss$ has a natural interpretation.
On the Coulomb branch, we might expect a minimum charge 't Hooft operator to have two
possible  states,
with its magnetic charge being aligned or anti-aligned with the symmetry breaking.  The two
states correspond to a zero in $P$ or a zero in $Q$.  In the dual description by Chern-Simons
theory with Wilson operators, a minimum charge 't Hooft operator corresponds to a Wilson
operator in the two-dimensional representation of $SU(2)$; such an operator again
represents two quantum states, with positive or negative electric charge along the axis
of symmetry breaking.

To confirm that the $2^\ss$ solutions correspond to two possible choices of the magnetic
charge for each 't Hooft operator, let us compute the magnetic charges of these solutions.
Suppose that $P$ is of degree $\ss_1$ and $Q$ of degree $\ss_2$, where $\ss_1+\ss_2=\ss$.
The ratio $P/Q$ does not depend on the normalization of the small section $s$.  This ratio
has electric charge $-1$ in the low energy abelian gauge theory; it is a section of the
line bundle $\M^{-1}$.  On the other hand, concretely, $P/Q$ has $\ss_1$ zeroes and $\ss_2$ poles
on the $z$-plane.  ($P/Q$ has neither a pole nor a zero at $z=\infty$, when understood as a section
of $\M^{-1}$.  In fact,
by a complex gauge transformation, the solution can be put near $z=\infty$ in
the  form of the original Nahm pole solution (\ref{nahmpole}),
and in particular is independent of $z$.  This trivializes
$\M$ near $z=\infty$ and makes $P/Q$ independent of $z$.)
So the line bundle $\M$ has degree $\ss_2-\ss_1$ and hence
\begin{equation}\label{melf}\m = \ss_2-\ss_1.\end{equation}
From this, we see that the 't Hooft operator at $z=z_a$ contributes either $-1$ or $+1$
to $\m$, depending on whether we place the factor of $z-z_a$ in $P$ or in $Q$.

To get some insight into why there are more solutions for $\ca\not=0$ than there are
for $\ca=0$, consider a slightly more general ansatz for $\varphi$:
\begin{equation}\label{pony}\varphi=\begin{pmatrix} \ca & p(z)\cr 0 & -\ca\end{pmatrix}.
\end{equation}
For $\ca\not=0$, the polynomial $p(z)$ can be removed by an upper
triangular gauge transformation, but now there is a smooth limit for $\ca=0$.
 The condition $s\wedge \varphi s=K$ becomes
\begin{equation}\label{turbo}  2\ca PQ+pQ^2 = \prod_{a=1}^\ss(z-z_a).\end{equation}
For $\ca\not=0$, $p$ is irrelevant, as it can be eliminated by $P\to P-pQ/2\ca$.
For $\ca\not=0$, the equation (\ref{turbo}) has $2^\ss$ solutions (modulo a complex
gauge transformation that preserves the form of $\varphi$), corresponding to factorizations of
$K(z)$ as $(2\ca P+pQ)Q$.   But all of these solutions have
$P\sim 1/\ca$ except for the one solution with $Q=1$ (and $P=0$) that we found already
in section \ref{solutions}.

We interpret this as follows.
In the presence of a minimum charge 't Hooft operator at a boundary point $z=z_a$, there is
always a magnetic charge $-1$ localized near the boundary.
The magnetic charge that is localized near the boundary is the same for all solutions because
the form of the solution near the boundary is always given by a standard model solution (the
one described in section 3.6 of \cite{fiveknots}),
independent of everything else.
In the case of a solution
of (\ref{counting}) with $Q(z_a)=0$ and (therefore)
$P(z_a)\not=0$, there is in addition a smooth BPS monopole, with magnetic charge $2$, located at
$z=z_a$ and at a value of $y$ that depends on $P(z_a)$. So the total magnetic charge
associated to $z=z_a$ is $-1+2=1$.   As $\ca\to 0$, $P(z_a)\to\infty$ and
the smooth monopole disappears to $y=\infty$.  In this way, all of the $2^\ss$ solutions become
equivalent for $\ca\to 0$, even though they are different for $\ca\not=0$.  As always, such a
description in terms of smooth BPS monopoles is only precise in the limit that the monopoles
are located at large values of $y$, so that the Bogomolny equations are a good approximation
near their positions.  As we explain most fully in section \ref{supermon}, this is  the case
exactly when $\ca$ is very small (compared to the inverse distances $1/(z_a-z_b)$ between
the 't Hooft operators) so
in particular the description by BPS monopoles becomes precise for $\ca\to 0$.

In the opposite case that $\ca$ is large
compared to the inverse distances, the symmetry breaking is strong and the different
singular monopoles on the boundary are so far separated that they do not significantly influence
each other.  In this case, the essential statement is simply that a single 't Hooft operator of minimal
charge, in the presence of symmetry breaking with $\ca\not=0$,
has two possible states, in which it looks like
an 't Hooft operator  of charge 1 or $-1$ in the effective
low energy theory.  Given this,
a system of $\ss$ widely
separated 't Hooft operators of minimal charge naturally has $2^\ss$ possible states.

\subsubsection{Implications}\label{physhilb}

The counting of solutions for $\ca\not=0$
circumvents a difficulty that we encountered in section \ref{solutions}.
With $2^\ss$ classical solutions,  the physical Hilbert
space in the presence of $\ss$ 't Hooft  operators of minimal charge
will have dimension $2^\ss$.  These $2^\ss$ states are potentially
dual to the states of $\ss$ Wilson operators labeled by the two-dimensional representation
of $SU(2)$.  And the representations of the braid group on $\ss$ strands
that are associated to the Jones
polynomial can certainly be realized in a vector space of dimension $2^\ss$.

Given this, it is reasonable to expect that the stretching strategy sketched in fig. \ref{stretching}
of section \ref{methods} can work if $\ca$ is generic, but that for $\ca=0$, an attempt to simplify
a classical  solution by stretching in one direction leads to behavior
along the lines suggested in fig. \ref{failure}.

More generally, in the presence of an 't Hooft operator on the boundary of
charge $k_a$, we would hope to find $k_a+1$ states of magnetic charge
$-k_a,-k_a+2,-k_a+4,\dots,k_a$.  Indeed, such an 't Hooft operator in $SO(3)$ gauge theory
is dual to a Wilson operator in $SU(2)$ gauge theory associated to the representation of
spin $k_a/2$; this representation has dimension $k_a+1$, and its weights are as indicated.
We get the right number of solutions
with the right magnetic charges if in
solving eqn. (\ref{counting}), we allow arbitrary factorizations of $K(z)$
with the zeroes split between $P$ and $Q$ in an arbitrary fashion.  Unfortunately,
we do not have a simple interpretation of the factorizations for which both $P$ and $Q$ vanish
at $z=z_a$.  After all, $s=\begin{pmatrix}P \cr Q\end{pmatrix}$ is supposed to be everywhere
nonzero.  It seems possible that the factorizations in which $P$ and $Q$ have a common
zero should be interpreted in terms of monopole bubbling,
that is,  as solutions of the equations in the presence of an 't Hooft operator of reduced charge
$k_a-2$, $k_a-4$, etc.   To develop this idea in detail
is beyond the scope of the present paper.

Going back to the simpler case that all $k_a$ are equal to 1, since the space of physical
states has a promising dimension $2^\ss$, the next step could be to try to compute the
action of the braid group and to extract the Jones polynomial.  Before attempting that,
we will describe a further deformation of the problem, which turns out to be illuminating.
Among other things, with this deformation, we will get a nice behavior regardless of the charges
of the 't Hooft operators.

\section{Analysis At General $\t$}\label{analog}

\subsection{Some Basics}\label{motivation}

Still searching for a useful description for 't Hooft operators of arbitrary magnetic charges,
we consider the possibility of deforming the equation that we are trying to solve to one that might
behave more conveniently upon stretching along one direction.

A strong
hint that there is a useful deformation comes by considering the relation of the
extended Bogomolny
equations (\ref{eqform}), (\ref{mmtmap}) to Hitchin's equations \cite{Hitchin}.
Hitchin's equations are equations on an oriented two-manifold $C$ for a
connection $A$ and an adjoint-valued one-form $\phi$:
\begin{align}\label{hitcheq} F-\phi\wedge\phi & = 0 \cr \d_A\star\phi& = 0 \cr
\d_A\phi&=0.\end{align}
The moduli space of solutions of these equations, up to gauge transformation, is a hyper-Kahler
manifold $\M_H(G,C)$.  As a hyper-Kahler manifold, $\M_H(G,C)$ has a family of
complex structures
parametrized by a copy of $\Bbb{CP}^1$.  The complex structures of $\M_H(G,C)$ have a simple
interpretation. They correspond to ways of splitting the three real equations in (\ref{hitcheq})
into two parts: two  real equations, which can be combined to a single complex equation; and
a third real equation that is
 ``orthogonal'' to the first two.  The complex equation then describes
the holomorphic data parametrizing $\M_H(G,C)$ in one of its complex structures, and
the remaining
real equation is a moment map condition.

An exceptional split corresponds to the case that the complex equation is made by combining
together the last two equations in (\ref{hitcheq}) to get the Higgs bundle equation
\begin{equation}\label{higgs}\bar\partial_A\varphi=0.\end{equation}
A more generic split involves a complex parameter $\zeta$.  For any $\zeta$, set
\begin{align}\label{generic} \D_z^\zeta& = \frac{D}{D z}-\zeta^{-1}[\phi_z,\,\cdot\,]\cr
                \D_{\bar z}^\zeta &=\frac{D}{D\bar z}+\zeta[\phi_{\bar z},\,\cdot\,].\end{align}
(We write here $\phi_z$ and $\phi_{\bar z}$ instead of $\varphi$ and $\bar\varphi$.)
For any $\zeta$, the equation
\begin{equation}\label{nurco}[\D_z^\zeta,\D_{\bar z}^\zeta]=0 \end{equation}
is equivalent to  two real linear combinations of the Hitchin equations (\ref{hitcheq}).
The possible complex structures on $\M_H(G,C)$ are parametrized by $\zeta$,
where we add a point at
infinity to the $\zeta$ plane to make $\Bbb{CP}^1$.  For every $\zeta$, one
defines a complex structure $I_\zeta$ in which the equation (\ref{nurco}) is
regarded as an equation
governing holomorphic data; in this complex structure, the holomorphic variables are $\A^\zeta_z=A_z-\zeta^{-1}\phi_z$ and $\A_{\bar z}^\zeta=A_{\bar z}+\zeta \phi$, and the equation
(\ref{nurco}) is holomorphic in those variables.  The third linear combination of the equations
is regarded in complex structure $I_\zeta$ as a moment map condition.

For generic $\zeta$, the equation (\ref{generic}) simply says that the complex connection
$\A^\zeta=\A_z^\zeta \d z+\A_{\bar z}^\zeta \d \bar z$ is flat. Once the equation (\ref{generic}) is
expressed in terms of $\A^\zeta$, it has no explicit dependence on $\zeta$.  Thus, $\M_H(G,C)$ when regarded
simply as a complex manifold in complex structure $I_\zeta$, without worrying about its Kahler metric, is independent of $\zeta$ for generic $\zeta$.  The exceptional values of $\zeta$ are 0 and $\infty$.
For example, for $\zeta\to 0$, we must multiply $\D_z^\zeta$ by $\zeta$, whence it reduces
to $-[\phi_z,\,\cdot\,]$.  Meanwhile, for $\zeta=0$,  $\D_{\bar z}=D_{\bar z}$.  So the
$\zeta\to 0$ limit of eqn. (\ref{nurco}) is the Higgs bundle equation (\ref{higgs}).  By similar
reasoning, the $\zeta\to\infty$ limit of (\ref{nurco}) is the complex conjugate of (\ref{higgs}).

The Higgs bundle equation has played a prominent role in our analysis because in section (\ref{preliminaries}), we derived the holomorphic data at $y\not=0$ from the equation $[\D_1,\D_2]=0$, which was none other than the Higgs bundle equation.  (The moment map condition in that analysis  condition was not the usual moment map equation for Higgs bundles
-- namely  the first equation in (\ref{hitcheq}) --  but rather it was the three-dimensional equation
(\ref{mmtmap}).) At this point, we are led to wonder whether we could modify the construction that led to (\ref{eqform}) so that the holomorphic data at $y\not=0$ will be given not by a Higgs bundle
but by the solution of a different complex linear combination of Hitchin's equations.
It is in fact possible to do so.  Indeed the path to doing so is not quite uniquely determined.

There is one particularly nice parameter by which we can vary the underlying four-dimensional equations.  This is simply the parameter $\t$ in eqn. (\ref{bpseqns}).  In addition
to changing the equations, we should consider the possibility of changing the boundary conditions.
Of the six fields $(\vec A,\vec\phi)$ (that is, the components of $A$ and $\phi$ that are tangent
to the boundary), only $\vec\phi$ has a singularity at $y=0$ in the basic Nahm pole solution
(\ref{nahmpole}). However, once one drops the dependence on the spatial coordinates $\vec x$, $\N=4$
super Yang-Mills theory has an $SO(6)$ symmetry that rotates the six components of $\vec A$ and $\vec \phi$.  Therefore, one can obey the classical Yang-Mills equations with a solution obtained by
 applying an $SO(6)$ rotation to the Nahm pole (such a rotated Nahm pole was studied in \cite{gw}, section 4).  Of course, the rotation in general will change the unbroken supersymmetry.  It turns out, however (see Appendices \ref{fromsix} and \ref{foureight} for more detail), that as long as the rotation matrix is contained in a certain $SU(3)$ subgroup of $SO(6)$, the
unbroken supersymmetry is unchanged, and if it  is contained in a certain $U(3)$ subgroup,
then the unbroken supersymmetry changes in a way that corresponds to a change in the parameter $\t$ in the four-dimensional equation (\ref{bpseqns}).  So there is considerable freedom in rotating the boundary condition,
with or without a change in $\t$.

After making a choice along the lines just described,
the deformed equations and boundary conditions have
a three-dimensional reduction that takes the familiar form of (\ref{commuting}) and (\ref{momentmap}), but with a different definition of the $\D_i$.  Deferring most of the details to the appendices,
we will summarize some formulas that arise in an illuminating special case.  If one wishes
to preserve the $SO(3)$ symmetry of rotations of the boundary, then the rotation of the Nahm pole
can only depend on a single parameter: the polar part of $\vec A$ must be a constant $\zeta$ times
the polar part of $\vec\phi$.  It turns out that such a boundary condition is compatible with
the four-dimensional equations (\ref{bpseqns}), with a modified value of $\t$ that depends on $\zeta$.

In the reduction to three dimensions, we must now impose $A_1-\zeta\phi_1=0=\phi_y$.
The three commuting differential operators $\D_i$ then take the form
\begin{align}\label{neqformo}\D_1 & =2\frac{D}{D\bar z}  +2 \zeta [\bar \varphi,\,\cdot\,] \cr
                                                              \D_2 & = - 2 \zeta \frac{D}{D z} + 2[\varphi,\,\cdot\,]\cr
                                                               \D_3& = \frac{\D}{\D y}. \end{align}
Here, as described in the appendix, $\A_y$ is an appropriate linear combination of $A_y$ and $\phi_1$.
The condition that the $\D_i$ commute must be supplemented by a moment map condition, which is also
described in the appendix.  As one would anticipate for a geometry that preserves four supercharges,
the commutativity constraint $[\D_i,\D_j]=0$ can be derived from a superpotential, which in fact
is a multiple of the Chern-Simons function:
\begin{equation}\label{supertwo}\W=\frac{1}{4\pi i}\int_{W}\Tr\,\left(\A\wedge\d \A+\frac{2}{3}\A\wedge
\A\wedge\A\right).\end{equation}
To be more exact, in varying $\W$ to derive the equations $[\D_i,\D_j]=0$, one imposes a constraint
on the variation of $\A$ at $y=0$, so as to avoid delta function terms in the variation of $\W$.

The holomorphic data away from $y=0$ are now easy to describe.  We recover the picture studied in
section \ref{teqone} if $\zeta=0$, but as soon as $\zeta$ is nonzero (and not infinite),
 the three commuting operators
$\D_i$ simply describe a complex flat connection on $C\times \R_+$.  The covariant derivatives
for this flat connection are  $\D_z=\D_1/2$, $\D_{\bar z}=-\D_2/2\zeta$, $\D_y=\D_3$.
So commuting operators $\D_i$ simply describe a complex flat connection.  We will call the flat connection $\A$, irrespective of how it was defined in terms of $A$ and $\phi$.  $\A$ is constrained by a moment map condition, which does not quite coincide
with the most commonly studied moment map condition for a complex flat connection \cite{Corlette},
though it is qualitatively similar; we expect it to have a unique solution for any $\zeta$.

The holomorphic data away from $y=0$ are hence simply a complex flat connection on $C\times\R_+$,
or equivalently, since this space is contractible to $C$, a complex flat connection on $C$.
We will mainly be interested in the case that $C$ is $\R^2$ or $\Bbb{CP}^1$.  In either case,
$C$ is simply-connnected, so a complex flat connection on $C$ is trivial.  One may wonder therefore
how anything of interest can happen.  The answer is that most of the  structure of interest
will come from the boundary condition at $y=0$.  In the case of symmetry breaking, there is also
some interesting structure in the behavior at $y=\infty$.

\subsection{Nahm Poles and Opers}\label{flatpole}

The first point is to understand how a complex flat connection can have a Nahm pole.
The answer is that the pole appears in $\A_y$ and $\A_z$, but  not in $\A_{\bar z}$.  The model
example of a flat connection with a Nahm pole is
\begin{align}\label{modsol}\A_z&=\frac{\ct_+}{y}\cr
\A_{\bar z} & = 0 \cr
\A_y & = \frac{\ct_3}{y}.   \end{align}
This describes a flat connection as long as $[\ct_3,\ct_+]=\ct_+$.  We are interested in the case that
$\ct_+=\ct_1+i\ct_2$, where the $\ct_i$, $i=1,2,3$ generate a principal $\frak{su}(2)$ subalgebra of $\frak g$.

Being flat, this connection can be described by a formula $\d+\A=g\d g^{-1}$.
For example, for $\frak{su}(2)$, we can take explicitly
\begin{equation}\label{inggauge}
g =\begin{pmatrix} y^{-1/2} & -zy^{-1/2} \cr 0 & y^{1/2} \end{pmatrix},
\end{equation}
which leads to
\begin{align}\label{odsol} \A_z& = \begin{pmatrix}0&1\cr 0 & 0 \end{pmatrix}\frac{1}{y}\cr
                            \A_{\bar z}&=0 \cr
                            \A_y& = \begin{pmatrix}1 & 0 \cr 0 & -1\end{pmatrix}\frac{1}{2y}.\end{align}
Alternatively, the model solution can be generated from the non-singular flat connection
\begin{equation}\label{nodsol}\A_z=\begin{pmatrix}0 & 1\cr 0 & 0\end{pmatrix}, ~\A_{\bar z}=\A_y=0
\end{equation}
by the same singular gauge transformation as in (\ref{singgauge}):
\begin{equation}\label{winggauge}
g =\begin{pmatrix} y^{-1/2} & 0 \cr 0 & y^{1/2} \end{pmatrix}.
\end{equation}

We now proceed rather as we did in the Higgs bundle case to explain the condition that must be
placed on a complex flat bundle $E$ so that it can be placed in the form (\ref{odsol}) near $y=0$,
modulo less singular terms.
We write simply $E$ for the restriction of $E$ to $C=C\times y$ for
some fixed $y>0$.
 We consider solutions $s$ of the equation $\D_ys=0$ that vanish
as $y^{1/2}$ for $s\to 0$.  Sections $s$ obeying these conditions span a rank one sub-bundle
$\L\subset E$.  In the case of the model solution, any such $s$ is a multiple of
\begin{equation}s=y^{1/2}\begin{pmatrix}0 \cr 1\end{pmatrix},\end{equation}
so as in the Higgs bundle case,
 $\L$ is simply spanned by sections whose upper component vanishes.

Also as before, if we regard $E$ as a flat bundle over $C$, then $\L$ is a
holomorphic sub-bundle; indeed, the object $s$ that we have just defined
is a holomorphic section of $\L$,
since it is certainly annihilated by $\D_{\bar z}$.
However, it is not true that $s$ is annihilated by $\D_z$.  On the contrary, a look at the previous
formulas shows at once that
\begin{equation}\label{tomox}  s\wedge \D_z s = 1.    \end{equation}

This brings us to the mathematical notion of an ``oper.'' (For an explanation of
this notion as well as a review of many related ideas that will enter our story
later, see \cite{Frenkel} or \cite{Teschner}.) For $G=SU(2)$, an oper is a flat rank two complex bundle
$E$  bundle over a Riemann surface $C$, with structure group $SL(2,\C)$, together with a holomorphic
line sub-bundle $\L\subset E$ with the following property: $\L$ is nowhere invariant
under parallel transport by $\D_z$.  The last statement means the following.
If $s$ is a local nonzero holomorphic section of $\L$, then $\D_zs$, which
will be $E$-valued since $\D_z$ is a connection on $E$, is nowhere $\L$-valued.
An equivalent statement, since $\L$ is spanned by multiples of $s$, is that $\D_z s$ is nowhere a
multiple of $s$.  Alternatively, $s\wedge \D_z s$ is everywhere nonzero.  The
last statement does not
depend on the choice of the nonzero section $s$, since if we replace $s$ by $f s$ (where $f$
is a nonzero local holomorphic function on $C$), we have $s\wedge \D_zs \to f^2 s\wedge \D_zs$.

We have extracted an oper structure from the Nahm pole boundary conditions; conversely let
us see that given an oper
 on $C$, that is a pair $(E,\L)$ obeying the  conditions just described, we
can construct a
solution of the Nahm pole boundary conditions.
Go to a gauge in which $\L$ is spanned by vectors whose
upper component vanishes.  Holomorphy of $\L$ means that $\A_{\bar z}$
is lower triangular in this gauge.  The oper condition $s\wedge\D_z s\not=0$ for any nonzero
local section of $s$ implies that in this gauge, the upper right matrix element of $\A_z$ is
nonzero.  By a further diagonal gauge transformation, we can set this matrix element to 1.
Then we pull back the flat bundle $E$ with connection $\A$ from $C$ to $C\times \R_+$
(to get a flat connection on $C\times\R_+$ with no dependence on $y$)
and make the singular gauge transformation (\ref{winggauge}).  Having an upper right matrix
element of 1 means that after the singular gauge transformation, $\A_z$ has the singular
behavior of the model solution (\ref{odsol}); being lower triangular, $\A_{\bar z}$ acquires
no singularity.  Finally, the gauge transformation gives $\A_y$ precisely the form of the model
solution.

So we have shown, at least for the case that $G$ has rank 1, that two-dimensional
opers correspond precisely to solutions of the Nahm pole
boundary conditions in three dimensions modulo less singular terms.  Conjecturally,
by a further smooth complex-valued gauge transformation, one can satisfy the moment
map condition.

\subsubsection{Some Further Remarks}\label{somefurth}
We add the following technical remarks. Since we want to be
able to consider 't Hooft operators of minimum charge,
we will take the gauge group in the rank 1 case to be $G=SO(3)$, rather than
$SU(2)$.  Accordingly, we should
restate the above derivation in terms of the adjoint bundle $\ad(E)$ rather than $E$.  Because
our considerations have been local on $C$,
rewriting the construction in terms of the adjoint bundle does not change very much
and we will omit it.  (The main difference is that what can be naturally defined
globally is in general not $\L$ but $\L^2$, which is a sub-bundle of $\ad(E)$.)

Also, everything we have said for $G$ of rank 1 has an analog for any
semi-simple $G$, somewhat as
we indicated in the Higgs bundle case at the end of section \ref{bc}.  For example, for $G=SU(n)$,
an oper is a flat complex bundle $E\to C$ of rank $n$ with $SL(n,\C)$-valued holonomies
together with a line sub-bundle $\L\subset E$
that is holomorphic and has the property that  if $s$ is a local nonzero holomorphic section of $\L$,
then $s,\,\D_z s,\dots,\D_z^{n-1}s$ furnish a local trivialization of $E$.
The equations $[\D_i,\D_j]=0$ together with the Nahm pole boundary condition
determine such an oper
structure, by arguments similar to those we have already given.

\subsection{Opers With Singularities}\label{opsing}

Now we would like to modify the Nahm pole boundary condition
to incorporate additional singularities
-- which we will associate with 't Hooft operators -- on the boundary at $y=0$.

The type of singularity that we want can be guessed by analogy with the discussion of
Higgs bundles.  We will still have a flat $G_\C$ bundle $E\to C\times \R_+$, and
near a generic boundary point, the flat connection will look like the model solution
(\ref{modsol}), up to a unitary (that is, $G$-valued rather
than $G_\C$-valued) gauge transformation.  We can still define a
holomorphic line sub-bundle $\L\subset E$ by considering sections $s$ obeying
$\D_ys=0$ and vanishing for $y\to 0$; and we still require that if $s$ is a local holomorphic
section of $s$, then $s\wedge \D_z s$ is generically nonzero.

The only difference is that now we assume the existence of exceptional points
$z_a,~a=1,\dots,\ss$, at
which $s\wedge \D_z s$ vanishes.  In fact, we specify positive integers $k_a $ and require
that $s\wedge \D_z s$ vanishes in order $k_a $ for $z\to z_a$:
\begin{equation}\label{badpoint} s\wedge \D_z s\sim (z-z_a)^{k_a}.\end{equation}
This is analogous to requiring $s\wedge\varphi s\sim (z-z_a)^{k_a}$ in the Higgs bundle case.

To get a precise problem of classical or quantum gauge theory with this sort of boundary
behavior, what remains is to specify precisely what sort of singularity a solution of the moment
map condition $\sum_i[\D_i,\D_i^\dagger]=0$ (or the four- or five-dimensional equations
that can be
dimensionally reduced to it) is supposed to have at $z=z_a$.   As in most such problems,
to do this one finds a model solution for the case of only one singularity, at, say, $z=0$ (and $y=0$) and
with an arbitrary $k$.  Then one asks that the singular behavior of
the solution near each of the points $z=z_a$, $y=0$
should coincide with that of the model solution, for $k=k_a$.  For the case of Higgs
bundles, the appropriate
model solutions were found (for $G$ of rank 1) in section 3.6 of \cite{fiveknots}, but for the
generalization considered here, at present we are only able to find the
model solutions numerically.   They are described in  Appendix \ref{bco}.

The objects that we have described so far correspond to solutions of the flatness and moment
map conditions on $C\times \R_+$ with boundary conditions associated to Nahm poles or opers,
except at finitely many boundary points where the oper condition is corrected.  In particular,
as soon as one gets away from $y=0$, one simply has a flat bundle
(with a moment map condition).   The monodromy of the flat bundle around the
points $z=z_a$ is therefore trivial,
so the exceptional behavior at the points $z=z_a$ only affects the oper property of the pair
$(E,\L)$, not the flatness of $E$.  Singularities of this kind are called oper singularities with
trivial monodromy.

\subsection{Oper Singularities And Bethe Equations}\label{opbethe}

Let us now make concrete (referring to \cite{Frenkel} for much more detail)
what sort of an object is an oper with monodromy-free singularities.
We will make this analysis for the case that $C$ is simply-connected, so that there are no
moduli in the choice of the flat bundle $E\to C$.   So $C$ will be either $\R^2$ or
$\Bbb{CP}^1$; that is, it will be the complex $z$-plane with or without an added point at infinity.
There are two reasons for assuming $C$ to be simply-connected: this is the most relevant
case for understanding the Jones
polynomial; and also, eliminating the choice of $E$ from the discussion will make it easier to
focus on the opers and their singularities.

Since $C$ is simply-connected, a flat bundle over $C$ is trivial.
So we can go to a gauge with $\A_z = \A_{\bar z}=0$.
The line sub-bundle $\L$ of $E$ is inevitably trivial for $C=\R^2$, or trivial after omitting
the point $z=\infty$ for $C=\Bbb{CP}^1$.  So it is globally generated by a section $s$,
but we cannot put $s$ in a simple form while also setting $\A_z=\A_{\bar z}=0$.
Instead, we
take \begin{equation}\label{nuptriz}s=\begin{pmatrix} P(z)\cr Q(z)\end{pmatrix},\end{equation}
with polynomials $P$ and $Q$.
$P$ and $Q$ are uniquely determined up to a linear transformation
\begin{equation}\label{helf}\begin{pmatrix}P\cr Q
\end{pmatrix}\to M\begin{pmatrix}P\cr Q
\end{pmatrix},~~M\in GL(2,\C).\end{equation}

Now $s\wedge \D_z s$ reduces to $P\partial_zQ-Q\partial_zP$.  So if the polynomial
$K(z)=\prod_{a=1}^d(z=z_a)^{k_a}$ encodes the positions and charges of the oper
singularities, then the equation we would like to solve is
\begin{equation}\label{omicro} P\partial_z Q-Q\partial_zP = K(z),\end{equation}
modulo the action of $SL(2,\C)$. (Choosing $K$ to have leading coefficient 1 has
reduced $GL(2,\C)$ to $SL(2,\C)$.)  It is convenient to fix the action of two of the
three generators of $SL(2,\C)$ by requiring that
the degree  of the polynomial $Q$ is less\footnote{Later on, in the presence of symmetry
breaking, we will have to relax this condition.} than the degree   of $P$ (if two polynomials have the same
degree, a linear combination of them has smaller degree and we call this $Q$), and that $Q$ has leading coefficient 1,
\begin{equation}\label{zeroes}Q(z)=\prod_{i=1}^q (z-w_i),\end{equation}
for some $w_i$.  These conditions leave only the freedom to add to $P$ a multiple of $Q$.

We can recast (\ref{omicro}) as
\begin{equation}\label{operbo}
\partial_z \frac{P}{Q}  =- \frac{K(z)}{Q^2}
\end{equation}
The left hand side of this equation
has zero residues at the zeroes $w_i$ of $Q(z)$.
The right hand side must also have zero residues.
This gives the constraints
\begin{equation}\label{bethe}
\sum_a \frac{k_a}{w_i - z_a} = \sum_{j\not=i} \frac{2}{w_i - w_j},~~i=1,\dots,q.\end{equation}
Vice-versa, given a solution of these equations, the residues of $K/Q^2$ are zero; hence
$\int K/Q^2\, \d z$ is a rational function $P/Q$, and $P$ is fixed up to a
constant multiple of $Q$, which is the expected indeterminacy.

If $C=\R^2$,  opers with the desired monodromy-free singularities
simply correspond to the solutions of the equations (\ref{bethe}).  For $C=\CP^1$, we must
further ensure that the oper does not have an additional
singularity at infinity.  The condition for this turns out to be that the degree $q$ of the polynomial
$Q$ is just one-half of the degree of $K$:
\begin{equation}\label{truy} q=\frac{k}{2},~~k=\sum_a k_a. \end{equation}
To determine whether the oper has a singularity at infinity, let $p$ be the degree of $P$
and define $\begin{pmatrix}\tilde P\cr \tilde Q\end{pmatrix}=
z^{-p}\begin{pmatrix}P\cr Q\end{pmatrix}$.  We view $\tilde P,\,\tilde Q$
as polynomials in $v=1/z$. The condition that
the degree  of $P$ exceeds the degree $q$ of $Q$ implies  that
there is no cancellation of the leading power of $z$ on the left hand
side of (\ref{operbo}) and hence that $p+q=k+1$.
Since $q<p$, it follows that
\begin{equation}\label{ruy} q\leq \frac{k}{2}\end{equation}
in general.  The condition that the oper has no singularity at $z=\infty$ or $v=0$ is
that $\left(\tilde P\partial_v\tilde Q-\tilde Q\partial_v\tilde P\right)_{v=0}\not=0$, and
it is not hard to see that this coincides with (\ref{truy}).

To get farther, we need the theory of integrable systems.
Rather ``miraculously,'' the equations (\ref{bethe}) are the Bethe equations of an integrable
model, which is the Gaudin model or a certain large impurity limit of the XXX spin chain.
(The connection between opers with monodromy-free singularities and the Gaudin model
is reviewed in \cite{Frenkel}, following  earlier developments such as \cite{BF,FFR}.  
For more on this, see section \ref{applic}.)
For $a=1,\dots,d$, let $R_a$ be a copy of the representation of $SU(2)$ of spin $j_a=k_a/2$, and
let $\H=\otimes_{a=1}^d R_a$.  The Hamiltonians of the Gaudin model are the commuting
operators on $\H$ given by
\begin{equation}\label{gaudham}H_a=\sum_{b\not=a}\frac{\vec T_a\cdot \vec T_b}{z_a-z_b}.\end{equation}
Here $\vec T_a$ are the generators of $\frak{su}(2)$ acting on $R_a$, and 
$\vec T_a\cdot\vec T_b$
is the inner product of $\vec T_a$ and $\vec T_b$ (defined with the quadratic form such
that $\vec T_a\cdot \vec T_a=j_a(j_a+1)$).    Actually, what we have written in (\ref{gaudham})
are the Hamiltonians for the Gaudin model for $G^\vee=SU(2)$.  (We call this group $G^\vee$
as it is naturally dual to the gauge group $G$ that appears in the rest of our analysis.)  There is
a Gaudin model for any $G^\vee$, and it bears the same relation to opers that we are about
to describe for
$SU(2)$, but if $G^\vee$ has rank bigger than 1, then the $H_a$ are only part of a complete set of commuting Hamiltonians.

Since the Gaudin Hamiltonians commute, they can be simultaneously diagonalized.
Moreover, since they commute with the action of $G^\vee$, their joint eigenvectors can
be organized in irreducible representations of $G^\vee$.  Because of the $G^\vee$ action,
to understand all of the joint eigenvectors of the Gaudin Hamiltonians, it suffices to
understand those joint eigenvectors that are also highest weight vectors for the action of
$G^\vee$.

In the theory of the Bethe ansatz for the  Gaudin model of $G^\vee=SU(2)$, it is shown that solutions of the Bethe
equations (\ref{bethe}) correspond to the joint eigenvectors that are also
highest weight vectors for the action
of $G^\vee$.  In this correspondence, the weight $\cmmib w$ is related to the degree $q$ of $Q$ by
\begin{equation}\label{duflo} \cmmib w=\frac{k}{2}-q.\end{equation}
In particular, if we want $G^\vee$-invariant joint eigenvectors of the Gaudin Hamiltonians,
we need $\cmmib w=0$ and $q=k/2$; but as we observed in (\ref{truy}), this is the condition
that the corresponding oper extends over $\CP^1$ with no singularity at infinity.

So the number of opers on $\CP^1$ with monodromy-free singularities is the same
as the number of $SU(2)$-invariant joint eigenvectors of the Gaudin Hamiltonians.
But  the joint eigenvectors of the commuting Gaudin Hamiltonians are a basis for $\H$,
and similarly the $G^\vee$-invariant joint eigenvectors are a basis for
 $\H^{G^\vee}$, the $G^\vee$-invariant
part of $\H$.  So the number of opers that obey the conditions that we have imposed is
precisely the dimension of $\H^{G^\vee}$.

This result is our first concrete success in comparing the counting of BPS solutions in
$G$ gauge theory to Chern-Simons theory with gauge group $G^\vee$.  Consider
Chern-Simons theory on $\Bbb{CP}^1$ with charges in the representations $R_a$,
$a=1,\dots,d$.  We place
these charges at points
$z_a\in \Bbb{CP}^1$.
In the classical limit, the space of physical states is just the $G^\vee$-invariant
part of $\H=\otimes_a R_a$; the restriction to $G^\vee$-invariant states is the Gauss
law constraint.  This also gives the right answer for the dimension of the physical
Hilbert space of Chern-Simons theory if the Chern-Simons coupling parameter $\k^\vee$
is generic.  On the other hand, in the dual description in which the Hilbert space is constructed
starting with time-independent solutions in $G$ gauge theory, the states
should correspond,\footnote{We explain in section \ref{fourdintro} why time-dependent
instanton corrections in the
$G$ gauge theory do not affect this counting of states.}
from the arguments we have given, to opers on $\Bbb{CP}^1$ with singularities of charge $k_a$
at the $z_a$.  Since the number of these opers is the same as the dimension
of $\H^{G^\vee}$, we have at least succeeded in reconciling the dimensions
of the spaces of physical states in the two
descriptions.  This gives an indication that with the help of the deformation that we have
exploited in the present section to $\zeta\not=0$, the program of section \ref{methods}
based on stretching a knot in one direction can actually work.

The counting of states is less transparent if we take $C=\R^2$ rather than $\CP^1$.
Qualitatively, it is clear that the number of physical states in Chern-Simons theory is larger on $\R^2$ than on
$\CP^1$, because, as the flux can escape to infinity, a physical state need not be
completely gauge-invariant.  However, to understand the condition that should be imposed at infinity
 is rather delicate, and
it is hard to understand in $G^\vee$ Chern-Simons
theory the result that seems to come from the opers: physical states on $\R^2$ correspond to highest weight vectors in $\H$.  After incorporating symmetry breaking in section \ref{moresymbr},
the comparison between the two descriptions will be simpler.

\subsubsection{Relation To Conformal Field Theory}\label{relconf}

In arriving at the Gaudin model, we have accomplished much more than
simply getting a number of classical solutions that is reminiscent of known constructions
of the Jones polynomial.  The Jones representations of the braid group
can be described \cite{TK}  as the monodromy
of the Knizhnik-Zamolodchikov equations \cite{KZ}.  These equations are as follows.
Express the usual parameter $q$
that enters the Jones polynomial as $q=\exp(2\pi i/(\k^\vee+2))$.   Let $\eusm B$
be the space of distinct $d$-plets $z_1,\dots,z_d\in\C$.  And let $\H^*$ be the trivial bundle
over $\eusm B$ with fiber $\H=\otimes_{a=1}^d R_a$.    The Knizhnik-Zamolodchikov equations
are the following system of equations for a section $\Theta$ of $\H^*$:
\begin{equation}\label{kz}\left(\frac{\partial}{\partial z_a}+\frac{H_a}{\k^\vee+2}\right)\Theta=0.
\end{equation}
The $H_a$ are the Gaudin Hamiltonians (\ref{gaudham}).
The Knizhnik-Zamolodchikov equations describe parallel transport of the section $\Theta$
of $\H^*$ with respect to a certain flat connection, which is implicitly defined in (\ref{kz}).
To verify flatness of the connection, one uses the fact that
the $H_a$ commute and also the relation $\partial H_b/\partial z_a=\partial H_a/\partial z_b$.
The solutions $\Theta$ of the Knizhnik-Zamolodchikov equation are conformal blocks of
two-dimensional current algebra with symmetry group $G^\vee$; they are important
in two-dimensional conformal field theory.

Since opers with monodromy-free singularities correspond to a basis for $\H$, we will, in our
approach to the Jones polynomial,  eventually
be using gauge theory to construct a flat connection on the bundle $\H^*$; moreover,
as we hope to recover the Jones representations of the braid group, this flat connection
should be gauge-equivalent to the one defined by the Knizhnik-Zamolodchikov equations.
Actually, this tempting-sounding route is not the one we will follow.  The very same Jones
representations of the braid group have another  realization in conformal field theory in terms
of Virasoro conformal blocks for correlators  of a product of
degenerate fields and this will prove more useful.

\subsection{Symmetry Breaking Again}\label{moresymbr}

Just as in section \ref{symbr}, we can gain some further clarity by moving away from the
origin of the Coulomb branch.    As always, we do so by turning on constant and
commuting expectation values for $\vec \phi$ near $y=\infty$.
In the present context, this means that the connection
form $\A$ does not vanish at infinity, but is a one-form with constant coefficients; moreover,
these coefficients commute with each other.

\subsubsection{``Real'' Symmetry Breaking}\label{realsym}

First we consider the case that only $\A_y$ has an expectation value at infinity.
This expectation value arises from the value of $\phi_1$ at infinity, and so just  as in (\ref{gauga})
we have
\begin{equation}\label{gaugan}
\A_y = \begin{pmatrix} \ca_1 & 0 \cr 0 & -\ca_1\end{pmatrix},~~y\to\infty,
\end{equation}
where we can take $\ca_1>0$.

The condition
that $\A_z$ and $\A_{\bar z}$ should have no exponential growth at infinity tells
us that they must be upper triangular in a gauge in which $\A_y$ looks like (\ref{gaugan}) for
$y\to\infty$.
Thus, near $y=\infty$, the real symmetry breaking gives us a natural way to put the whole
flat connection in a triangular form.

An oper that is endowed with a covariantly constant reduction of its structure group to the
group of upper triangular matrices -- that is, to a Borel subgroup -- is called a Miura oper.
This notion is described in detail
in \cite{Frenkel}.  Any oper bundle without monodromy can be given
a Miura oper structure; in fact, there is a one-parameter family of ways to do so.  Concretely,
if the rank two flat bundle $E\to C$ has trivial monodromy,
then the associated bundle of $\Bbb{CP}^1$'s
(whose fibers are obtained by projectivizing the fibers of $E$) also carries a  flat
connection without monodromy.
 Let us call this bundle $\B$. Picking an arbitrary section
 of  $\mathcal B$ over some given point $p\in C$
and parallel transporting it, we get a covariantly constant section of $\mathcal B$ which turns
the  underlying oper into a
Miura oper.  This procedure
introduces one complex modulus --
the choice of a point in the fiber of $\B$ over the starting point $p$.  This means that a Miura
oper without monodromy depends on a complex modulus.  (When -- as in section
\ref{compsym} -- we introduce symmetry breaking in a complex direction, this modulus will
disappear, because there will be no freedom to make a gauge rotation of $\A_y$ at infinity
relative to $\A_z$.)

The Bethe roots have a particularly nice interpretation in the case of a Miura oper.
To explain this most simply, let us go back to the case that $C=\R^2$
and use a gauge with $\A_z=\A_{\bar z}=0$.  The behavior for $y\to\infty$ singles out a sub-bundle
$\tilde\L$ of the rank 2 bundle $E$ that is invariant under parallel transport; we may call it a flat
sub-bundle.
In a gauge with the asymptotic behavior (\ref{gaugan}), $\tilde \L$ is generated by a covariantly constant
section $\tilde s$ that vanishes for $y\to\infty$.
After a complex gauge transformation to set
$\A_z=\A_{\bar z}=0$, $\tilde s $ is simply constant; we may as well take
\begin{equation}\label{rofor}\tilde s=\begin{pmatrix} 1 \cr 0 \end{pmatrix}.\end{equation}
On the other hand, the behavior for $y\to 0$ determines a holomorphic (not flat) sub-bundle
$\L\subset E$, generated as before (in a gauge $\A_z=\A_{\bar z}=0$) by
\begin{equation}\label{wagan}s=\begin{pmatrix} P \cr Q\end{pmatrix}.\end{equation}
The choice of the Miura structure $\tilde s$ gives a way to pick a natural linear combination
of the components of $s$, namely
\begin{equation}\label{agan}\tilde s \wedge s=Q. \end{equation}
The $Q$ determined this way is not necessarily the one that we used in section \ref{opbethe},
where we took $Q$ to be a linear combination of components of $s$ with minimum degree;
now $Q$ is simply determined by the Miura  structure.  With our new choice,
the zeroes of $Q$  have a simple interpretation: they are the points at which $s$ is
a multiple of $\tilde s$.  In other words, the sub-bundle determined by the behavior for $y\to\infty$
is generically different from the sub-bundle determined by the behavior for $y\to 0$.   The
zeroes of $Q$  -- which are called Bethe roots -- are precisely the points at which these coincide.   Another way to say the
same thing is that the Bethe roots are the values of $z$ at which there is a solution of $\D_ys= 0$
that vanishes for both $y\to 0$ and $y\to\infty$.  As we discussed in section \ref{symbr},
in the context of the Bogomolny equations one would say that there are smooth BPS
monopoles at those values of $z$ (and some values of $y$).

For a concrete example, suppose that there are no 't Hooft operators at all.
The flat bundle $E\to \R^2$ is completely trivial and it has up to isomorphism a unique
oper structure with
\begin{equation}\label{only} s=\begin{pmatrix} z \cr 1\end{pmatrix}.\end{equation}
In the absence of symmetry breaking, our convention that $Q$ is the linear combination
of components of $s$ with smaller degree leads to $Q=1$, and hence (up to the freedom
of adding to $P$ a multiple of $Q$ and rescaling it) $P=z$.

In the presence of real symmetry breaking, the Miura structure gives a distinguished
choice   (\ref{agan}) of
$Q$ which has no reason to be a constant.  If $Q$ is not constant, then by adding to $P$
a multiple of $Q$ and rescaling it, we can set $P=1$, so
\begin{equation}\label{onl} s=\begin{pmatrix}P \cr Q \end{pmatrix}=\begin{pmatrix}1 \cr z-w
\end{pmatrix},\end{equation}
for some $w$.
The polynomial $K=PQ'-QP'$ is 1, consistent with the absence of any 't Hooft operators.
We have found, in the presence of real symmetry breaking, holomorphic data corresponding to
a one-parameter family of
solutions depending on the choice of a point $w\in \R^2$.  This is a solution with no 't Hooft
operator and a single Bethe root.  We will call it a bare Miura oper.
 In the right context, when lifted back to four dimensions,
we will interpret this solution later as a ``string'' that is localized at $z=w$ and at  a value
of $y$ that depends on $\zeta$.

In general, the degree of $K$ is at most one less than the
degree of $Q$ (this bound is achieved precisely if $P=1$), so the number of Bethe roots is at most one
more than the degreee $k=\sum_ak_a$ of $K$.

\subsubsection{``Complex'' Symmetry Breaking}\label{compsym}

If we  give expectation values at infinity to all components of $\vec\phi$, while requiring
$A$ to vanish at infinity,
then the complex connection $\A$ is constant and  diagonal for $y\to\infty$,
\begin{align}\label{const}
\A_z &\sim \frac{1}{\zeta} \begin{pmatrix} \ca & 0 \cr 0 & -\ca \end{pmatrix} \cr
\A_{\bar z} &\sim \zeta \begin{pmatrix} \bar \ca & 0 \cr 0 & - \bar \ca \end{pmatrix} \cr
\A_y &\sim \begin{pmatrix} \ca_1 & 0 \cr 0 & -\ca_1 \end{pmatrix} ,\cr
\end{align} where $\ca$ is a complex number.
The factors of $\zeta$ arise in the change of variables from $\varphi$ to $\A$.

At infinity in the $z$ direction, for any $y$, the solution reduces to the unique solution \cite{Kronheimer} of Nahm's equations which has a
Nahm pole at $y=0$ and behaves as
(\ref{const}) at $y=\infty$.   In particular, the connection form is constant
for $z\to\infty$ with fixed $y$.
For this form of the connection, we can write the small holomorphic section $s$ as
\begin{equation}\label{goodform}
s = \begin{pmatrix} e^{- \zeta \bar \ca'\, \bar z} & 0 \cr 0 &
e^{ \zeta \bar \ca'\, \bar z}  \end{pmatrix} s_0
\end{equation}
where $\ca'$ (which equals $\ca$ for $y\to\infty$) is the constant value of $\A_{\bar z}/\zeta$
at large $z$ with fixed $y$,
and $s_0$ has a finite limit at large $z$ and fixed $y$.

In order to analyze the holomorphic data in such solutions, it is unnatural to gauge
the connection away. The information we want would be hidden in the
behavior of the necessary gauge transformation at infinity.
 Rather, we will pick a complex gauge transformation which brings
the connection exactly (not just asymptotically) to the form
\begin{align}\label{consto}
\A_z &=\frac{1}{\zeta} \begin{pmatrix} \ca & 0 \cr 0 & -\ca \end{pmatrix} \cr
\A_{\bar z} &= \zeta \begin{pmatrix} \bar \ca & 0 \cr 0 & - \bar \ca \end{pmatrix} \cr
\A_y &= \begin{pmatrix} \ca_1 & 0 \cr 0 & -\ca_1 \end{pmatrix} \cr
\end{align}
As we will now see, this can be done by a relatively simple type of gauge transformation.  Consider any gauge field on $\R^2\times \R_+$, such as one that
has the asymptotic form discussed above, such that the connection form approaches
a nonzero constant for $z\to\infty$ with fixed $y$.
If we were to compactify $\R^2$ to $\CP^1$, we would say that such a connection
has an irregular singularity at $z=\infty$ with a double pole.
Connections with irregular singularities have Stokes phenomena. If one only considers gauge
transformations with no essential singularities (that is, with polynomial growth
only at infinity), the Stokes data is gauge-invariant.  Two flat  connections with an irregular
singularity are gauge-equivalent by a gauge transformation with only polynomial growth
at infinity
if and only if they have the same monodromy and the same Stokes data.

As our connection has  only has a double pole at $z=\infty$,
it has only two Stokes sectors. Stokes theory tells us that the monodromy around $z=\infty$
can be decomposed into the product of a diagonal formal monodromy matrix, and
a sequence of Stokes matrices, which are alternatingly upper and lower triangular
with ones on the diagonal.  With only two Stokes sectors, the expression for the monodromy
is
\begin{equation}\label{monodr}M=
\begin{pmatrix}1 & b \cr 0 & 1 \end{pmatrix}\begin{pmatrix}1 & 0 \cr \tilde b & 1 \end{pmatrix}\begin{pmatrix}\mu & 0 \cr 0 & \mu^{-1} \end{pmatrix} ,\end{equation}
with constants $b,\tilde b,$ and $\mu$.
As we are on $\R^2$, which is simply-connected, the monodromy at infinity must be $M=1$.
This together with the form (\ref{monodr}) of the monodromy
implies that the three factors in (\ref{monodr}) -- the Stokes matrices and the
formal monodromy -- must all
equal 1. Hence we can bring our connection to the constant diagonal form (\ref{consto})
by a gauge transformation
which grows only polynomially at infinity.

In particular, $s$ will now take the form
\begin{equation}
s = \begin{pmatrix} e^{- \zeta \bar \ca\, \bar z} & 0 \cr 0 & e^{ \zeta \bar \ca\, \bar z}  \end{pmatrix} \begin{pmatrix}P(z) \cr Q(z) \end{pmatrix}
\end{equation}
with polynomials $P$ and $Q$.
The equation $s \wedge \D_z s=K(z)$ gives
\begin{equation}\label{normo}
\left(P \partial_z Q - Q \partial_z P\right) - \frac{2 \ca P Q}{\zeta} = K(z)
\end{equation}
which can be converted to the requirement that
$Ke^{2 \ca z}/Q^2 $ has no residues at the zeroes $w_i$ of $Q(z)$. This becomes
\begin{equation}\label{bethetwist}
\frac{2 \ca}{\zeta} + \sum_a \frac{k_a}{w_i - z_a} = \sum_j \frac{2}{w_i - w_j}
\end{equation}
We are interested in solutions of (\ref{normo}) modulo a rescaling $P\to\lambda P$, $Q\to
\lambda^{-1}Q$ (which corresponds to an automorphism of the diagonal flat connection), so
we actually only care about the zeroes of $Q(z)$.

These are, again,  Bethe equations, this time for the Gaudin model with an irregular singularity
at $z=\infty$  \cite{Ry,FFT,FFRy}.  It is shown in  \cite{FFRy} that solutions of these Bethe
equations are in one-to-one correspondence with eigenvectors of the spin chain, 
making in all $\prod_a(k_a+1)$ solutions.  It may also be possible to extract
this result from the theory of the 
 XXX spin chain, which is more familiar than the Gaudin model. The Bethe equations of
the Gaudin model arise from
a specific ``large impurity limit'' of those of the XXX chain.
The deformation parameter $\ca$
maps to the twist of the XXX chain, which is known to break the global $SU(2)$
symmetry to $U(1)$, and simplify the counting of solutions to Bethe equations:
instead of a solution
for each eigenvector of the
spin chain Hamiltonian
that is an $SU(2)$ highest weight,  one gets a solution for each eigenvector.

As a simple example, consider the case of a single 't Hooft operator of charge 2,
so $K(z)=z^2$.  Just as in section \ref{physhilb}, there is a solution of (\ref{normo})
with $P=1$ and $Q$ a quadratic polynomial in $z$, and a solution with $Q=1$ and
$P$ a quadratic polynomial.  What happens if $P$ and $Q$ are both linear in $z$?
Again as in section \ref{physhilb}, we can solve (\ref{normo}) with $P=Q=z\sqrt{-\zeta/2\ca}$,
but this solution does not correspond to an oper, since $P$ and $Q$ have a common zero.
The novelty is that there is also an acceptable solution with $P=-(\zeta/2\ca)(z-\zeta/\ca)$, $Q= z+\zeta/\ca$,
corresponding to a solution of the Bethe equations (\ref{bethetwist}) with a single Bethe root.
This gives a total of $2+1=3$ opers obeying the necessary conditions.
The last solution disappears if we take $\ca=0$ and has a common zero for $P$ and $Q$ if
we take $\zeta\to 0$.

\subsection{Opers And Stress Tensors}\label{opstress}

In our analysis of opers with trivial monodromy, we have used a gauge in which the connection
is trivial, $\A_z=\A_{\bar z}=0$.  Correspondingly, we had to make
a general ansatz for the small section $s$ that generates
the holomorphic sub-bundle $\L\subset E$:
\begin{equation}\label{poxo} s=\begin{pmatrix}P\cr Q\end{pmatrix}.\end{equation}
Here we will make a gauge transformation to put $s$ in the standard form with
upper component vanishing, and see what we can say about $\A_z$.
This will have two benefits.  We will begin to understand the relation of opers to
conformal field theory.  And we will get a description that is more general, not limited to the
case  (which however is particularly important in the present paper)  of oper bundles of trivial monodromy.

To put $s$ in a simple form by a smooth gauge transformation would make $\A_z$ and
$\A_{\bar z}$ both nonzero.   It turns out to be more helpful to keep $\A_{\bar z}=0$.
To also keep $\A_z$ regular would  require that our gauge transformation should be holomorphic, which is too restrictive
a condition.  Instead we will consider meromorphic gauge transformations, which will keep
$\A_{\bar z}=0$, put $s$ in a standard form, and generate poles in $\A_z$.

The most obvious meromorphic gauge transformation that puts $s$ in a standard form is
\begin{equation}\label{helm} h=\begin{pmatrix}Q & -P\cr 0 & Q^{-1}\end{pmatrix}.\end{equation}
We have
\begin{align}\label{elm}  hs& = \begin{pmatrix}0 \cr 1\end{pmatrix} \cr
          h\partial_z h^{-1} & = \partial_z+
          \begin{pmatrix} -\frac{Q'}{Q} & -K  \cr 0 & \frac{Q'}{Q}\end{pmatrix},\end{align}
where as before $K=PQ'-QP'$.  It turns out to be more convenient to
go to a gauge in which the upper right matrix element of $\A_z$ is $-1$, at the cost of
mapping $s$ to a multiple of itself (this leaves unchanged the line bundle generated by $s$).  We make a further gauge transformation by
\begin{equation}\label{furth}\tilde h=
\begin{pmatrix}1/\sqrt{K} & 0 \cr 0 & \sqrt{K}\end{pmatrix}.\end{equation}
The possible double-valuedness of $\sqrt{K}$ is of no concern, for the following reason.
If $G^\vee=SO(3)$, so that $G=SU(2)$, then $K$ is a perfect square as all its zeroes are of even
degree.  If instead $G^\vee=SU(2)$, then $G=SO(3)$, and we should really be writing all
formulas in the adjoint representation, rather
than the two-dimensional representation; accordingly,
 the sign of a gauge transformation is irrelevant.
After a gauge transformation by $\tilde h$,  $\A_z$ takes the form
\begin{equation}\label{omogro}\A_z=
\begin{pmatrix} -v & -1\cr 0 & v \cr\end{pmatrix}\end{equation}
where we have set
\begin{equation}\label{tomogro}v=-\frac{ K'}{2K}+\frac{Q'}{Q}=-\sum_a \frac{ k_a/2}
{z-z_a}+\sum_i\frac{1}{z-w_i}.\end{equation}
In the last step, we used $K=\prod_a (z-z_a)^{k_a}$, $Q=\prod_i(z-w_i)$.
Finally, a lower triangular gauge transformation 
\begin{equation}\label{morgo} \begin{pmatrix} 1 & 0 \cr v & 1\end{pmatrix}\end{equation}
leads to our final result for $\A_z$:
\begin{equation}\label{torog} \A_z=\begin{pmatrix} 0 & -1 \cr \bt & 0 \end{pmatrix},\end{equation}
with
\begin{equation}\label{orog} t= -v'-v^2.\end{equation}
In general, $\bt$ has poles at both the $z_a$ and the $w_i$.  Near $z=z_a$,
\begin{equation}\label{roog}t\sim-\frac{ j_a(j_a+1)}{(z-z_a)^2}+\frac{c_a}{z-z_a}+\dots, ~ j_a=k_a/2.\end{equation}
Near $z=w_i$,
\begin{equation}\label{frog}t\sim  \frac{1}{z-w_i}
\left(\sum_a\frac{k_a}{w_i-z_a}-\sum_{j\not=i}\frac{2}{w_i-w_j}\right).\end{equation}
Thus, $\bt$ has no singularity at $z=w_i$ if and only if the Bethe equations (\ref{bethe}) are satisfied.

To get some more insight, set $\D_z=\partial_z+[\A_z,\,\cdot\,]$ and look for a flat section,
that is a holomorphic solution of  $\D_z\begin{pmatrix}f\cr \tilde f\end{pmatrix}=0$.
We find that $\tilde f=f'$ and
\begin{equation}\label{opeq}\left(\frac{\partial^2}{\partial z^2} +t(z)\right)f = 0.\end{equation}
We have carried out this derivation using a particular local coordinate $z$, but the notion
that we started with -- a flat bundle with an oper structure -- did not depend on the local coordinate.
So eqn. (\ref{opeq}) must be covariant under a change of the local coordinate, with a suitable
transformation for $\bt$.  A short calculation (or a more careful study of the above derivation)
shows that under a change of local coordinate $z\to \tilde z$, and a suitable transformation of 
$\bt$, 
the object $\bt$ transforms like a stress tensor in two-dimensional conformal field theory.
In other words, it transforms not as a quadratic differential, as one might naively think
from its pairing with the second derivative $\partial^2/\partial z^2$ in (\ref{opeq}), but with
an ``anomalous'' term involving the Schwarzian derivative $\frac{1}{2} \{z,\tilde z\}$.
The double pole in $t$ at $z=z_a$ is as if there is a primary field inserted at $z_a$.

If we had started from $\A_z = \zeta^{-1} \mathrm{diag}(\ca,-\ca)$ and used the same sequence of gauge transformations,
with $K=PQ'-QP' - 2 \ca \zeta^{-1} PQ$, we would have arrived to the same formulas, but with an extra constant term  in $v$:
\begin{equation}\label{tomogro2}v=- \frac{\ca}{\zeta}-\sum_a \frac{ k_a/2}
{z-z_a}+\sum_i\frac{1}{z-w_i}.\end{equation}
Now $\bt(z)$ has a pole of order four at $z=\infty$. This would correspond in conformal field theory to the insertion at infinity
of a somewhat unusual operator \cite{G}.

Finally, the Bethe equations have the following interesting property. They describe stationary points at fixed $z_a$ (and $\ca$)
of a Yang-Yang function:
\begin{align}\label{yy}
\W(w_i,z_a) = -  \sum_{i < j} \log((w_i - w_j)^2) + \sum_{i,a} k_a \log(w_i - z_a) & -
 \frac{1}{4} \sum_{a < b} k_a k_b \log( (z_a - z_b)^2) \cr & +\frac{2 \ca}{\zeta} \sum_i w_i - \frac{\ca}{\zeta} \sum_a k_az_a
\end{align}
In other words, the Bethe equations can be written as
\begin{equation}
\frac{\partial \W}{\partial w_i} =0.
\end{equation}

The Yang-Yang function has another interesting property: the coefficients of the single poles in $\bt$ at $z=z_a$ -- sometimes called accessory parameters --   are given by
\begin{equation}
c_a = \frac{\partial \W}{\partial z_a}.
\end{equation}
Some terms in $\W$ which are independent of the $w_i$ have been included to insure that this relation is satisfied.
A similar relation holds for $\ca \partial_\ca \W$.

\subsection{Opers and Virasoro Conformal Blocks}\label{opbl}

In this section, we will show how these formulae arise naturally in the semiclassical limit of
Virasoro conformal blocks. The semiclassical limit is defined as a limit in which the central charge $c$ of the Virasoro algebra goes to infinity,
while the conformal dimensions of operators also scale in the same way as $c$.

There is a useful way to parametrize the central charge: $c=1+6 Q^2$, with $Q=b + b^{-1}$. In this parametrization,
we take $b \to 0$ to get a semiclassical limit. The conformal dimensions of operators are conveniently parametrized as $\Delta = \alpha(Q-\alpha)$.
The parameter $\alpha$ is often referred to as ``momentum.''
We keep $b \alpha=\eta$ fixed as $b\to 0$.  Then the insertion of an energy-momentum tensor $T$
in a correlation function scales as $b^{-2}$, and we can define
the finite limit $\bt = b^2 T$.
We propose to identify this $\bt$,  inserted in certain conformal blocks, with the $\bt$ of section
\ref{opstress}.

The quantum stress-tensor $T(z)$ has an
anomaly under conformal transformations; it shifts by a multiple $\frac{c}{12} \{z,\tilde z\}$
of the Schwartzian derivative.
Hence $\bt=b^2T$ has a conformal anomaly that is independent of $b$ for $b\to 0$.
The behavior of $\bt$ near $z=z_a$ in the previous section corresponds to the behavior near a
Virasoro primary field with $\alpha_a = - \frac{k_a}{2b}$.
These operators are very special:
correlation functions and conformal blocks which involve only operators of this type
can be described very easily by a
free-field realization. We can describe
such a  realization in close parallel to the discussion in the previous section.

Let $\chi$ be a two-dimensional free field with two-point function
$\langle\chi(z)\chi(z')\rangle=-\frac{1}{2} \ln(z-z')$.
A standard way to construct an energy-momentum tensor of central charge $c=1+6 Q^2$
is to take
\begin{equation}
T = - :\partial \chi \partial\chi: + Q \partial^2 \chi.
\end{equation}
If we define $v=-b \partial \chi$, this definition reduces to (\ref{orog})
in the limit $b \to 0$. This is the first hint that a free-field realization
can be useful for us. Operators of dimension
$\Delta = \alpha(Q-\alpha)$ can be readily described as
normal-ordered exponentials of the free boson,
\begin{equation}
V_{\alpha}(z) = :e^{2\alpha \chi(z)}:\, .
\end{equation}

A second hint comes from (\ref{tomogro}): the quantity $v$ is the semiclassical limit of the expectation
value of $-b \partial \chi$ in the presence of chiral vertex operators
of
momenta $- {k_a}/{2b}$ at $z=z_a$ and of momenta
$1/b$ at $z=w_i$. The operators $V_{1/b}(w_i)$ have dimension $1$. They are
usually called  screening operators in free-field realizations \cite{DF,Felder}, and are naturally
integrated over curves. The singular part
of the stress tensor near the location of a screening operator
\begin{equation}
T(z) V_{1/b}(w_i) \sim \frac{1}{(z-w_i)^2} V_{1/b}(w_i)  +
\frac{1}{z-w_i} \partial_{w_i} V_{1/b}(w_i) +\cdots = \partial_{w_i}
\left( \frac{1}{z-w_i} V_{1/b}(w_i) \right) \cdots
\end{equation}
is a total derivative, and drops off upon integrating over the position of the screening
operator.

We can easily compute the following free field correlation function:
\begin{align}\label{easy}
\left\langle \prod_i  V_{1/b}(w_i)
\prod_a  V_{-k_a/2b}(z_a)  \right\rangle_{\mathrm{free}}& = \prod_{i < j}(w_i - w_j)^{-\frac{2}{b^2}
}\prod_{i,a}(w_i - z_a)^{\frac{k_a}{b^2}}
\prod_{a <b}(z_a - z_b)^{-\frac{1}{2b^2}k_a k_b}\cr & = \exp\left({\frac{1}{b^2} \W(w,z)}\right).
\end{align}
The exponent on the right is the Yang-Yang function (\ref{yy})! (For the moment,
the terms proportional to $\ca$ are absent as we  have not included symmetry breaking.)

Consider the integral
\begin{equation} \label{freeinto}
\left\langle  \prod_a  V_{-k_a/2b}(z_a)
\right\rangle_{\Gamma} = \int_\Gamma \left\langle \prod_i
 V_{1/b}(w_i)  \prod_a  V_{-k_a/2b}(z_a)  \right\rangle_{\mathrm{free}}
\prod_i \d w_i,
\end{equation}
where $\Gamma$ is any integration cycle for which the integral converges.
Because of (\ref{easy}), this is equivalent to
\begin{equation} \label{freeint}\left
\langle  \prod_a  V_{-k_a/2b}(z_a)\right
\rangle_{\Gamma} = \int_\Gamma \,\,\exp(\W(z_a,w_i)/b^2)\,\,\prod_i \d w_i.
\end{equation}

A Virasoro conformal block for the expectation value of a product of primary fields
is a candidate correlation function that is compatible with the Virasoro Ward identity:
\begin{equation}\left
\langle  T(z) \prod_a  V_{-k_a/2b}(z_a)  \right\rangle_{\Gamma} =
\left( \sum_a \frac{\Delta_a}{(z-z_a)^2} +
\frac{1}{z-z_a} \frac{\partial}{\partial z_a} \right)\left \langle
\prod_a  V_{-k_a/2b}(z_a) \right \rangle_{\Gamma}
\end{equation}
The functions defined in (\ref{freeinto}) or (\ref{freeint}) have this property
for an arbitrary choice of the number of $w$'s and the integration cycle $\Gamma$; this is
proved using the definition (\ref{freeinto}) and the fact that the screening charges are
primary fields of dimension 1.
(The function $\left\langle  T(z) \prod_a  V_{-k_a/2b}(z_a) \right \rangle_{\Gamma} $
is defined by the integral (\ref{freeinto}) with an insertion of $T(z)$
in the free field correlation function on the right hand side.)  What we have just
described is the free-field realization of the conformal blocks \cite{DF,Felder}.

The space of possible integration
cycles $\Gamma$ for the integral (\ref{freeint})  has a natural
flat connection (the Gauss-Manin connection) as the points $z_a$, $a=1,\dots,d$ vary.
Hence the functions $\langle  \prod_a  V_{-k_a/2b}(z_a)
\rangle_\Gamma$ --
for any fixed number of $w$'s --  furnish a representation of the braid group on
$d$ strands.
It is known \cite{Lawrence,SV} that these are precisely the representations of the braid
group that are associated to the Jones polynomial and its generalizations.
This relation of the Jones polynomial to
conformal field theory will be more useful for the present paper than
the relation via the Knizhnik-Zamolodchikov equation, which was noted in
section \ref{relconf}.

The functions  $\langle  \prod_a  V_{-k_a/2b}(z_a)
\rangle_\Gamma$ are not all possible conformal blocks for a product of primary
fields with the dimensions of the $V_{-k_a/2b}$; rather, they are all such conformal
blocks if the operators $V_{-k_a/2b}$ are degenerate primary fields in the sense
introduced in \cite{BPZ}.
Alternatively, these functions are all possible conformal blocks
if the oper derived from the small $b$ limit of $\bt=b^2T$ is supposed to have trivial
monodromy at the points $z=z_a$.    These concepts and the relation between them
are described in section \ref{degtriv}.

\subsubsection{The Irregular Case}\label{irreg}

Now let us incorporate the
complex symmetry breaking parameter $\ca$ in the
above discussion.  We can certainly in the integral (\ref{freeint}) over the $w$'s modify
the exponent to include the terms proportional to $\ca$ in
the Yang-Yang function (\ref{yy}).
But what does this mean in conformal field theory?  We need to replace the free
field correlation function (\ref{easy}) by
\begin{align}\label{oddball}
&\left\langle \prod_i
V_{1/b}(w_i)  \prod_a   V_{-k_a/2b}(z_a) \right \rangle_{\mathrm{free}}  \cr & =
\prod_{i \neq j}(w_i - w_j)^{-\frac{2}{b^2}}\prod_{i,a}(w_i - z_a)^{\frac{k_a}{b^2}}
\prod_{a \neq b}(z_a - z_b)^{-\frac{1}{b^2}k_a k_b}
e^{\frac{\ca}{\zeta b^2}
 \left( 2 \sum_i w_i - \sum_a z_a  \right)}.
 \end{align}
What is the conformal field theory interpretation of this formula?

Almost by construction, the right hand side is the free-field correlation
function of the given product of fields with peculiar boundary conditions for $\chi$ at infinity,
$\chi \sim {\ca z}/{b \zeta} $. Alternatively, we have inserted at infinity an ``irregular vertex operator'', i.e. the $L \to \infty$ limit of 
\begin{equation}
\exp\left({-\frac{2\ca}{\zeta b} L^2 \partial \chi(L) }\right).
\end{equation}

In the presence of such an irregular vertex operator, the stress-tensor
has the expected degree four pole.
Nothing changes in the above formulae, except that the choice of
possible integration contours is enlarged.
The result of the integral is a conformal block with an irregular puncture at infinity,
as defined in \cite{G}, in addition to the standard punctures of momenta $-{k_a}/{2 b}$.
In the context of free fermions, operators associated to irregular singularities
were originally defined in \cite{Miwa,Moore}.

\subsubsection{Degenerate Primary Fields And Trivial Monodromy}\label{degtriv}

Now we will review some standard facts about representations of the Virasoro algebra.
This will enable us to explain what is special about the particular conformal blocks that
come from the free field representation.

For a generic value of the conformal dimension $\Delta$, the Verma module
defined as the span of
all possible Virasoro descendants of a highest weight vector of dimension $\Delta$ is irreducible.
For a set of special values
\begin{equation} \alpha = \alpha_{r,s} = - \frac{(r-1) b}{2} - \frac{s-1}{2 b},~~~r,s=1,2,3,\dots \end{equation}
or $\alpha = Q- \alpha_{r,s}$, this is not true: a certain descendant
at level $r s$ is again a highest weight vector, and has zero norm.
In a unitary conformal field theory, the descendant in question will vanish, and even in
a non-unitary theory, it might vanish.  The primary field whose descendant vanishes
is called a degenerate primary field.  We write $\Phi_{r,s}$ for such a field.

The vanishing descendant of a degenerate primary field will certainly decouple in
correlation functions.
   We call conformal blocks obeying
such a relation degenerate conformal blocks.
They satisfy a condition known as
 a ``degenerate fusion rule.''  (It can be proved using the differential equation that
 we mention shortly.)
In the OPE of an operator  $\Phi_{r,s}$ and an operator of momentum
$\alpha$, only operators of
momentum $\alpha - \frac{r' b}{2} - \frac{s'}{2 b}$  appear, with
\begin{equation}
r' = r-1 , r-3, \cdots ,1-r \qquad s' = s-1, s-2 \cdots ,1-s.
\end{equation}

A conformal block with an insertion of momentum $\alpha_{r,s}$ satisfies
the null-vector decoupling condition if and only if
its correlation functions  satisfy a certain differential equation of order $rs$.
An important special case is  $r=2$, $s=1$; the equation is
\begin{equation}\label{opeq2}
\partial^2 \Phi_{2,1}(z) + b^2 : T(z) \Phi_{2,1}(z): =0.
\end{equation}

This reduces   in the semiclassical limit to the differential equation
(\ref{opeq}) associated to an oper.   (For $b\to 0$, the normal ordering in (\ref{opeq2}) is
irrelevant; the only part of $b^2T$ that survives for $b\to 0$ is the response to the ``heavy''
fields with momenta of order $1/b$.)
Moreover, in the semiclassical limit, the expectation value of 
$\bt=b^2T$ will always be such that the monodromy
of the differential equation around points with additional $\Phi_{1,s}$ insertions is trivial.
Indeed, the degenerate fusion rule
implies that  the OPE of $\Phi_{2,1}$ and $\Phi_{1,s}$ contains only one primary field
$\Phi_{2,s}$; from this, it follows that
the monodromy of $\Phi_{2,1}$ around a $\Phi_{1,s}$ puncture is trivial.
This is a quantum version of the
trivial monodromy condition on the differential equation (\ref{opeq}) associated to the oper.
  On the other
hand, the singularity of $\bt$ near a $\Phi_{1,s}$ insertion is precisely that which we have
exhibited in (\ref{roog}) (with $j=s/2$).  The upshot of this is that the semiclassical
limit of a conformal block for a correlation function
$\left\langle \prod_{a=1}^d\Phi_{1,k_a}(z_a)\right\rangle$ determines an oper with precisely the sort
of monodromy-free singularities that we extracted from three-dimensional gauge theory
in section \ref{opsing}.

The  conformal blocks constructed from the free field formula (\ref{freeinto})
describe correlation functions of degenerate primary fields, simply
because the free field vertex operators of momenta $-\frac{k_a}{2b}$ are
degenerate.   Related to this, it
is possible to show that the conformal blocks which are given by the free-field realization
do satisfy the degenerate fusion constraints.
The corresponding
differential equations are equivalent to the Picard-Fuchs equations satisfied by the
free-field integrals, or to the natural flat connection on the space of integration cycles.
They are a close analogue to the Knizhnik-Zamolodchikov equations.  The free field
realization gives all the conformal blocks for the correlation function
$\left\langle \prod_{a=1}^d\Phi_{1,k_a}(z_a)\right\rangle$ that are allowed by the fusion rules,
so there are no more to be had.

The interpretation of opers with monodromy-free singularities  in terms of correlation
functions of degenerate conformal fields gives an intuitive explanation to the Bethe
equations.  Naturally, $\bt$ should have no poles at the points $w_i$, because no
conformal fields are inserted there.

\section{Four-Dimensional Solutions and Parallel Transport}\label{fourd}
\subsection{Introduction}\label{fourdintro}

We now turn to the problem of analyzing time-dependent solutions of 
the original BPS equations (\ref{bpseqns}).
Even though this is a problem of classical partial differential equations, a 
quantum mechanical view is helpful.

\begin{figure}
 \begin{center}
   \includegraphics[width=3.5in]{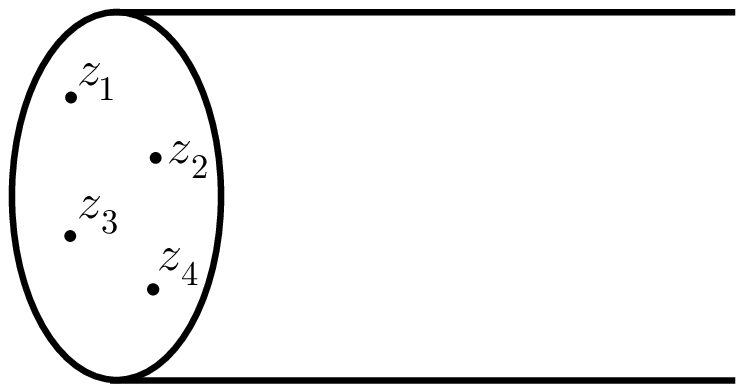}
 \end{center}
\caption{\small  A snapshot at fixed time of a time-independent situation.  In the three-manifold $M_3=C\times \R_+$, knots
are present at boundary points $z_1,\dots,z_4$.}
 \label{again}
\end{figure}
We start with a time-independent situation -- twisted $\N=4$ super Yang-Mills 
on a four-manifold $\MM=\R\times \Mthree$,
where $\Mthree$ is a three-manifold and we think of $\R$ as the time direction. 
In our application, $\Mthree=C\times \R_+$, with $C$ a Riemann
surface.  If knots are present, we assume initially that they are time-independent.
In this situation,
which is depicted again for convenience in fig. \ref{again},
we want to find the BPS states -- quantum ground states.

The first approximation, already analyzed in sections \ref{teqone} and \ref{analog}, is to find time-independent classical solutions.
In going from classical solutions to BPS states, we will ignore the noncompactness of $\Mthree$.
This means that we will ignore the existence of a continuum of non-BPS excitations.

 If there is only a finite set of classical solutions of the BPS equations (and they
 are nondegenerate -- there
are no zero modes in expanding around such a solution), then the classical
approximation to the space of BPS states is very simple.
Let $\I$ be the set of classical solutions.  Then for each $I\in \I$, there is in perturbation
theory a quantum ground state $\psi_I$ that
is localized near $I$.   In perturbation theory, the $\psi_I$ form a basis for the space
$\H$ of BPS states.

Nonperturbatively, in problems of this general type, instanton effects might lift some of these approximate ground states
away from zero energy.  However, we are
actually here dealing with a problem in which this does not occur.  This is because in the
time-independent case, even with knots present,
our problem has four supercharges, and  an instanton (that is, a classical solution with non-trivial
time dependence) violates at least
two of them.  This leads to the existence of two fermion
zero modes in an instanton background, which is one too many to contribute to a
matrix element of the supercharge $Q$ between approximate
ground states $\psi_I$ and $\psi_J$.

\subsubsection{Relation To Morse Theory}\label{relm}

A more explicit understanding of why instantons do not lift the classical vacuum
degeneracies comes from the relation of this problem to Morse
theory.\footnote{For a relatively accessible introduction to the relevant aspects of Morse theory,
see \cite{hutchings}.}  Supersymmetric quantum mechanics related to Morse theory \cite{Morse} is, in general,
a theory of maps from $\R$ to $\U$, where $\U$ is a Riemannian manifold with metric tensor
$g$  endowed with a real-valued
function $h$ that we call the superpotential.  
For generic $\U$ and $h$, the model has
two supercharges, one of which is conjugate to the exterior derivative:
\begin{equation}\label{zolmo} Q=e^h\d e^{-h}.\end{equation}
The classical vacua correspond to critical points of $h$.    If $h$ is a Morse function
-- that is, its critical points are all isolated and nondegenerate -- then in perturbation theory,
each critical point $I$ corresponds to an approximate quantum ground state $\psi_I$.
The fermion number $q_I$ of $\psi_I$ is equal to the Morse index of the critical point $I$
(the number of negative eigenvalues of the matrix of second derivatives of the function $h$
at $I$).  Since $Q$ increases the fermion number by one unit, quantum corrections inducing
non-zero matrix elements $\langle \psi_J|Q|\psi_I\rangle$ are possible only if
\begin{equation}\label{yre} q_J=q_I+1.\end{equation}
Such nonzero matrix elements can be computed by counting, in a suitable sense,
the instanton solutions that interpolate between the critical point $I$ in the far past and
the critical point $J$ in the far future.
The relevant ``instanton'' equations, in other words the conditions for a map $\R\to \U$ to
be $Q$-invariant, are the gradient flow equations of Morse theory:
\begin{equation}\label{orlox} \frac{\d x^i}{\d t}=-g^{ij}\frac{\partial h}{\partial x^j}.\end{equation}

The problem we are studying of twisted $\N=4$ super Yang-Mills theory on
$\R\times \Mthree$ (we primarily take $\Mthree=C\times\R_+$ but the following remarks
are more general) is an infinite-dimensional problem of this sort,\footnote{\label{zort}
To be more precise, our problem is a gauge-invariant version of a such a problem -- corresponding to a supersymmetric sigma-model with target $\U$
coupled to gauge fields that gauge a symmetry of $\U$.  The gauge group in our case is the group
of maps from $\Mthree$ to the finite-dimensional group $G$, while $\U$ is the space of complex-valued connections on $\Mthree$.  However, in our problem the gauge group acts freely
on $\U$; this is ensured by the Nahm pole boundary condition.   As a result,
the gauge-invariance will not play a major role.
In effect,  for our purposes, we can replace $\U$ by its quotient by the group of complex
gauge transformations and reduce to the case that there are no gauge fields.}
with $\U$ being the space of complex-valued connections on $\Mthree$.  We view the
Chern-Simons function
\begin{equation}\label{csfun}\W=\frac{1}{4\pi i }\int_{\Mthree}\,\Tr\,\left(\A\wedge\d \A+\frac{2}{3}
\A\wedge\A\wedge\A\right)\end{equation}
as a holomorphic function on the complex manifold $\U$.  Holomorphy means that the
one-dimensional sigma model with target $\U$ and superpotential $\W$
has four supercharges (this actually depends on the fact that the metric of $\U$ is Kahler and is also true in the gauge-invariant
case mentioned in footnote \ref{zort}).   We actually want to study this model in the context of
a twisting of $\N=4$ super Yang-Mills theory in which a particular supercharge
$Q$ is distinguished.  This supercharge depends on a twisting parameter $\t$ \cite{KW}
and is an infinite-dimensional version of $Q=e^h\d e^{-h}$
where $h$ is the ordinary Morse function
\begin{equation}\label{morsefn}h=\mathrm{Re}\,(e^{i\alpha}\W),\end{equation}
and
\begin{equation}\label{alphaq} \t = \frac{1- \sin \alpha}{\cos{\alpha} }\end{equation}
For a Morse function of this type, the gradient flow equation becomes
\begin{equation}\label{lphaq} \frac{\d \bar w^{\bar j}}{\d t}=-g^{\bar j i}\frac{e^{i\alpha}}{2}
\frac{\partial\W}{\partial w^i},\end{equation}
where the $w^i$ are local holomorphic coordinates on $\U$.

A down-to-earth manifestation of the relation of our problem to Morse theory is
that the underlying four-dimensional supersymmetric equations (\ref{bpseqns})
are the gradient flow equations (\ref{orlox}) for the Morse function $h$ on the infinite-dimensional
manifold $\U$.  (This is one of the main ideas in \cite{Analytic,NewLook},
where the gauge invariance mentioned in footnote \ref{zort} has been taken into account.)

Now we can give a more explicit explanation of why nonperturbative effects in our problem
will not spoil the supersymmetry of any of the approximate quantum ground states
$\psi_I$.  In general, for a Morse function that is the real part of a holomorphic function,
isolated critical points all have the same (middle-dimensional) Morse index and thus
the same value of the fermion number $q$.  Hence the condition (\ref{yre}) is never
satisfied.

Another route to the same result is as follows.  In general,
gradient flow for a Morse function such as $h$ that is the real part of a holomorphic function
has a conserved quantity, namely the imaginary part of the relevant holomorphic function,
in our case
\begin{equation}\label{conserved}  j = \mathrm{Im}\,(e^{i\alpha}\W). \end{equation}
For generic $\alpha$, all critical points have distinct values of $j$ and hence distinct critical
points cannot be connected by  a solution of the gradient flow equation.  Hence there are no instantons that
might spoil the supersymmetry of the states $\psi_I$.

\subsubsection{Time-Dependence}\label{time}

As explained in section \ref{methods}, we do not literally want to
consider a time-independent situation; rather, we want to allow
for a slow time-dependence of the positions of the knots.  Let
$\ZZ$ be the space of distinct points $z_1,\dots,z_d\in C$.   The space $\H$ of BPS
states is the fiber of a bundle $\hat\H$ over $\ZZ$.   This bundle carries a natural flat connection.
This is a general property of topological field theory, but
the Morse theory interpretation leads to a particularly nice description.

In supersymmetric quantum mechanics related to Morse theory,
since the supercharge $Q=e^h\d e^{-h}$ is conjugate
to the exterior derivative $\d$,  the ground states $\psi_I$ associated to critical
points must have an interpretation in terms of the cohomology or dually the homology
of $\U$.  There is a standard way to understand this in Morse theory.  To a critical point
$I$, one associates the downward-flowing cycle $\J_I$ consisting of all points in $\U$
that can be reached by gradient flow starting at $I$.  In other words, one considers
solutions of the gradient flow equation on a half-line $(-\infty,0]$, with initial conditions
that the flow starts at $I$ at $t=-\infty$.  $\J_I$ parametrizes the values at $t=0$ of such
flows.

The case that $h$ is the real part of a holomorphic function has special features,
and is particularly simple, so let us focus on that case.  We make the further simplifying assumption that the critical points
of $h$ are isolated and irreducible.  $\U$ is inevitably not compact
(or it would not admit a non-constant holomorphic function).  In this very special situation, the $\J_I$ are
called Lefschetz thimbles.  The thimbles $\J_I$ are closed
if the angle $\alpha$ used in defining $h$ is sufficiently generic (to prevent the existence
of gradient flows between distinct critical points), but they are not compact.
So they do not represent classes in the ordinary homology of $\U$.  However, as $\J_I$
is defined by downward gradient flow with respect to $h$, one has $h\to -\infty$ at infinity
along $\J_I$.  As a result, the $\J_I$ are elements of a certain relative homology group
-- the homology $H(\U,\U_<)$ of $\U$ relative to the region $\U_<$ where $h$ goes to
$-\infty$.  (In the notation, we do not indicate the dimension of a homology cycle, because
this relative homology is nonzero only in the middle dimension.  That is related to the
fact that the critical points all have a middle-dimensional Morse index.)

The space $\hat\H$ of supersymmetric ground states can be identified with the relative
homology $H(\U,\U_<)$.  In this correspondence, the quantum ground state $\psi_I$
associated to a critical point $I$ maps to the element $\J_I$ of $H(\U,\U_<)$.  For
an explanation of this from a physical point of view (in the context of
supersymmetric quantum mechanics related to Morse theory), see \cite{FLN} or \cite{NewLook}.

The interpretation in terms of relative homology means that $\hat\H$ has an integral
structure and hence a natural flat connection. To give it a fancy name, this
flat connection is the Gauss-Manin connection on the relative homology.
This connection is trivial
for generic values of $\alpha$ and the $z_i$: the
$\psi_I$ are flat sections, and the connection on $\hat\H$ is
fully described by the smooth evolution of the classical critical points and corresponding
thimbles.  (In transporting the thimbles, one must keep track of their orientations;
the sign of the relative homology class associated to $\J_I$ depends on the orientation
of $\J_I$.)

Crucially, there are codimension one walls in the space $S^1 \times \ZZ$ of parameters
$\alpha$ and $z_i$ where the thimbles fail to be closed (and so do not define elements
of the relative homology), and
the map from the critical points to quantum states jumps discontinuously.
This can occur if there are gradient flow lines from $I$ to $J$; in this case, $\J_I$ is not
closed as it contains points arbitrarily close to $J$, but not $J$ itself.

We write $\ell_I^J$ for the locus in $S^1\times \ZZ$ on which the following necessary
conditions are obeyed for flows from $I$ to $J$: the value of the conserved quantity
$j$ is equal at $I$ and $J$, while $h(I)>h(J)$.  The first condition is a single real condition,
while the second is just an inequality.  So $\ell_I^J$ is of real codimension 1, and we call
it a Stokes wall.

In crossing a Stokes wall $\ell_I^J$, only the thimble $\J_I$ becomes
ill-defined.  It  jumps by a multiple of $\J_J$:
\begin{equation}\label{stokesS}
\J_I \to \J_I + \frak{m}_{IJ} \J_J,
\end{equation}
where $\frak{m}_{IJ}$ is the ``number'' of  gradient flow lines
from $I$ to $J$ counted in an appropriate sense.   A given line contributes 1 or $-1$ to the sign
 depending on the direction in which the difference between the values of
  $j=\mathrm{Im}(e^{i \alpha} \W(w,z))$ at $I$ and $J$ passes through zero.
For an elementary explanation of such matters, see section 2 of \cite{Analytic}.

The correspondence between states $\psi_I$ and thimbles $\J_I$ means that
the $\psi_I$ have the same jumping in crossing Stokes walls.
In the $\psi_I$ basis, the connection is trivial except
across the Stokes walls, where the transport matrix
is a triangular ``Stokes factor''
\begin{equation}\label{sf}
S[\ell_I^J] = 1 + \frak{m}_{IJ} e^J{}_I
\end{equation}
Here $e^J{}_I$ is the matrix whose only non-zero element is $1$ at position $J$, $I$.

In particular, the parallel transport along a path $\P$ in $S^1 \times \ZZ$
is  a path-ordered product of factors of the following kind: (a)
between two Stokes walls, one has only the ``formal monodromy'' which
expresses the permutations of the classical critical points, with minus signs
that keep track of the orientations of the thimbles; (b) every time
one crosses a Stokes wall, the monodromy acquires a corresponding Stokes factor.

One can visualize the matrix elements $\frak{n}_I^J[ \P ]$  of the transport matrix for the path
$\P$
as counting paths from a critical point $I$ to a critical point $J$, where away from Stokes walls, one
has to follow  a critical point continuously, but in crossing a wall, one is allowed
to ``jump'' along a gradient flow trajectory from one critical point to the next.
A matrix element of the transport matrix is given by a sum of contributions of hybrid
paths of this type, with each path contributing 1 or $-1$ depending on how the orientation
of a thimble evolves along the given path.

Incidentally, the fact that the monodromy for parallel transport along a path $\P$
depends only on the homotopy class of $\P$ implies wall-crossing
formulas for the numbers $\frak{m}_{IJ}$ that control the jumping.
More generally, the whole picture can be interpreted in terms of
BPS states in an LG model based on $\W$, but we will not pursue that interpretation in this paper. 

Let us collect a few properties of the matrix elements $\frak{n}_I^J[ \P ]$:
\begin{itemize}
 \item $\frak{n}_I^J[ \P ]$ only depends on the homotopy class of  $\P$;
  \item The composition of paths maps to matrix multiplication:
  $\frak{n}_I^K[ \P_1 \circ \P_2 ]= \sum_J\frak{n}_I^J[ \P_1 ]\frak{n}_J^K[ \P_2 ]$;
  \item For an infinitesimal path $\delta \P$ from $z$ to $z + \delta z$
  it is almost always true that $\frak{n}_I^J[ \delta \P ] = \delta_I^J$ --  this fails
  only in crossing a  Stokes wall;
  \item  In crossing a Stokes wall, $\frak{n}_I^J[ \delta \P ] - \delta_I^J = \frak{m}_{IJ}$, where
  $\frak{m}_{IJ}$ is computed by a count of flow lines.
\end{itemize}

\subsubsection{The Dual Basis}\label{dual}

If $\Gamma$ is any cycle in the relative homology $H(\U,\U_<)$, then as the thimbles are a basis
for this relative homology, $\Gamma$ is equivalent
in relative homology to a linear combination of thimbles,
 \begin{equation}\label{holm} \Gamma=\sum_I\frak c^I \J_I.\end{equation}
How can the coefficients $\frak c^I$ be determined?  If we had, in some sense, a dual basis $\K^I$
to the $\J_I$'s with pairings
\begin{equation}\label{oggo} \langle \K^J,\J_I\rangle=\delta^J{}_I,\end{equation}
then we would identify the coefficients in (\ref{holm}) as
\begin{equation}\label{ggo} \frak c^I=\langle \K^I,\Gamma\rangle.\end{equation}

The Poincar\'e dual of the relative homology $H(\U,\U_<)$ is the opposite relative
homology $H(\U,\U_>)$ of $\U$ relative to the region $\U_>$ with $h\to+\infty$.  A natural
basis of $H(\U,\U_>)$ is given by the upward-flowing thimble $\K^I$ associated to
critical points. For each critical
point $I$, $\K^I$ is  defined
as the boundary values at $t=0$ of solutions of the gradient flow equation on the half-line
$[0,\infty)$ that approach the point $I$ for $t\to+\infty$.   Since $h$ decreases along gradient flow lines, the smallest value it assumes
along such a flow is its value at $t=\infty$,  which is its value at the critical point $I$.
So $h$ is bounded below along $\K^I$, but possibly not bounded above, and $\K^I$ takes
values in $H(\U,\U_>)$.   As for the pairing (\ref{oggo}), from the definitions of the $\K$'s and
$\J$'s,
$\langle\K^J,\J_I\rangle$ counts flows on the full real line $(-\infty,\infty)$ that start
at $J$ at $t=-\infty$ and end at $I$ at $t=+\infty$.  For $J\not=I$ there are no such flows
(for generic $\alpha$ where the thimbles are well-defined). For $J=I$, since $h$ strictly
decreases along a nonconstant flow, the only flow is the constant one that sits at $J$ for all times.   For a certain natural relative orientation of the $\J$'s and $\K$'s, the contribution of the constant flow to the pairing is $+1$.

\subsubsection{Non-Single Valued Superpotentials and $q$-Grading}\label{nonsingle}

In the framework that we have presented so far, the
matrices which represent the action of the braid group
on $\H$ have integer-valued entries, since the relative homology has an integral
structure.  Now we want to consider a situation where $\W$
is not single-valued. To be more precise, we will consider
a holomorphic function like the Chern-Simons functional $\W$, whose real part is single-valued,
but whose imaginary part is well defined only modulo $2 \pi \Z$.
We can reduce to the framework which we have employed so far by replacing $\U$
by the smallest cover $\hat \U$ on which $\W$ is single-valued.

Passing to $\hat\U$ comes at the cost that now the number of critical points
will be infinite, since each critical point in $\U$ has infinitely many preimages in $\hat \U$.
A critical point $\hat I$ in $\hat \U$ is the same as a critical point $I$ in
$\U$ together with a choice of a branch  of $\W$ at $I$.
Locally, we can pick an arbitrary preimage $\hat I_0$ of a critical point $I$, and denote
as $\hat I_n$ the critical point for which the value of $\W$ is shifted
by $-2 \pi i n$ compared to the value
at $\hat I_0$.

The Stokes factors and transport matrices are now matrices of infinite size, but
their matrix elements  can be computed by the techniques we have described, and are integer-valued.
As the calculations only depend on the gradient of $\W$, these matrices
commute with the deck transformation $\hat I_n \to \hat I_{n+1}$.
It is convenient to introduce a variable $q$ taking values in $\C^*$.
Then if   we simply write $q^n\psi_I$ as a symbolic shorthand for $\psi_{\hat I_n}$, we can replace
infinite-dimensional matrices whose entries are integers with
finite-dimensional matrices whose entries are  Laurent polynomials in $q$ with integer
coefficients. We denote such a matrix  as $S[\ell_I^J;q]$. In order to keep track of the lift of
$\W$
as we move in $\ZZ$, the formal monodromy matrices which encode the permutation
of critical points and the Stokes matrices associated to gradient flows
are valued in powers of $q$,
to keep track of the change of $\W$ along a path.   The matrix elements of the
transport matrix are now obtained by summing over hybrid paths (continuous evolution away
from Stokes walls and gradient flow across Stokes walls) with weight  $\pm q^n$, where $-2\pi i n$
is the change of superpotential along the path and as usual the sign involves the orientation
of the thimbles.

An alternative description of all this is as follows.
Since $\U$ admits the non-single-valued superpotential $\W$, its first Betti number
is positive and one can introduce a ``theta-angle'' $\theta$
in the supersymmetric quantum mechanics, weighting by $e^{in\theta}$ a path in
which $\W$ jumps by $-2\pi i n$.
This has the effect of replacing the relative homology of
$\U$ with a twisted version of the relative homology, valued in
a flat line bundle of monodromy $q=e^{i\theta}$.    Then we consider the Gauss-Manin
connection for homology twisted by this flat bundle.  The holonomy matrices
for this connection have entries that are Laurent polynomials in $q$ with integer coefficients
and they can be computed as just described.

\subsection{Classical Description of Counting of Four-Dimensional Solutions}\label{clco}

Now, let us consider the problem that we are really interested in -- time-dependent
solutions of the supersymmetric equations (\ref{bpseqns}) on $\R\times C\times \R_+$.
The solutions will be time-dependent because the boundary conditions are time-dependent --
we allow the positions $z_i$ of singular monopoles on the boundary of $\Mthree=C\times \R_+$
to vary with time.  Although we typically assume an adiabatic evolution of the monopole positions,
the counting of four-dimensional solutions is topological,
and the adiabatic assumption is not necessary.

In the simplest setup, with singular monopole strands at the boundary braided in time,
the superpotential depends holomorphically on some parameters (the positions of the strands
in $C$), and the parameters evolve in time.
Schematically, the equations take the form of a ``forced gradient flow''
\begin{equation} \label{flow}
\frac{\d \bar w_{\bar i}}{\d t} = -e^{i \alpha} g^{\bar i j}  \frac{\partial \W(w,z(t))}{\partial w_j}
\end{equation}

We suppose that the singular monopoles begin at positions $\vec z_i=(z_{1},\dots,z_{k})$ 
near time $t=-\infty$ and end at positions $\vec z_f=(z'_{1},\dots,z'_{k})$ near $t=+\infty$.
In fact, we can assume that the positions $z_i(t)$ of the singular monopoles
are constant except in a bounded
interval $-T<t<T$, for some $T$, during which they follow a path $\P$
in their parameter space $\ZZ$.  In such a situation, we can look for solutions of the forced
gradient flow equation that begin at a specified critical point $I$ of $\W(w,\vec z_i)$, and end at
a specified critical point $J$ of $\W(w,\vec z_f)$.   The ``number'' of such solutions, with each
solution weighted by the sign of the fermion determinant, is a topological invariant -- unchanged
under deformations of the path $\P$ or the metric on $\U$.  We will call this invariant $\frak N^J_I$. 
(For a reason that will be clear momentarily, we really only want to define $\frak N^J_I$ if
$\vec z_i$ and $\vec z_f$ are not on Stokes walls.)

In section \ref{time}, we already associated an integer invariant $\frak n^J_I[\P]$ to this situation.
$\frak n^J_I[\P]$ was a matrix element of the Gauss-Manin connection for transport along
the path $\P$ from $\vec z_i$ to $\vec z_f$.  One can think of $\frak n^J_I[\P]$ as the expansion
coefficients when a thimble $\J_I$ in the relative homology $H(\U,\U_<)_{\vec z_i}$ is transported
along the path $\P$ using the Gauss-Manin connection, and then expressed in terms
of the thimbles $\J'_J$ that furnish a basis of $H(\U,\U_<)_{\vec z_f}$:
\be\label{zelx}\J_I=\sum_J\frak n^J_I \J'_J.\ee
We claim that in fact
\be\label{elx}\frak N^J_I=\frak n^J_I.\ee
The importance of this relation is that $\frak N^J_I$ is what we want,
the counting of time-dependent solutions, while $\frak n^J_I$ is more easily computed,
since this requires only the study of time-independent problems.

One explanation of (\ref{elx}) is as follows. By definition, $\J_I[\vec z_i] $ is the set 
of points in $\U$ which can be reached by flows
\begin{equation}
\frac{\d \bar w_{\bar i}}{\d t} = -e^{i \alpha} g^{\bar i j}  \frac{\partial \W(w,z_i)}{\partial w_j}
\end{equation}
which asymptote to $I$ in the past.  Here the $z_i$ are regarded as constants.
Now consider the equation (\ref{flow}) for forced gradient flow on the semi-infinite
interval $(-\infty,t_0]$.  For $t_0\leq -T$, the values at $t_0$ of a solution of
this equation parametrize $\J_I[\vec z_i]$, but for $t_0>-T$, 
they  parametrize a $t_0$-dependent continuous deformation  of this space that we will call 
$\tilde \J_I[\vec z_i;t_0]$.

We saw in section \ref{dual} that that for any given cycle $\Gamma$ in
$H_m(\U ,\U_<)_{\vec z_f}$, the coefficients $\frak{c}^J $ in the expansion
\begin{equation}
\Gamma =\sum_J\frak{c}^J \J_J[\vec z_f]
\end{equation}
count, in the sense of an index, the number of flows
\begin{equation}
\frac{d \bar w_{\bar i}}{dt} = -e^{i \alpha} g^{\bar i j}  \frac{\partial \W(w,z_i)}{\partial w_j}
\end{equation}
which start from $\Gamma$ and asymptote to the critical point $J$ in the future.

Hence the number of four-dimensional solutions  which flow from $\tilde \J_I[\vec z_i;t_0]$ 
at some given time $t_0$ after the braiding occurs to the critical point $J$ in the future 
are the coefficients
$\tilde{\frak N}^J_I$ in the expansion
\begin{equation} \label{ntop}
\tilde \J_I[\vec z_i;t_0] =\sum_J\tilde {\frak N}_I^J\J_J[z_f].
\end{equation}
But since $\tilde\J_I[\vec z_i;t_0]$ parametrizes flows on the interval $(-\infty,t_0]$ that
start at $I$, a flow from $\tilde\J_I[\vec z_i;t_0]$ to $J$ on the interval $[t_0,\infty)$ is
equivalent to a flow from $I$ to $J$ defined on the whole real line.  So the 
$\tilde{\frak N}^J_I$ are the same as the desired invariants
$\frak N^J_I$:
\begin{equation} \label{nzop}
\tilde \J_I[\vec z_i;t_0] =\sum_J\frak N^J_I  \J_J[z_f]
\end{equation}
The relative homology is defined over $\Z$, and an integral relative homology class
such as $\tilde\J_I$ has no continuous deformations.   So
clearly, as long as the continuous deformation  from $\J_I[\vec z_i]$ to 
$\tilde \J_I[\vec z;t_0]$ induced by the flow equations (\ref{flow}) lives at any given 
time in $H_m(\U ,\U_<)_{\vec z(t)}$, it
coincides with the natural transport along $\P$ by the Gauss-Manin connection.
In this case, (\ref{nzop}) is equivalent to the desired result  $\frak{N}^J_I = \frak{n}^J_I$.  

To show that 
$\tilde\J_I[\vec z;t_0]$  lies in $H_m(\U,\U_<)_{\vec z(t_0)}$ for any $t_0$, we are supposed
to prove that $\mathrm{Re}\,\W(w,\vec z(t_0))$ goes to $-\infty$ at infinity  along
$\tilde \J_I[t_0]$.  Indeed, if a sequence of forced gradient flows
on the semi-infinite interval $(-\infty,t_0]$ goes to infinity, it does so by diverging for $t\to t_0$,
in which case $\mathrm{Re}(\W(w,z(t_0))$ (whose gradient drives the flow for $t\to t_0$)
must go to $-\infty$.

An alternative approach to  (\ref{elx}) is the following.  Suppose that $\vec z_f=\vec z_i$ and $\P$
is the trivial path between them.  Then $\frak N^J_I=\frak n^J_I=\delta^J_I$.  As we
vary $\vec z_f$, both $\frak N^J_I$ and $\frak n^J_I$  may jump in and only in crossing Stokes
walls; they jump in exactly the same way, so they remain equal.  We have
already describing the jumping of $\frak n^J_I$.  The jumping of $\frak N^J_I$ occurs
because in crossing a Stokes wall, a time-dependent solution may disappear to infinity,
 as follows.  Suppose that, for $\vec z_f$ on some Stokes wall, there is
a jump in $\frak n^J_I$, resulting from an ordinary  gradient flow from some critical point $J'$ of 
$\W(w,\vec z_f)$ to $J$.  Such a flow produces a jump
\be\label{jumpo} \frak n^J_I\to \frak n^J_I\pm \frak n^{J'}_I,\ee
where the sign depends on the direction in which one crosses the Stokes wall.
To see a corresponding jump in $\frak N^J_I$, one looks for forced gradient trajectories
from $I$ to $J$ that consist of a forced gradient trajectory from $I$ to $J'$ followed, at some
time very far in the future, by the same ordinary gradient trajectory from $J'$ to $J$ that causes
the jump of $\frak n^J_I$.  A two-step
forced trajectory of this kind exists if $\vec z_f$ is near the Stokes wall and on the proper side of
it; the time at which the second step of the flow occurs diverges as $\vec z_f$ crosses the Stokes
wall.  This leads to the disappearance of the two-step solution and a jump of $\frak N^J_I$
that just matches the jump of $\frak n^J_I$.

\section{From Braiding Of Thimbles To Free Field Integrals}\label{fthimconformal}

According to the reasoning in section \ref{fourd},
to understand the braid group representations associated to the Jones polynomial,
we are supposed to compute a natural monodromy action on the middle-dimensional
relative homology of an infinite-dimensional space $\U$ of connections on a three-manifold
$\Mthree=C\times \R_+$.    This may sound hopelessly abstract.  We will now show how it can be turned into something concrete and calculable.

\subsection{From Thimbles to Integrals}\label{fromth}

A convenient way to describe the evolution of states in $\H$, including
the $q$-grading, is to view the homology cycles
as integration cycles.  Instead of looking at the evolution of the homology cycles,
it is equivalent to look at the evolution of the integrals. Of course, we
need an integrand which
can be integrated on the thimbles $\J_I$, which are not compact.
A function that fills the bill is $e^{\W/\varepsilon}$, where $\varepsilon$ is chosen
in a suitable half-plane. Since
\begin{equation}
h = \mathrm{Re} \left( e^{i \alpha} \W \right)
\end{equation}
goes to minus infinity along a thimble, the condition we want is $\mathrm{Re} \left( e^{i \alpha} \varepsilon\right)>0$.
This will ensure the convergence of the integrals
\begin{equation}\label{thelko}
\I_\Gamma = \oint_\Gamma e^{\W(w)/\varepsilon} \d \Omega
\end{equation}
where $\d\Omega$ is a holomorphic volume form on $\U$ (which will be kept fixed in
what follows) and the integration
cycle $\Gamma$ is a thimble, or more generally any cycle
in the relative homology $H(\U,\U_<)$.

The thimbles are
particularly nice integration cycles, because the $\varepsilon\to 0$ limit of the integral
over a thimble is very simple.  On a thimble $\J_I$ defined by gradient flow from a
critical point $I$, the function $h$ has a unique maximum, namely the critical point.
So for $\varepsilon\to 0$, the integral over a thimble is
\begin{equation}\label{thelk}
\I_I:= \oint_{\J_I} e^{ \W(w)/\varepsilon}\, \d \Omega\sim \exp(\W_I/\varepsilon)\left(\varepsilon^{-\mathrm{dim}\,\J_I/2}c_0+\dots\right),
\end{equation}
where $\W_I$ is the value of $\W$ at the critical point $I$.

This formula is valid throughout the half-plane $\mathrm{Re} \left( e^{i \alpha} \varepsilon\right)>0$,
but actually as long as $\alpha$ is not on a Stokes wall, this asymptotics holds in a slightly larger sector in the complex plane.\footnote{\label{breakdown}For
the stated asymptotics to break down, the first step is to cross a Stokes wall, so that the thimble we started with evolves into a linear
combination of thimbles with at least two terms.  Initially, the asymptotics (\ref{thelk}) remain valid, as any extra thimbles that
appear at the Stokes wall initially make exponentially small contributions.  If one varies $\alpha$ further, one of the extra thimbles
may eventually become dominant.  The combined process always involves varying $\alpha$ by an angle strictly greater than $\pi/2$
from its initial value. To show this, one just compares the values of $\W$ at the two critical points; these values have equal imaginary
parts at the Stokes wall, and equal real parts when the two critical points exchange dominance.}
This property uniquely characterized the basis of integrals
$\I_I$ among all the possible $I_\Gamma$: if we were to take a linear combination of
several $\I_I$, the asymptotics would fail at some ray in the extended half-plane where two
critical points exchange dominance. This characterization is familiar in Stokes theory,
and motivated the terminology ``Stokes walls.''

As we vary the parameters of $\W$, the homology $H(\U,\U_<)$ will vary continuously.
If we vary the integration cycle $\Gamma$ continuously, the integral will vary holomorphically
in the parameters of $\W$. The monodromy of the cycles $\Gamma$ is the same as
the monodromy of the integrals $\I_\Gamma$.

\subsection{Two Chern-Simons Theories}
In our present context, the thimble integral (\ref{thelko}) is a Chern-Simons path integral on the three manifold $\Mthree = C \times \R_+$, 
albeit on an unusual integration cycle. 
Such an integral can be concretely expressed in terms of $\N=4$ gauge theory on $\Mthree \times \tilde \R_+$.  (In this section only,
we write $\tilde \R$ or $\tilde\R_+$ for the $x^1$ direction to distinguish it from the $y$ direction $\R_+$.) This statement was one of the main
conclusions of \cite{Analytic, NewLook}.  In our case, since $\Mthree = C \times \R_+$,
the four manifold is $M_4 = C \times \R_+ \times \tilde \R_+$. 

This is the second Chern-Simons theory to appear in this paper, at least implicitly. 
Our whole analysis concerns the calculation of the Jones polynomial, in a gauge theory setup which 
is $S$-dual to a setup which computes the Jones polynomial by Chern-Simons theory 
on $W = C \times \tilde \R$. In that ``original'' Chern-Simons theory, the knot is 
a  Wilson loop, the gauge group is $G^\vee$, and the coupling parameter is $\k^\vee$. 
The Jones polynomial is a Laurent polynomial in 
\begin{equation}\label{torg} q=\exp(2\pi i/(\k^\vee+2)),\end{equation}
 where $2$ is the dual Coxeter number of $SU(2)$.  

As explained in \cite{fiveknots},
Chern-Simons theory on a three-manifold $W$ can be computed via
 topologically twisted $\N=4$ super Yang-Mills theory on $W\times \R_+$ if one relates $\k^\vee$ to the
twisting parameter $\Psi^\vee$ of the $\N=4$ theory by 
\begin{equation} \label{psikvee} \Psi^\vee=\pm (\k^\vee+2). \end{equation}
  In this description
one uses a D3-NS5 boundary condition at the origin in $\R_+$. The sign $\pm$ depends on the relative choice of orientation 
between $M_4$ and $W$. 

One can also apply $S$-duality, converting the gauge group from $G^\vee=SU(2)$ to $G=SO(3)$ and converting
the D3-NS5 boundary condition to a D3-D5 boundary condition; this boundary 
condition involves a Nahm pole,
as explained in \cite{gw}.  In this new description, which has been the starting
point of the present paper, the twisting parameter is $\Psi=-1/\Psi^\vee$.    Because 
this dual description
is difficult, we have tried to simplify it, as first explained in section \ref{methods}, by 
``stretching'' $W$ in one direction.
Thus we approximated $W$ by $\tilde \R\times C$, and looked for solutions of the BPS 
equations on $W\times \R_+$ that are ``pulled back''
from $M_3=C\times \R_+$.   If this dual description is formulated as supersymmetric 
quantum mechanics, with $\R$ as the time
direction and the field variables being gauge fields on $M_3=C\times \tilde \R_+$, then 
the superpotential is $\Psi$ (not $\Psi^\vee$) times
the Chern-Simons function.  

We then relate the braiding of solutions 
on $M_3$ to the braiding of thimbles for  
Chern-Simons theory on $M_3$, which we express by  $\N=4$ theory on $M_3\times \tilde \R_+$.  In this
description, we impose  
D3-NS5 boundary conditions at the origin of $\tilde \R_+$ (and D3-D5 at the origin of  $\R_+$). 
The level $\k$ of this Chern-Simons description is related to the twisting parameter 
of the $\N=4$ theory by
\begin{equation} \label{psik} \Psi=\mp (\k+2). \end{equation}
The two Chern-Simons descriptions are related by S-duality, together with the exchange of the roles of $\R_+$ and $\tilde \R_+$.
The opposite sign in (\ref{psikvee}) and (\ref{psik}) is due to the fact that exchanging $\R_+$ and $\tilde \R_+$ 
reverses the orientation of $M_4$. 

 Combining this with $\Psi=-1/\Psi^\vee$ and $\Psi^\vee
=\k^\vee+2$, we find that the relation
between the level $\k^\vee$  in the Chern-Simons description that is related to the Jones polynomial 
in the traditional way and the level $\k$  in the Chern-Simons description
that relates the Jones polynomial  to Nahm poles and opers is \begin{equation}\label{sdualk} \k+2=1/(\k^\vee+2).\end{equation}

This four-dimensional setup, with the $\N=4$ theory on a manifold $M_4 = C \times \R_+ \times \tilde \R_+$ with a
``corner,'' 
is rather interesting, and we believe it deserves to be explored further. We will not do so in this paper. 

\subsection{From Chern-Simons to Conformal Blocks}

For this paper, more useful than the relation of Chern-Simons theory on $\Mthree=C \times \R_+$ to four dimensions is its relation to conformal blocks on $C$. 
The most familiar version of this statement \cite{WittenJones} is that a path integral with
$\A_{\bar z}$ fixed and $\A_z$ varying gives a WZW conformal block (that is,
a conformal block in two-dimensional
current algebra, with the symmetry group $G$ being the same as the gauge group of the Chern-Simons theory).  Local variations of the fixed value of $\A_{\bar z}$
insert a holomorphic current $J(z)$ in the conformal block:
\begin{equation}
\delta \log \cZ = \int \mathrm{Tr} \langle J_z \rangle \delta A_{\bar z}
\end{equation}

 This statement has an analog \cite{HVerlinde}
 that leads to Virasoro conformal blocks in the case of $SU(2)$ or $SO(3)$ gauge theory,
 or to more general $W$-algebra conformal blocks in the case of gauge groups of higher rank \cite{Bilal,DeBoer}.  This analog involves a different boundary condition in which, at the price of breaking some gauge symmetry at the boundary, one  fixes some parts of $\A_z$, and some parts of $A_{\bar z}$. For gauge group $SU(2)$, the boundary condition which leads to Virasoro conformal blocks
is simply stated:
\begin{align}\label{bcon}
\A_z = \begin{pmatrix} * & 1 \cr * & *\end{pmatrix} \qquad \A_{\bar z} = \begin{pmatrix} \times & 0 \cr * & \times\end{pmatrix}
\end{align}
Here we denote as $*$ the elements which are free to vary, and as $\times$ elements which are fixed.

The connection just described is an oper! (Since $\A_{\bar z}$ is lower-triangular, a bundle with this connection has a holomorphic
sub-bundle $L$ whose sections are of the form $\begin{pmatrix}0\cr *\end{pmatrix}$, and because the upper right matrix element of 
$\A_z$ nowhere vanishes, this sub-bundle is nowhere preserved
by $\D_z=\partial_z+[\A_z,\,\cdot\,]$.)   So the complex boundary condition (\ref{bcon})
is the one that is induced by the  Nahm pole boundary condition studied in section \ref{analog}.
To explain the
relation to Virasoro conformal blocks, we note that the Nahm pole boundary condition
depends on a choice of complex structure on $C$. Once a complex structure
is picked with a local complex coordinate $z$, nearby complex structures
can be described by a Beltrami differential $\mu_{\bar z}^z$.  The relation is that
in the new complex structure, the holomorphic fields on the space of complex connections 
are not $\A_{\bar z}$ but $\A_{\bar z}-\mu_{\bar z}^z\,\A_z$. 
Making this deformation is equivalent to replacing the ``$0$'' in the  boundary condition for
$\A_{\bar z}$ in (\ref{bcon})  with $\mu_{\bar z}^z$ (so that now $\A_{\bar z}-\mu_{\bar z}^z\A_z$ is lower triangular).
Hence a local variation of the boundary condition associated to a change in complex structure
inserts a  holomorphic stress tensor
\begin{equation}
\delta \log \cZ = \int \langle T_{zz} \rangle \delta \mu_{\bar z}^z,
\end{equation}
and this leads to the relation between Chern-Simons theory with the oper boundary condition and Virasoro conformal blocks.

In the semiclassical limit, the operator $T_{zz}$ reduces to the classical stress tensor $\bt(z)[\A]$ of the oper. More precisely, the identification of parameters from 
\cite{HVerlinde} is that if we define \begin{equation}\label{bpar}-b^{-2} =\k+2,\end{equation}
 with $\k$ the Chern-Simons level, then the stress-tensor has central charge $c=1+6(b+b^{-1})^2$ and in the semiclassical limit, 
$b^2 T(z) \to \bt(z)[\A]$. 

If we combine (\ref{bpar}) with  (\ref{sdualk}) and (\ref{torg}), we find the relationship between the variable $q$ usually used in describing the Jones polynomial
and the parameter $b$ used in describing Virasoro conformal blocks:
\begin{equation}\label{zorg}q=\exp(-2\pi i/b^2).\end{equation}

We have here assumed that $G$ is $SU(2)$ or $SO(3)$.
For general gauge group $G$, both Nahm poles and $W$-algebras are
labeled by an $\frak{su}(2)$ embedding in the Lie algebra $\frak{g}$ of $G$.
Inspection confirms that the boundary conditions used to define
a general $W$-algebra conformal block are induced by the corresponding Nahm pole.

\subsubsection{Analog For Liouville}\label{analiouville}

Though we will not need this fact in the present paper, we should remark that the relation
between  Virasoro conformal blocks and Chern-Simons theory  has a simple extension
to a relation beween Liouville theory and Chern-Simons theory.  To do Liouville
theory on a Riemann surface $C$, one considers Chern-Simons theory on $C\times I$
where $I$ is a unit interval.  At one end of $I$, one imposes the Nahm pole boundary condition
and at the other end, one imposes a variant of the Nahm pole boundary condition with $z$
and $\bar z$ exchanged.  Liouville partition functions and correlation functions
are built by combining holomorphic and anti-holomorphic Virasoro conformal blocks, which
arise naturally from Chern-Simons on $C\times I$ with boundary conditions just stated.
(For the case of a compact symmetry group, it is
already known that Chern-Simons on $C\times I$ reproduces the WZW model on $C$.)   By slightly extending arguments that
we present presently, light degenerate fields and generic primary fields of Liouville theory, inserted at a point $p\in C$,
 correspond to Wilson operators or monodromy defects on $p\times I$.

In the classical limit, the correspondence between Chern-Simons and Liouville theory means the following.  A classical solution of Chern-Simons theory on $C\times I$ with
boundary conditions  as above is a flat  bundle  on $C$ whose holomorphic and antiholomorphic
structures both obey the oper condition.  Indeed, a classical solution of Liouville theory corresponds to
a metric on $C$ of constant negative curvature.  If $\omega$ and $e$ are the vierbein and spin connection of this metric,
then we can define a corresponding $SL(2,\R)$ flat connection $\A=\omega {\ct}_3+
e_{ z} \ct_++e_{\bar z}\ct_-$.
With a standard representation of the $\ct_i$, $\A_z$ is upper triangular with an upper right matrix element that is everywhere nonzero,
and $\A_{\bar z}$ is lower triangular with a lower left matrix element that is everywhere nonzero.  So both the holomorphic and
antiholomorphic structures defined by this flat connection satisfy oper conditions, as expected in the Chern-Simons description.  
(The antiholomorphic oper structure is defined
with the roles of ``upper triangular'' and ``lower triangular'' matrices reversed.) 
 So this gives the mapping between the two theories
at the classical level.

\subsection{Wilson Line Operators}\label{wilson}

To further understand the mapping from three-dimensional 
Chern-Simons theory with oper boundary
conditions to Virasoro conformal blocks in two dimensions,
 we will explore the interpretation  of Wilson line
operators.  First let us recall what happens
if one uses standard boundary conditions that relate Chern-Simons theory to current algebra.  
In this case, a Wilson line operator ending on the boundary of a three-manifold $M_3$
represents insertion of a conformal primary field at that boundary point in the WZW conformal
block.   If the Wilson line operator transforms in a finite-dimensional representation $R$ of $G$,
then the corresponding conformal primary field transforms in the same representation.
This is consistent with the fact that a Wilson line operator ending on the boundary
is not gauge-invariant, but transforms in the representation $R$,
just like the corresponding primary field of the WZW model.

What is the analogous interpretation of a Wilson line operator that ends on a boundary
at which one imposes Nahm pole boundary conditions?
The Nahm pole breaks the gauge symmetry at the boundary, so we have
to pick a component of the Wilson line operator. As we discussed in section \ref{flatpole},
a generic vector diverges
when parallel transported to the boundary.
Given a Wilson line operator ending at $y=0$, the most easily
defined gauge invariant information is the coefficient of the most negative power
of $y$. This is extracted simply by contracting with
an appropriate power of the small section $s$. Actually, we will find  useful 
a rescaled version of $s$, namely $\hat s = K(z)^{-1/2} s(z)$,
which satisfies
\begin{equation}
\left( \D_z^2 + t(z) \right) \hat s=0 \qquad \D_{\bar z} \hat s =0 \qquad \D_{y} \hat s = 0
\end{equation}
and has definite conformal dimension $-1/2$.  In the gauge (\ref{torog}),
$\hat s=\begin{pmatrix}0\cr 1\end{pmatrix}$.

In the two-dimensional representation of $SU(2)$, we would consider an operator
\begin{equation}\label{zello}
P \exp \left (-\int_\gamma \A\right)  \hat s(z),
\end{equation}
where $\gamma$ is a path ending on the boundary at $y=0$. 
In the classical limit, under conformal transformations of the boundary, this has the same conformal dimension as $\hat s$, i.e. $-1/2$.
For a spin $k/2$
representation, one must contract with $k$ powers of $\hat s$ and the classical limit of
the dimension is $-k/2$.

We want to argue now that a Chern-Simons path integral with a spin
$k/2$ Wilson operator ending on the boundary gives a Virasoro conformal block
with the insertion of a  ``light'' degenerate field of Liouville momentum $-b k/2$.
Such a field has the  correct classical dimension $-k/2$ in the $b \to 0$ limit,
and furthermore the exact formula for its  quantum dimension
\begin{equation}
-\frac{k}{2} b\left(b+\frac{1}{b} + \frac{k}{2} b\right) = -\frac{k}{2} - \frac{k(k+2)}{4}b^2=-\frac{k}{2}
+\frac{k(k+2)}{4(\k+2)}
\end{equation}
is the sum of the classical dimension of $\tilde s^k$ and the dimension of a spin $k/2$ operator
in a WZW model (of level $\k+2=-1/b^2$ as in (\ref{bpar})). Furthermore, the operator (\ref{zello}) 
satisfies classically the correct differential equation:
the $k=1$ operator is annihilated by $\partial_z^2 + t(z)$, etc.

 Part of what makes possible  the correspondence between spin $k/2$ Wilson
lines and degenerate conformal fields
possible is that  the degenerate fields satisfy fusion rules which coincide with the
fusion rules of spin $k/2$ operators in the WZW model.  This last fact is part of the input in the
statement that the braid group representations associated to the Jones polynomial can
be computed by the braiding of either primary fields of the WZW model or degenerate conformal
fields of the Virasoro algebra.

\subsection{Singular Monopoles and ``Heavy'' Degenerate Fields}\label{hdf}
Next, we would like to identify in the Chern-Simons description of Virasoro conformal blocks
the ``heavy'' degenerate fields of Liouville momentum $-k/2b$. These are the degenerate fields whose
conformal dimension diverges for $b\to 0$. We claim that
they correspond to the insertion of singular monopoles at the boundary.

The main insight of section \ref{opsing} was
that at a Nahm boundary with singular monopoles (and generic $\zeta$),
 the connection is an oper with singularities of trivial monodromy. We observed
that the classical stress
tensor of such an oper has poles that agree with the semiclassical limit of
the quantum stress tensor in the presence of a heavy degenerate field.
Moreover, the trivial monodromy condition holds quantum-mechanically
as well: a light degenerate field
has no monodromy around a heavy degenerate field, as they fuse in a unique channel.

In the Chern-Simons setup, the classical trivial monodromy condition follows
naturally from the fact that
the singular monopole does not extend in the bulk. This was part of our derivation in
section \ref{opsing}.  Quantum mechanically, we need to consider the behavior when
a light degenerate field -- represented in three dimensions by an expression such as $P\exp\left(-\int_\gamma\A\right) \hat s$
-- approaches the singular monopole.
The Wilson loop itself is topological,
and the small section $\hat s$ has trivial monodromy around the singular monopole.

\subsection{Putting The Pieces Together}\label{punchline}
We can now finally establish a link between the
solutions of the four-dimensional BPS equations (\ref{bpseqns}) that we started with
and the braid group representations associated to the Jones polynomial.

The time-independent solutions
of the BPS equations correspond to opers with trivial monodromy.
We have identified the braiding of the corresponding quantum states with the braiding
of complex integration cycles for Chern-Simons theory, and then with the braiding
of degenerate Virasoro conformal blocks. These are known \cite{Lawrence,SV} 
to be the braid group
representations associated to Jones polynomials.   So we have arrived at our
goal, though in a form that may sound a little abstract.

We can put this result in a perfectly concrete form using the free field representation of
the conformal blocks.
Opers with trivial monodromy are also associated to critical points of the Yang-Yang
function for the Bethe equations; this is the logarithm of
the integrand in the free field realization of conformal blocks.
We can derive a degenerate Virasoro conformal block either from an infinite-dimensional
thimble associated to an oper with trivial monodromy, or from a finite-dimensional thimble
associated to a critical point of the Yang-Yang function.  
Either way, we get a conformal block with definite and
uniform semiclassical limit in a sector of angular width
greater than $\pi$ in the $b^2$ plane. As those are unique, the two bases
of conformal blocks must coincide.

Hence the braiding representations associated to the four-dimensional gauge
theory coincide with the braiding
representations of integration cycles in the space of Bethe parameters
$w_i$. This is not as surprising
as it may seem if we turn on a symmetry breaking parameter:
then we have interpreted the $w_i$ as positions of bulk BPS monopoles, and our
claim possibly amounts to the statement
 that the  four-dimensional nonabelian gauge theory on the Coulomb branch
reduces to a theory of massive monopoles and abelian gauge fields. We will
develop this point of view further in section \ref{supermon}.

\section{Braiding Representations of Integration Cycles}\label{brrep}

\subsection{Overview}\label{moreover}

A highlight of what we have learned so far is the existence of a natural map from
the braid group representations derived from the four-dimensional gauge theory equations
(\ref{bpseqns}) to the braid group representations associated to correlation functions of
Virasoro degenerate fields.  Those braid group representations can be effectively studied using
the free field representation, which we reviewed in section \ref{opbl}.  Making this
explicit will be our goal here.

We consider a degenerate correlation function $\left\langle \prod_{a=1}^d V_{-k_a/2b}(z_a)
\right\rangle$.
We assume that the $z_a$ are distinct points in $\C$.  To represent conformal blocks,
we introduce $q$ variables $w_i\in \C$, which we assume to be distinct from each other
and from the $z_a$. The allowed values of $q$ have been analyzed in section \ref{analog}.
Moreover, we consider the $w_i$ to be indistinguishable, in the sense
that  configurations
that differ by permuting them are equivalent.
 We write $\M$ for the space of such distinct
and indistinguishable variables $w_i\in \C\backslash\{z_1,\dots,z_d\}$.  We also write
$\hat \M$ for the smallest cover of $\M$ on which the Yang-Yang function $\W$ of eqn.
(\ref{yy}) is single-valued.

In the free field representation, degenerate conformal blocks are written in the form
\begin{equation}\label{dolf}\int_\Gamma \exp\left(\W/b^2\right)\,\d w_1\dots
\d w_q.\end{equation} $\Gamma $ is a middle-dimensional cycle in $\hat\M$,
chosen so that the integral converges.\footnote{As explained in \cite{Lawrence}, the cycle $\Gamma$
should actually be odd under the exchange of any pair of $w$'s, to compensate for the sign
change of the differential form $\d w_1\wedge\dots\wedge\d w_q$ under permutations.  This
means that the appropriate relative homology is actually the part that is antisymmetric
under permutations of the $w$'s.  To be concrete, suppose that there
are two $w$'s and we find a solution of the Bethe equations at which the $w$'s  equal $\alpha$ and $\beta$ up to permutation.
Then we can define a cycle $\CC'$ associated to the critical point $w_1=\alpha$, $w_2=\beta$, and a cycle
$\CC''$ associated to the critical point $w_1=\beta$, $w_2=\alpha$.  The difference $\CC'-\CC''$ is an
element of the antisymmetric part of the homology.  In practice, we can omit to explicitly form such differences
and also ignore minus signs arising from permutations of the factors in $\d w_1\wedge\dots\wedge \d w_n$.}

Morse theory offers a systematic way
to produce all such integration cycles: a basis of integration cycles is given by the
thimbles associated to critical points of $\W$.
Cycles of this kind are never  compact; they have noncompact ends on which
 the Morse function $h = \mathrm{Re} \,\W$ goes to $-\infty$.  In our problem,
 this happens when
 one of the $w_i$ either approaches one of the $z_a$ or, in
 the presence of symmetry breaking, goes to infinity in the correct direction.
In simple situations, instead of using Morse theory, one can describe integration cycles by
hand. In constructing a cycle $\Gamma$ by hand, one has to make sure that
the Morse function really goes to $-\infty$ at infinity along $\Gamma$.  For example,
this will fail if too many $w_i$ approach simultaneously the same $z_a$.

With symmetry breaking, some of the important integration cycles have ends at $w=\infty$
and the use of noncompact integration cycles is unavoidable.  However, in the absence
of symmetry breaking, the noncompact integration cycles produced by Morse theory have
their ends at $w_i\to z_a$, for various $i$ and $a$, and are
equivalent in the appropriate twisted relative homology to compact
cycles in which the $w_i$ wrap around the $z_a$ in a suitable fashion. (For an example,
see fig. \ref{compare} below.)  In the 
extensive  literature on integration cycles in free-field realizations of
conformal blocks \cite{DF,Felder}  and their application to the 
Jones polynomial \cite{Lawrence,SV},
compact integration cycles are often used.
Symmetry breaking, or in other words the introduction of  an irregular
singularity at infinity, has not been considered in this context, as far as
we know.

The use of Morse theory has advantages and disadvantages.
The main disadvantage is that the thimbles do not
correspond to a standard BPZ basis of conformal blocks defined by fusing
the degenerate fields in specific channels.
The main advantage is that in the basis of thimbles,
the braid group is manifestly represented by
matrices whose entries are Laurent
polynomials in $q$ with integer coefficients.   This property, which was explained in
section \ref{nonsingle}, is important vis-a-vis the Jones polynomial
and Khovanov homology.

In what follows, we will first analyze a few important examples with a small number
of degenerate insertions $z_a$ and Bethe roots $w_i$, with or without symmetry breaking.   Then in section \ref{genpicture},
we analyze the general case in the presence of symmetry breaking.  From that analysis,
we get the experience we need to deduce a general description of the Jones polynomial
-- not just the associated braid group representations.  This is presented in section
\ref{abdesc}.

\subsubsection{$SU(2)$ Versus $SO(3)$}\label{contrast}

We pause for a technical remark concerning the assertion that the entries of the braiding matrices
are Laurent polynomials in $q$.

If the gauge group is $G=SU(2)$, meaning that the dual group is $G^\vee=SO(3)$ and the charges
$k_a$ of the singular monopoles are all even, then the Yang-Yang function $\W$ as defined in (\ref{yy}) is
well-defined mod $2\pi i$.  Hence a change of branch of $\W$ multiplies $\exp(\W/b^2)$ by
an integer power of $q=\exp(-2\pi i/b^2)$, and the braiding matrices are Laurent polynomials in
$q$ with integer coefficients.

If instead $G=SO(3)$, $G^\vee=SU(2)$, then some of the $k_a$ may be odd.  (Indeed, we will do our
detailed computations for the case that all $k_a$ are 1.)  Then $\W$ is well-defined mod $2\pi i$
if the $w_i$ are varied for fixed positions $z_a$ of the knots, but is only well-defined mod
$2\pi i/4$ when the $z_a$ are varied.  Consequently, for $G=SO(3)$, the braid matrices will actually
be Laurent polynomials in $q^{1/4}$.

Of course, we could eliminate this by writing the formulas in terms of $\tilde q=q^{1/4}$, but we prefer
not to do so since $q$ as we define it is the natural instanton counting parameter in four dimensions.  The underlying reason for the difference between $G=SU(2)$ and $G=SO(3)$ is that
the Chern-Simons function, normalized as we have done in (\ref{supertwo}), is gauge-invariant
mod $2\pi i$ in $SU(2)$ gauge theory, but gauge-invariant mod $2\pi i/4$ in $SO(3)$ gauge theory.
The last statement holds on any sufficiently rich three-manifold $\Mthree$, or on any $\Mthree$ if singular monopoles of odd charge are present.  In the latter case, one considers only gauge transformations that are trivial at the position of the singular monopole.

\subsection{A Single Critical Point}\label{scp}

We begin with the two examples in which the Yang-Yang function has only a single critical point.

\subsubsection{ Braiding of Two Primaries With Minimal Charge}\label{brmin}

The first example arises  in the absence of
symmetry breaking, with two $z_a$ of charge $k=1$ and one $w$.
An obvious integration cycle $\CC_{12}$ is a segment joining the two $z_a$. Now, let us
compare it with the thimble. The Bethe equation
\begin{equation}
\frac{1}{w - z_1} + \frac{1}{w - z_2} =0
\end{equation}
has a unique solution $w = \frac{1}{2}(z_1 + z_2)$. The thimble flows down from $w$ to the $z_i$
along a straight line, and coincides with the obvious cycle $\CC_{12}$.
The critical value of the Yang-Yang function
\begin{equation} \W = -\frac{1}{2} \log(z_1 - z_2) + \log(w-z_1) + \log(w-z_2) =
\frac{3}{2} \log(z_1 - z_2) + \mathrm{const.} \end{equation}
is such that $\exp(\W/b^2)$
agrees with the  OPE coefficient of the two degenerate fields in the
identity channel $V_{-1/2b}(z_1)V_{-1/2b}(z_2)\sim (z_1 - z_2)^{\frac{3}{2 b^2}}$.
(Recall that for small $b$, the dimension of $V_{-1/2b}$ is $-3/4b^2+\mathcal{O}(1)$.)
This is expected, because with two $z_a$ and one $w$ the oper has no singularity
at infinity, in view of the discussion of eqn. (\ref{ruy}), and hence describes the fusing
of two fields to the identity.

In order to fuse the two degenerate fields in a channel of momentum $-b$
we would consider something even simpler: a case without any $w$'s.
In this channel, the conformal block is simply
the free field correlation function $\left\langle V_{-1/2b}(z_1)V_{-1/2b}(z_2)\right\rangle$
with no integral at all.  With no $w$'s, the discussion of (\ref{ruy}) shows that the oper
does have a singularity at infinity -- corresponding to fusion of the two $k=1$ degenerate
primary fields to a $k=2$ degenerate primary field.

To be precise, in defining an integration cycle such as 
$\CC_{12}$, we should specify a choice of branch of $\W$. (Differently
put, the cycle is supposed to be defined in the covering space $\hat\M$.) If we 
braid $z_1$ around $z_2$, $\CC_{12}$ evolves continuously, but we may end up with
a different branch of $\W$. As the charges at $z_1$ and $z_2$ are identical, a basic 
braiding move is to exchange the position of $z_1$ and $z_2$, and then relabel them.
In the absence of $w$, we would only have the factor $(z_1 - z_2)^{-\frac{1}{2 b^2}}$ coming
from the part of $\W$ which only depends on $z_a$, so a braiding which exchanges $z_1$ 
and $z_2$ counterclockwise would give a factor of $e^{- \frac{i \pi}{2b^2}} = q^{\frac{1}{4}}$. 
In the presence of one $w$, we can get the result by following the value of
$\W$ at the midpoint of $\CC_{12}$:
\begin{equation}\label{cmono} \CC_{12} \to -q^{-\frac{3}{4}} \CC_{12}. \end{equation}
The minus sign follows from the change in orientation of $\CC_{12}$.

\begin{figure}
 \begin{center}
   \includegraphics[width=3.5in]{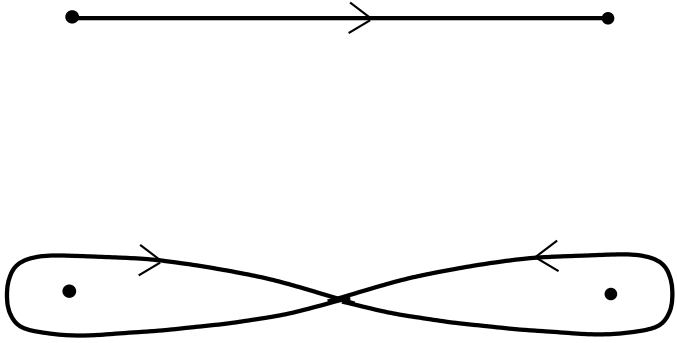}
 \end{center}
\caption{\small  The thimble $\CC_{12}$ (top) compared to a closed integration contour (bottom), which is equivalent to $(1-q^{-1})\CC_{12}$.  }
 \label{compare}
\end{figure}

It is interesting to compare the thimble with  closed contours 
which are commonly used in order to describe BPZ conformal blocks,
as depicted in fig. \ref{compare}. As illustrated in the picture, 
these contours are equivalent in homology to the thimble times a Laurent polynomial in $q$.
From the point of view of the Stokes matrices, they are not as elementary as the thimble.

\subsubsection{One Primary Field With Symmetry Breaking}\label{onesym}

\begin{figure}
 \begin{center}
   \includegraphics[width=3.5in]{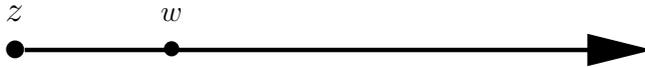}
 \end{center}
\caption{\small This ray parallel to the real axis
is the Lefschetz thimble for the case of one primary field
of minimal charge with symmetry breaking. }
 \label{linel}
\end{figure}
The other basic example with one critical point occurs in the presence of symmetry breaking
with a single degenerate field of $k=1$, and a single Bethe root. The Bethe equation reads
\begin{equation}
\frac{1}{w - z} =c  \qquad c = - \frac{2 \ca}{\zeta}.
\end{equation}
So $w=z + \frac{1}{c}$.
For convenience, we will take the constant $c$ to be real and positive. 
The thimble again coincides (fig. \ref{linel})
with the most natural integration
cycle $\CC$, along a ray starting at $z$ and parallel to the positive real axis, passing through
$w$.
This example illustrates that an integration cycle in the presence of symmetry breaking may end
at infinity.  The cycle that we have just described cannot be replaced with an equivalent compact
cycle.

\subsection{Two Critical Points}\label{twoc}
Next we can consider  examples with only one Bethe root $w$, but two critical points.
This occurs 
with two primaries of $k=1$ in the presence of symmetry breaking,
or with three such primaries in the absence of symmetry breaking.

\begin{figure}
 \begin{center}
   \includegraphics[width=3in]{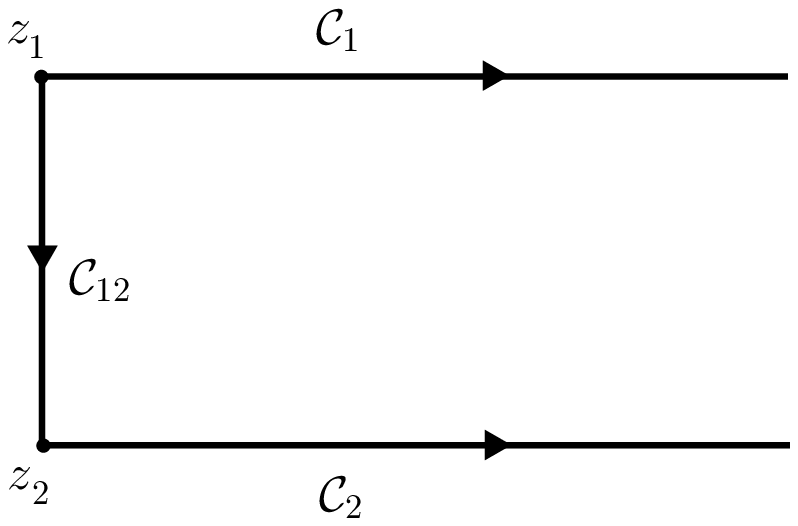}
 \end{center}
\caption{\small  The cycles $\CC_1$, $\CC_2$ and $\CC_{12}$   }
 \label{cyclesonetwo}
\end{figure}

\subsubsection{The First Example}\label{twop}

We consider first the case of two primaries  with symmetry breaking.  We denote
the positions of the primaries as $z_1$ and $z_2$ and we continue to assume that the
symmetry breaking parameter $c$ is positive. 
As long as $z_1-z_2$
is not real,
 there is a symmetric choice of basic
 integration cycles $\CC_a$, $a=1,2$: rays which start at $z_a$ and are 
 parallel to the positive real axis.
The difference $\CC_1 - \CC_2= \CC_{12}$ is a segment from $z_1 $
to $z_2$. See fig. \ref{cyclesonetwo}. Any two of these three cycles can be 
thimbles, depending on the
relative values of $c$ and $z_1 - z_2$.  Since there are always only two thimbles,
it is never the case that all three of $\CC_1$, $\CC_2$, and $\CC_{12}$ are thimbles.

In this simple example, we already have the basic ingredients of the
general braid group representation. The elementary move is to exchange the two $z_a$,
either clockwise or counterclockwise. We start with a configuration
in which the $z_a$ have distinct imaginary parts, so that the $\CC_a$ are well-defined.
We may as well take the real parts of the $z_a$ to be zero.
We want to define the branches of $\W$ along $\CC_1$, $\CC_2$, and $\CC_{12}$ so that it is
true that $\CC_{12}=\CC_1-\CC_2$.
Picking any branch of $\W$ on $\CC_{12}$, we define $\W$ on $\CC_1$ and on $\CC_2$ so
that at the unique point where $\CC_1$ intersects $\CC_{12}$ or where $\CC_2$ intersects
$\CC_{12}$, the definitions agree.  This will ensure that $\CC_{12}=\CC_1-\CC_2$.
Although $\CC_1$ and $\CC_2$ do not intersect, we can deform them slightly
 so that they meet at a reference point far to the right, and then the two values of
 $\W$ will agree at this reference point.  (This is ensured by the fact that one can 
 define $\W$ to be single-valued in the semi-infinite rectangle bounded by $\CC_1$,
 $\CC_2$, and $\CC_{12}$.)

Suppose that $\mathrm{Im}\,( z_1-z_2)>0$. Then we can cross to $\mathrm{Im}\,( z_1-z_2)<0$
in two
ways,  with $z_1$ passing either to the left or to the right of $z_2$.
The two operations are inverses, so it will suffice to consider one in detail.
If $z_1$ passes to the right of $z_2$, then
$\CC_1$ evolves continuously as a ray
parallel to the positive real axis.
On the other hand, $\CC_2$ does not. We can use instead $\CC_{12}$,
defined continuously as a segment from $z_1$ to $z_2$.  In the basis of $\CC_1$ and $\CC_{12}$,
this braiding move is diagonal: we only have to keep track of the branches
of $\W$ and orientation of cycles. We want to express the final result in the basis of $\CC_1$
and $\CC_2$.

In the half-braiding, $\CC_1$ is multiplied by a factor of $q^{-\frac{1}{4}}$: $z_1$ is transported
clockwise around $z_2$ and the reference point does not move significantly.
 After the half-braiding, we rename $\CC_1$ as $\CC_2$:
\begin{equation}\label{elfin}
\CC_1 \to q^{-\frac{1}{4}} \CC_2.
\end{equation}

\begin{figure}
 \begin{center}
   \includegraphics[width=3.5in]{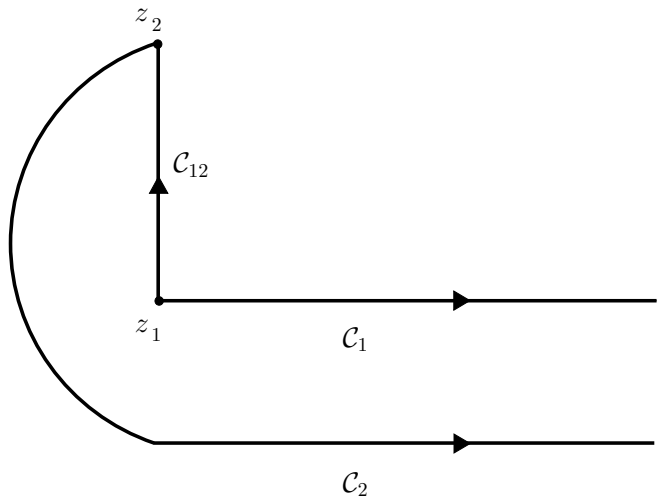}
 \end{center}
\caption{\small  
The cycles $\CC_1$, $\CC_2$ and $\CC_{12}$ after a braiding operation.}
 \label{cyclesonetwob}
\end{figure}

\begin{figure}
 \begin{center}
   \includegraphics[width=3.5in]{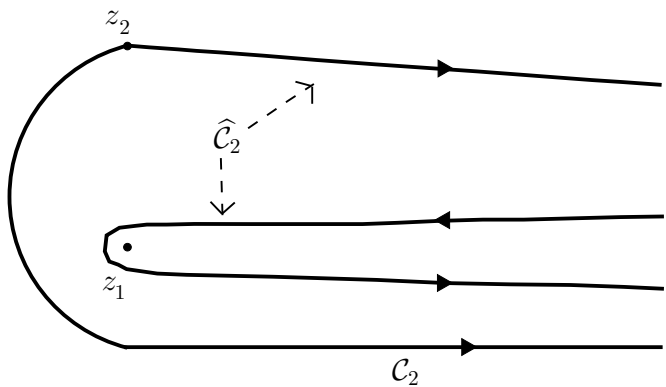}
 \end{center}
\caption{\small  The cycle $\CC_2$, after the braiding operation,  has been copied from fig. \ref{cyclesonetwob}:
it starts at $z_2$, curves below $z_1$ to the left and then goes to $\mathrm{Re}\,z=\infty$.
It is equivalent in homology to a zig-zag cycle, labeled $\widehat\CC_2$ in the figure, 
which, starting at $z_2$,
heads directly to $\mathrm{Re}\,z=\infty$ before doubling back around $z_1$ and
 returning to $\mathrm{Re}\,z=\infty$.  Thus
 $\widehat\CC_2$ is the sum of three pieces, each
of which heads to or from $\mathrm{Re}\,z=\infty$; each piece is equivalent
to a power of $q$ times an elementary cycle $\CC_1$ or $\CC_2$ (a ray starting at $z_1$ or $z_2$
and parallel to the positive $z$ axis).
 }
 \label{cyclesonetwoc}
\end{figure}

On the other hand, $\CC_2$ becomes the cycle in fig. \ref{cyclesonetwob}. It 
can be deformed to the
sum of three pieces  that zig-zag to and from $\mathrm{Re}\,z=\infty$, as 
in fig.  \ref{cyclesonetwoc}.  Each of the three pieces is equivalent to a power of $q$ times one
of the original cycles $\CC_1$ and $\CC_2$.  Just as in (\ref{elfin}), one piece (after
again exchanging the labels of $\CC_1$ and $\CC_2$) is
 $q^{-\frac{1}{4}} \CC_2$. The other two pieces are
images of $-q^{-\frac{1}{4}} \CC_2$ and $q^{-\frac{1}{4}} \CC_1$ under a deck transformation.
Hence the braiding transformation is
\begin{equation}
\CC_2 \to q^{-\frac{1}{4}} \CC_2-  q^{\frac{3}{4}} \CC_2 + q^{\frac{3}{4}} \CC_1.
\end{equation}

Notice that with these transformation rules,
\begin{equation}\label{zolt}
\CC_{12} \to -  q^{\frac{3}{4}} \CC_{12}.
\end{equation}
This is the same result that we found in the 
eqn. (\ref{cmono}) (the sign of the exponent is reversed because in
deriving (\ref{zolt}), we braided $z_1$ clockwise around $z_2$); 
symmetry breaking does not affect the fact that
$\CC_{12}$ represents the conformal block in which the two degenerate
fields fuse to the identity. The braiding matrix has eigenvalues
$-  q^{\frac{3}{4}}$ and $q^{-\frac{1}{4}}$, which correspond to the two possible fusion channels.
The linear combination $\CC_1 + q^{-1} \CC_2$ transforms as
\begin{equation}
\CC_1 + q^{-1} \CC_2 \to q^{-\frac{1}{4}} (\CC_1 + q^{-1} \CC_2).
\end{equation}
and hence it represents the fusion in the channel of momentum $-b$.

Now, other choices of normalization of $\CC_1$ and $\CC_2$ may occur more
naturally in various situations. If we change the relative normalization between
$\CC_1$ and $\CC_2$, say by setting $\tilde \CC_1 = q^{-s/2} \CC_1$
and $\tilde \CC_2 = q^{s/2} \CC_2$, we can  get braiding formulae
\begin{equation}\label{furthox}
B_{12} :~~ \tilde \CC_1 \to q^{-s-\frac{1}{4}} \tilde \CC_2
\qquad   \tilde \CC_2 \to \left( q^{-\frac{1}{4}} -  q^{\frac{3}{4}} \right)
\tilde \CC_2 + q^{s+\frac{3}{4}} \tilde \CC_1.
\end{equation}
We will find the choice $s=-\frac{1}{2}$ to be useful momentarily,
so whenever we write $\tilde \CC_a$
we assume that choice of $s$.

\begin{figure}
 \begin{center}
   \includegraphics[width=3.5in]{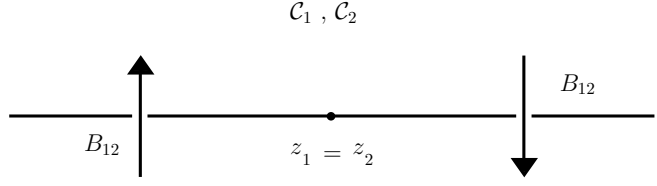}
 \end{center}
\caption{\small  The cycles $(\CC_1,\CC_2)$ (defined as rays in the direction of
$\mathrm{Re}\,cz$ that start at $z=z_1$ or $z_2$)
provide a local basis for homology, but the definition of the cycles jumps by the braiding 
matrix $B_{12}$ across the line on which $Z=c(z_1 - z_2)$ is real.}
 \label{braida}
\end{figure}

The behavior  of the $\CC_a$ as a function of $c(z_1 - z_2)$ is depicted in fig. \ref{braida}.
The $\CC_a$ fail to be well-defined when $c(z_1-z_2)$ is real.
In this very simple example, the integral over the $\CC_a$ can be expressed explicitly
in terms of familiar functions:
\begin{align}
\I_a &= \oint_{\CC_a} (w-z_1)^{\frac{1}{b^2}}(w-z_2)^{\frac{1}{b^2}}
(z_1-z_2)^{-\frac{1}{2b^2}} e^{\frac{c z_1}{2} + \frac{c z_2}{2} - c w} \d w \cr &= c^{-1 - \frac{3}{2 b^2}} \oint_{\CC_a} (W-Z)^{\frac{1}{b^2}}W^{\frac{1}{b^2}}Z^{-\frac{1}{2b^2}} e^{\frac{Z}{2} - W} \d W 
\end{align}
Here we defined $W = c(w-z_2)$ and $Z = c(z_1 - z_2)$. This integral can be explicitly
written in terms of Bessel functions. The integrals over $\CC_1$ and $\CC_2$ give a basis of
Bessel functions with specific asymptotic behavior at large $Z$, and the braiding matrix
$B$ captures the Stokes phenomena of Bessel functions:
\begin{align}
\I_1 &= \frac{1}{\sqrt{\pi } c^{1 + \frac{3}{2 b^2}}} Z^{\frac{1}{2 b^2}+\frac{1}{2}} \Gamma \left(1+\frac{1}{b^2}\right)
   K_{\frac{1}{2}+\frac{1}{b^2}}\left(\frac{Z}{2}\right) \cr 
   \I_2 &= \frac{q^{1/4}}{\sqrt{\pi } c^{1 + \frac{3}{2 b^2}}} Z^{\frac{1}{2 b^2}+\frac{1}{2}} \Gamma \left(1+\frac{1}{b^2}\right)
   K_{\frac{1}{2}+\frac{1}{b^2}}\left(-\frac{Z}{2}\right) 
\end{align}

In the large $Z$ limit, the integral
along $\CC_a$ is controlled by the region where $w$ is close to $z_a$. It is useful to
pick a branch of the logarithms such that for $w$ near $z_1$ (or $z_2$),
the sum of the logarithms in the
superpotential approaches $\frac{1}{4} \log(z_1 - z_2)^2$.  This is the same as the change in normalization between $\CC_a$ and $\tilde \CC_a$ for $s=-\frac{1}{2}$.
This basis of $\tilde \CC_a$ will be useful whenever we are at strong symmetry breaking. In this basis
\begin{equation}\label{nurth}
B_{12} := \tilde \CC_1 \to q^{\frac{1}{4}} \tilde \CC_2  \qquad   \tilde \CC_2 \to \left( q^{-\frac{1}{4}} -  q^{\frac{3}{4}} \right) \tilde \CC_2 + q^{\frac{1}{4}} \tilde \CC_1.
\end{equation}

So far, we have analyzed this problem using cycles $\CC_1$, $\CC_2$, and $\CC_{12}$
that are visible by inspection.  
Let us compare these cycles with the thimbles.
The Bethe equation with two $z$'s and one $w$ is
\begin{equation}
\frac{1}{w - z_1} + \frac{1}{w - z_2} =c
\end{equation}
and it has two solutions.

There are two regimes of interest. If $|Z|>>1$, then the critical points are approximately
$w = z_a + 1/c$, $a=1,2$. For each of the two critical points, assuming that $c>0$ 
and $\mathrm{Re}\,Z=0$, both
$w-z_1$ and $w-z_2$ have positive
real part, and it is natural in defining $\W$ to pick branches of $\log(w-z_1)$ and $\log(w-z_2)$
such that the imaginary parts are bounded by $\pm \pi/2$.  The thimbles defined this way
coincide with the $\tilde \CC_a$ we defined above (and not with the $\CC_a$).
The advantage of this choice is that it 
extends naturally to the general case of many fields with symmetry
breaking, which we will treat in section \ref{genpicture}. 
If $|Z|<<1$, then we have approximate critical points $w = \frac{1}{2}(z_1 + z_2)$ and $w=2/c$.
The first critical point sits between the two $z_a$,
and the associated thimble is $\CC_{12}$. The second critical point is associated to
$\CC_1$ if $\mathrm{Re}\, Z >0$, $\CC_2$ otherwise.

\begin{figure}
 \begin{center}
   \includegraphics[width=4in]{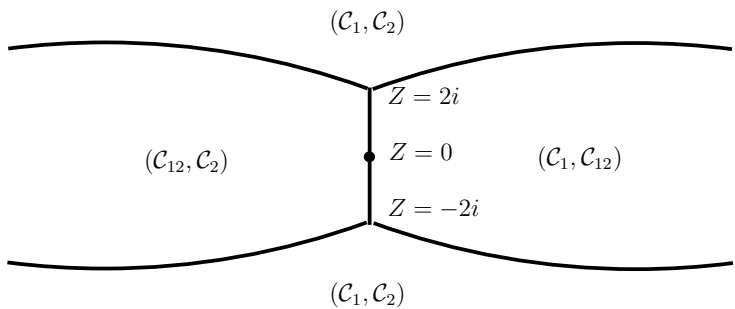}
 \end{center}
\caption{\small  The pattern of Stokes walls and the bases of thimbles in the $c(z_1 - z_2)$ plane }
 \label{braidb}
\end{figure}

We depict the Stokes walls for the system of thimbles in fig. \ref{braidb}. There are
regions in parameter space where the basis of thimbles consists of any two 
of $\CC_1$, $\CC_2$ and $\CC_{12}$, up to a choice of branches of the 
superpotential. Pairs of regions meet along Stokes walls,
and triples of regions meet at the points $Z=\pm 2i$  where the two critical points coincide.
This is a general feature; in a generic problem of this type, there will always be loci of
complex codimension $1$ 
where two critical points coincide. In a plane transverse to such a locus,
$\W$ can be modeled by a simple cubic function $\W = w^3 + \delta w$ of one variable $w$,
with a parameter $\delta$; this function has two critical points that coincide for $\delta=0$,
where three Stokes walls meet.  The
local Stokes behavior is universal; it corresponds to the behavior of the Airy function.

As we explained in our general discussion
of section \ref{fromth}, the asymptotic behavior of integrals for $b\to 0$ is clearest in the basis
of thimbles.  For the present problem, fig.  \ref{braidb} captures the relevant information.
There are four regions: the upper and lower regions correspond to the system of thimbles we saw in the $|Z|>>1$ limit, the two intermediate regions to the system of thimbles
which we saw in the $|Z|<<1$ limit.

\begin{figure}
 \begin{center}
   \includegraphics[width=2.5in]{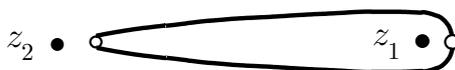}
 \end{center}
\caption{\small  The two Morse flows from the critical point (empty dot) near $z_2$ to the critical point (empty dot) near $z_1$.  The flows occur at slightly different values of
${\mathrm{Im}}\,Z$. }
 \label{braidc}
\end{figure}

The formulas (\ref{furthox}) or (\ref{nurth}) describe the braiding matrix $B_{12}$
that compares the region at the top of fig. \ref{braidb}, where the thimbles are $\CC_1$ and 
$\CC_2$, to the region at the bottom, where again the thimbles are $\CC_1$ and $\CC_2$.
We did this computation by inspection, not by analyzing the Stokes lines.  
 In deriving (\ref{nurth}), we assumed that
$z_1$ moves half-way around $z_2$ in a clockwise direction; this is equivalent to
moving from the top to the bottom of fig. \ref{braidb} with $\mathrm{Re}\,Z>0$.
From fig. \ref{braidb},
we see that in this process, we will cross two Stokes lines.
This means that
in the basis of thimbles, the braiding matrix $B_{12}$   ``decomposes'' into a 
sequence of two elementary moves,
each associated to one Stokes line. 
Each elementary move involves a gradient flow in which the Bethe root $w$
flows from a critical point just to the right of $z_2$ to a critical point just to the right of $z_1$.  
There are two such flows, differing by whether $w$ passes above or below $z_1$.  
The two flows, which occur at slightly different values of $\mathrm{Im}\,Z$, 
are sketched in fig. \ref{braidc}.  In the formula $B_{12}\tilde C_2=
(q^{-1/4}-q^{3/4})\tilde C_2+q^{1/4}\tilde C_1$ of eqn. (\ref{nurth}), the term $q^{1/4}\tilde C_1$
on the right hand side is the contribution of the formal monodromy alone, while the other
two terms are contributions from the two gradient flow solutions.

In the small $Z$ region, braiding around $Z=0$ is represented in the basis of 
thimbles by a triangular matrix: $\CC_{12}$ is an eigenvector. There is a general 
reason for this fact;  the integral over $\CC_{12}$ is smaller than the integral over
any other cycle either in the $Z \to 0$ limit, or in the $b \to 0$ limit. In either limit, the 
integral over $\CC_{12}$ is controlled by the usual saddle point approximation, and 
the  critical point associated to $\CC_{12}$ is 
the one at which the Morse function is the smallest.  So
$\CC_{12}$ must be an eigenvector of the monodromy.
With any number of $z$'s and $w$'s, the set of thimbles which has one $w$ in between a given
pair of very close $z_a$
span the ``small'' eigenspace of conformal blocks where the two degenerate fields of 
momentum $- \frac{1}{2b}$ fuse to the identity.

\subsubsection{The Second Example}\label{sec}

The second example with two critical points arises if there are three singular
monopoles of minimum charge and a single Bethe root $w$.
Placing the singular monopoles at $z_1,z_2,z_3$,  there are
 three obvious possible integration cycles: a straight path connecting $z_a$ to $z_b$ 
 for any $a,b$.
We can pick the branch of the superpotential in such a way that these
three cycles add to zero.

This definition makes sense if the three points are not aligned. The parameter space of $z_a$
is then split into two halves: either $z_1$, $z_2$ and $z_3$ form a triangle with positive orientation, or they form a triangle with negative orientation. We will denote the
three natural cycles in either case as $\CC_{ab}^\pm$. So
\begin{equation} \label{skeinc}
\CC^+_{12} + \CC^+_{23} + \CC^+_{31} = 0 \qquad \CC^-_{12} + \CC^-_{23} + \CC^-_{31} =0
\end{equation}
but the two bases are related in an interesting way across the loci where the $z_a$ are collinear.

There are three such loci, where one of the three $z_a$ passes between the other two.
For example, if $z_2$ passes between $z_1$ and $z_3$, $\CC^+_{12}$ and $\CC^+_{23}$ will be related to
$\CC^-_{12}$ and $\CC^-_{23}$ simply by a change of branch of $\W$, while the transformation
of $\CC^+_{13}$ then follows from (\ref{skeinc}).

 The Bethe equation
\begin{equation}\label{polz} \frac{1}{w-z_1}+\frac{1}{w-z_2}+\frac{1}{w-z_3} =0 \end{equation}
is equivalent to a quadratic equation for $w$, so it has two solutions, 
corresponding to two independent thimbles.  The thimbles are 
equivalent to two of the paths joining a pair of $z$'s, but
which pairs appear depends on the choice of the $z$'s.  If the 
$z$'s are collinear, then the Bethe roots are located in the segments between 
adjacent $z$'s, and the thimbles coincide with those segments. For example, 
if $z_2$ is between $z_1$ and $z_3$,
the two thimbles are  $\CC^\pm_{12}$ and $\CC^\pm_{23}$, up to powers of $q$.

As usual, the thimble joining $z_a$ and $z_b$ corresponds to 
the conformal block where the corresponding two degenerate 
fields fuse to the identity. The relations (\ref{skeinc}) correspond to the elementary 
``skein relation'' between the three different ways to fuse two of the three primary 
fields to the identity. In other words,  whenever $z_a$ and $z_b$ are close together, 
there is a thimble which joins them and is an eigenvector of the braiding of $z_a$ and 
$z_b$, and that
braiding  matrix is triangular.

\begin{figure}
 \begin{center}
   \includegraphics[width=3.5in]{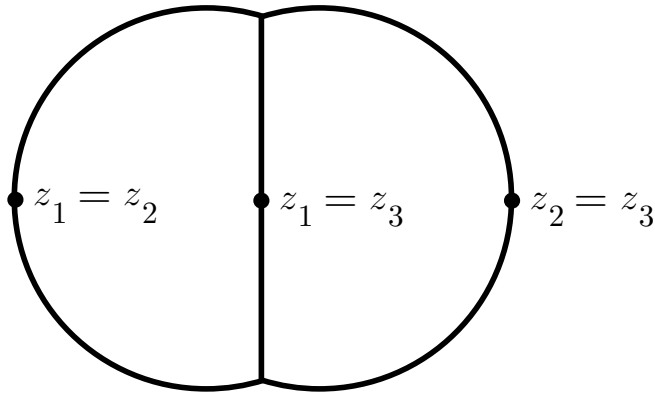}
 \end{center}
\caption{\small  The pattern of Stokes walls in the space of shapes of the triangle with vertices $z_1$, $z_2$, $z_3$.}
 \label{stok}
\end{figure}

We depict the Stokes walls in fig. \ref{stok}.
The integrals $I_{ab}$ can be readily evaluated in terms of hypergeometric functions. 

\subsection{A Final Example}\label{fin}
There is one more example that is both instructive and relevant to understanding the general picture. This
 is the case of two $z_a$ of charge $1$, accompanied by two $w_i$. This can only happen in the presence of complex symmetry breaking. 
It is easy to see that the  Bethe equations (\ref{normo}) have only a unique solution. 
 $Q$ and $K$ are both  of degree $2$, so   $P$ is of degree 0 and can be set to 1.
This leads to linear equations that uniquely determine the coefficients in $Q$.  Since the solution of the
Bethe equations is unique, there is only one conformal block and the monodromy in braiding the $z_a$
can only be multiplication by a function of $q$.

If $z_1$, and $z_2$ are well-separated, i.e.  $c(z_1 - z_2)$ has large absolute value, then 
the solution of the Bethe equations  is easily described: $w_1 \sim z_1 + 1/c$ and $w_2 \sim z_2 + 1/c$.
There is an obvious integration cycle, with $w_1$ and $w_2$  integrated respectively over the rays from $z_1$ and $z_2$
to infinity in the $c$ direction; these rays were labeled
$\CC_1$ and $\CC_2$ in fig. \ref{cyclesonetwo}.
 We can denote this cycle as $\CC = \CC_1 \times \CC_2$. 

The unique thimble for the problem is a small deformation of $\CC$ if the imaginary part of $Z = c(z_1 - z_2)$
is large. As we deform $Z$, the unique thimble will deform continuously. Here Morse theory is rather useful: it is rather tricky to 
verify by hand that $\CC$ goes back to itself under braiding of $z_1$ and $z_2$, as it requires a contour deformation which does not 
keep $\CC$ in a simple product form. But the evolution of the thimble provides us implicitly with such a deformation. 
It is likewise far from obvious at first sight that the conformal block produced by the contour integral 
over $w_1$ and $w_2$ will be a simple function with abelian monodromy in the $Z$-plane. 
Nevertheless, this must be the case. 

The effect of braiding is easily computed at large $c$, by looking at the saddle point estimate for the integral.
At the saddle, $w_1 \sim z_1 + 1/c$ and $w_2 \sim z_2 + 1/c$ remains true along the whole braiding, 
as long as it is executed at large $|Z|$. The Hessian of $\W$ is close to a large, $Z$-independent, multiple ($\sim| c |^2$) of the identity matrix,
so the braiding phase at large $Z$ is controlled by the   value of $\exp(\W/b^2)$ at the critical point. The relevant factors are
\begin{equation}
(z_1 - z_2)^{-b^{-2}/2} (w_1 - z_2)^{b^{-2}}(w_2 - z_1)^{b^{-2}}(w_1 - w_2)^{-2 b^{-2}} \sim (z_1 - z_2)^{-b^{-2}/2}.
\end{equation}
 Hence we recover the same braiding phase as in a setup with two $z_a$ of charge $1$ and no $w_i$.

This last statement will be part of the input for constructing an effective abelian description in section \ref{abdesc}.
In an abelian theory, one would expect that the braiding of two objects depends only on the product of their charges.  
So braiding of two objects both of charge $-1$ ($z$'s unaccompanied by $w$'s) or two objects both of charge 1
($z$'s accompanied by $w$'s) gives the same result.   Braiding of an object of charge 1 with an object of charge $-1$
is less simple, since non-trivial gradient flows enter the picture, as we saw in section \ref{twop}.

\subsection{General Picture With Symmetry Breaking}\label{genpicture}
In this section, we develop a general picture with symmetry breaking.
We take the symmetry breaking parameter $c$ to be real and positive; 
we consider any number of singular monopoles at positions
$z_a$, $a=1,\dots,d$, and any number of Bethe roots $w_i$, $i=1,\dots,q$.
 At first we will set all the charges $k_a$ to $1$.
 
We can produce a basis of integration cycles by hand.  We assume that the imaginary 
parts of the $z_a$ are distinct, and ordered so that 
$\mathrm{Im} (z_a- z_{a'}) >0$ if $a'>a$. We define the rays $\CC_a$ 
which start at $z_a$ and are parallel to the positive
real axis.

Our integration cycles will be products $\CC_{a_1 a_2\cdots a_q} := 
\CC_{a_1} \times\CC_{a_2} \times \cdots \times \CC_{a_q}$ for 
every subset of $q$ distinct $z$'s. This gives ${d \choose q}$ integration 
cycles, which in the region of large $c$ can be interpreted as thimbles.
Indeed the Bethe equations 
\begin{equation}
\sum_a \frac{1}{w_i - z_a} = c + \sum_{j \neq i} \frac{2}{w_i - w_j}
\end{equation}
have approximate solutions for large $c$ with each $w_i$ equal
approximately to $z_{a_i}+1/c$, with $a_i\not=a_j$ for $i\not=j$.  The corresponding
thimbles are precisely the $\CC_{a_1a_2\dots a_q}$.
Summing over all $q$, we get $2^d$ critical points or conformal blocks in all, as expected.

The braid group representation is rather simple in this basis. We let the $z_a$
move around in the complex plane.  The Stokes walls are approximately
at the locus where the real parts of two $z_a$ coincide.  Morse flows will occur if and only if
a $z_a$ unaccompanied by a $w_i$ passes to the right of a $z_{a'}$ accompanied
by a $w_i$. The resulting behavior involves only the two $z$'s that are crossing and
 is precisely what we analyzed in detail
in section \ref{twop}.  The braiding matrix when $z_a$ crosses $z_{a'}$ acts non-trivially
only on $\CC_a$ and $\CC_{a'}$ (and any cycle $\CC_{a_1a_2\dots a_q}$ that contains
one or both of these), and takes the same form as (\ref{nurth}).  As in fig. \ref{braidc},
the braiding matrix for this process can be interpreted as resulting from a pair of Morse
theory flows.
 It is tedious, but elementary, to check that the braid group relations
are satisfied, as they should be.

\subsubsection{Degenerate Fields Of Any Charge}\label{degany}

Now, still with complex symmetry breaking, we will relax the constraint that all $k_a$ equal $1$.
First, we can consider a single $z$ of charge $k$, with $q$ Bethe roots. The Bethe equations only have solutions if $k \geq q$. 
(This is clear from eqn. (\ref{normo}) for the opers.) Actually, according to the theory of Bethe equations, the solution is unique, for given $q$, and corresponds to a unique thimble or integration cycle.

\begin{figure}
 \begin{center}
   \includegraphics[width=3.5in]{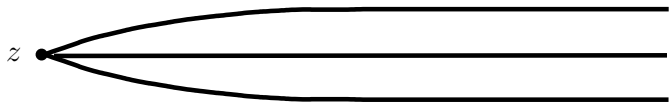}
 \end{center}
\caption{\small  An integration cycle for one singular monopole of non-minimal charge.}
 \label{thimone}
\end{figure}

\begin{figure}
 \begin{center}
   \includegraphics[width=3.5in]{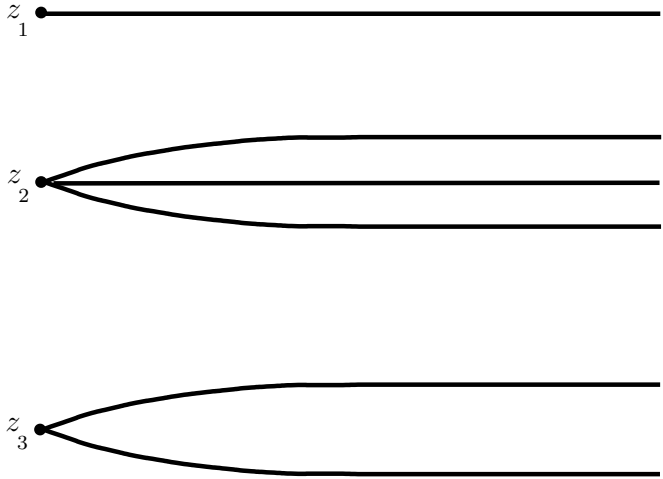}
 \end{center}
\caption{\small  An integration cycle for several singular monopoles of non-minimal charge. To $z_a$ we
attach $q_a$ of the $w$'s, for suitable $q_a$. In the picture,
the $q_a$ are $1,3$, and 2.}
 \label{thimtwo}
\end{figure}
Indeed, it is easy to describe this unique possible integration cycle $\CC^{(q)}$: one
integrates each $w_i$ from $z$ to $+\infty$, along distinct, non-intersecting paths,
as in fig. \ref{thimone}.

We can then immediately describe a basis of integration cycles
in the most general case, with any number of $z_a$ of charge $k_a$,
and the number of $w_i$ being $q \leq \sum_a k_a$. We get a unique cycle for 
each way to decompose
$q = \sum_a q_a$ with 
$q_a \leq k_a$: for each $a$, we integrate $q_a$ of the $w_i$ from $z_a$ to $+\infty$,
along distinct, non-intersecting paths, as in fig. \ref{thimtwo}.  This gives an
integration cycle  $\prod_a \CC^{(q_a)}_a$.

At strong symmetry breaking (or equivalently, if the $z_a$ are well 
separated along the imaginary axis), these integration cycles correspond 
to the thimbles associated to the unique solutions of the Bethe equations with 
$q_a$ of the $w_i$ near $z_a$. There is no obstruction, in principle, to derive the 
braid group representation in this
basis.

For a homological approach -- essentially corresponding to the free field
realization -- to the construction of braid group representations
for any $k_a$, though without symmetry breaking, see \cite{Ltwo}.

\subsection{Turning Off Symmetry Breaking}
Since symmetry breaking is so useful in simplifying 
our analysis, the question arises of verifying that symmetry
breaking does not affect the values of knot or link invariants.

The description of integration cycles in the previous section is applicable 
as long as the symmetry breaking parameter $c$ is non-zero. But for $c\to 0$,
we lose some integration cycles, as the $w_i$ cannot go to infinity any more. In terms of 
critical points of $\W$, the behavior for small $c$ is easy to describe.
For small $c$, for each solution of the Bethe equations,
the Bethe roots split naturally into two subsets. Some number  $q_0$ of 
Bethe roots,  
which we call  $w^{(0)}_i$, $i=1,\dots,q_0$, remain of order $1$ in the limit $c \to 0$, while
the remaining
$q_\infty$ Bethe roots, which we call $w^{(\infty)}_j,$ 
$j=1,\dots,q_\infty$, are of order $c^{-1}$.

In this situation, the Bethe equations for the $w^{(0)}_i$ are well approximated by the
Bethe equations for that number $w$'s in the absence of symmetry breaking.
On the other hand, the Bethe equations for the $w^{(\infty)}_j$ are well approximated by
the Bethe equations for that number of $w$'s, in the presence of a single
$z$ of charge $k_{\mathrm{eff}}=\sum_a k_a - 2 q_0$.  Eqn. (\ref{ruy}) ensures that $k_{\mathrm{eff}}$ is
non-negative, so in fact
 \begin{equation}\label{helpful}0\leq k_{\mathrm{eff}}\leq \sum_ak_a.\end{equation}
As we have discussed in section \ref{degany},
the Bethe equations for the $w^{(\infty)}_i$ have a single
solution if
$k_{\mathrm{eff}}\geq q_\infty$; otherwise, they have no solutions.
Summing over all decompositions $q=q_0+q_\infty$ and all solutions of the Bethe equations for
the $w^{(0)}_i$, and finally over all possible values of $q$,
one gets the expected number  $\prod_{a=1}^d (k_a+1)$ of solutions of the Bethe equations.

The counting can be carried out as follows, in terms of Wilson
operators of the dual $SU(2)$ gauge theory.  The singular monopoles of charge $k_a$
correspond to representations $R_a$ of $SU(2)$ of spin $k_a/2$ and dimension $k_a+1$.
The tensor product $R=\otimes _a R_a$ has dimension $\prod_a(k_a+1)$.
This representation can be decomposed as a direct sum of $SU(2)$ modules of spin
$k_{\mathrm {eff}}/2$ (where $k_{\mathrm{eff}}$ is bounded by $0\leq k_{\mathrm{eff}}\leq \sum_ak_a$, just as in
(\ref{helpful})).  The states that transform with this spin are as numerous as
 the solutions of the Bethe equations with
 $\sum_a k_a-2q_0=k_{\mathrm{eff}}/2$ (and all possible values of $q$).

Now consider braiding of the $z_a$.  For sufficiently small $c$, when
one crosses a Stokes wall, there are Morse flows in which
$q_\infty$ becomes smaller, but no such flows  in which
$q_\infty$ becomes larger.  The reason for this is that for
$c\to 0$,  the values of the Morse function $h=\mathrm{Re}\,\W$
at a critical point are greater the greater is $q_\infty$.
(This is because some contributions to $h$ are of order $\ln(1/|c|)$ for $c\to 0$. With the help of
(\ref{helpful}), one can show that the coefficient of $\ln(1/|c|)$ is an increasing function of $q_\infty$.)
As a result, decomposing the space $\H$ of physical states according to the value of $q_\infty$,
the monodromy matrix is block triangular:
\begin{equation}\label{blocks}B\sim \begin{pmatrix} *& *&* & * & * \cr  *&*&*&*&*\cr
                                                                                   0& 0 & *&* & *\cr 0&0&*&*&*\cr
                                                                         0&0&           0 & 0 & * \cr \end{pmatrix}.\end{equation}
The diagonal blocks (which are of rank $2,2,$ and 1 in the example given)
are the monodromy representations  that one would have in the absence of symmetry breaking for given $q_0=q-q_\infty$.

To the extent that one can compute knot or link invariants by taking traces of braid group representations, the off-diagonal blocks in (\ref{blocks})
are not important as they do not contribute to traces.  Actually, to compute the Jones polynomial and related invariants of knots and links,
one needs in addition to the braid group representations an
additional ``fusion'' operation in which a pair of $z_a$ of the same
charge is created or annihilated.  The additional information
that we need to ensure that knot invariants are unaffected by symmetry breaking
and do not change upon setting $c=0$ is that fusion never
involves creating or annihilating any $w$'s at infinity.  This is natural because of the local
nature of the fusion operation.

\subsubsection{A Clarification}\label{clarify}

A careful reader might notice a small sleight of hand in this derivation.  The inequality $k_{\mathrm{eff}}\geq 0$ was deduced
from (\ref{ruy}), but the original derivation of (\ref{ruy}) was based on picking $Q$ to have a smaller degree than $P$.
The Bethe equations at $c=0$ certainly have solutions in which this is not the case.  Why are we entitled to restrict to this case?

Suppose that at $c=0$, we find a pair $P_0,Q_0$ obeying the oper condition
\begin{equation}\label{opcon} P\frac{\d Q}{\d z}-\frac{\d P}{\d z}Q=K(z).\end{equation}
Now suppose that we turn on very weak symmetry breaking.  The oper equation becomes 
\begin{equation}\label{opconz} P\frac{\d Q}{\d z}-\frac{\d P}{\d z}Q-cPQ=K(z).\end{equation}
We hope that as $c$ is turned on, there is a pair $(P(z;c),Q(z;c))$ obeying (\ref{opconz})
and such that $Q$ has an expansion
\begin{equation}\label{turnox}Q(z;c)=Q_0+cQ_1+c^2Q_2+\dots.\end{equation}
This will ensure that the $Q(z;c)$ has roots that approach the roots of $Q_0$ as $c\to 0$,
plus possible additional roots that go to infinity for $c\to 0$.  One might expect
that $P$ would have an expansion of the same form, but this is not the case.  The degrees
$p$ and $q$ of polynomials $P,Q$ obeying (\ref{opcon}) satisfy $p+q=k+1$, but as soon as $c\not=0$,
the relation becomes $p+q=k$.  So the degree of $P$ must drop as soon as $c\not=0$.
The way that this happens is that the expansion for $P$ is actually
\begin{equation}\label{nopcon}P(z;c)=c^{-1}P_{-1}+\tilde P_0 +c P_1+\dots\end{equation}
where we write $\tilde P_0$ for the coefficient of $c^0$, as this polynomial does not coincide
with $P_0$.  Plugging (\ref{turnox}) and (\ref{nopcon}) in (\ref{opconz}), we learn
from the term of order $c^{-1}$ in the equation that $P_{-1}$ is a multiple of $Q_0$,
and this multiple must be nonzero or else the term of order $c^0$  in the equation would
force $p+q\geq k+1$.  So $p\geq q_0$, and this together with $p+q=k$  and $q\geq q_0$ implies
that $q_0\leq k/2$, as desired.  

 The moral of the story is that solutions of the Bethe
equations for $c=0$ with $q>k/2$ do exist, but they are unstable to symmetry breaking.
Various forms of this statement are known in the literature on integrable systems.

\subsection{Three-Dimensional Interpretation}\label{abdesc}
To apply our results to knots and not just to braids, it will help  to understand
the three-dimensional interpretation of what we have computed so far.  We consider
knots in $\R^3$, so the four-manifold on which we are trying to count solutions of
eqns. (\ref{bpseqns}) is $M_4=\R^3\times \R_+$.  We describe $\R^3$ with Euclidean
coordinates $x^1,x^2,x^3$.  The adiabatic evolution considered so far has been
in the $x^1$ direction, while we have combined the other coordinates to a complex variable
$z=x^2+ix^3$.  As usual, we take the gauge group to be $G=SO(3)$.

 We will focus on the case of strong symmetry breaking.
 The symmetry breaking involves
the choice of an expectation value $\vec\phi=\mathrm{diag}(\vec\ca,-\vec\ca)$, 
where $\vec\ca$ is a vector
in $\R^3$.   As long as the complex symmetry breaking is nonzero,
this vector does not point in the $x^1$ direction, that is, the direction that
we chose   for the
adiabatic evolution.  Topologically, if the directions are not the same, we may as well
think of them as  orthogonal: we 
consider adiabatic evolution in the $x^1$ direction, and symmetry breaking
with $\vec c = - 2 \vec\ca/\zeta$ pointing in the positive $x^2$ direction.  This will
be strong complex symmetry breaking with real, positive $c$, in the terminology that
we have used so far.

 We will concentrate on the case that the strands have minimum magnetic charge
 only, and thus are dual to the two-dimensional representation of $SU(2)$. 
At a generic time,  each strand has two possible states: 
it is or it is not accompanied by a Bethe root $w_i$. 
In the low energy effective abelian gauge theory, the strand has magnetic charge 1
if accompanied by a Bethe root, and otherwise $-1$. 

The magnetic charge of a given strand changes when the Bethe root accompanying
that strand moves to another strand.   In the context of adiabatic
evolution, this results from a Morse theory flow in which a Bethe root moves from
one strand to another.  
This happens at a value of $x^1$ at which one crosses a Stokes wall.
The lesson of section \ref{twop} is that (in the limit of strong symmetry breaking)
one crosses a Stokes wall at a time (that is a value of $x^1$) at which two strands have the 
same value of $x^3=\mathrm{Im}\,z$.  So the crossing of a Stokes wall occurs when
two strands have common values of $x^1$ and $x^3$, and thus differ only in $x^2$.
Differently put, this happens when the two strands are separated in the direction of
symmetry breaking.

\begin{figure}
 \begin{center}
   \includegraphics[width=3.5in]{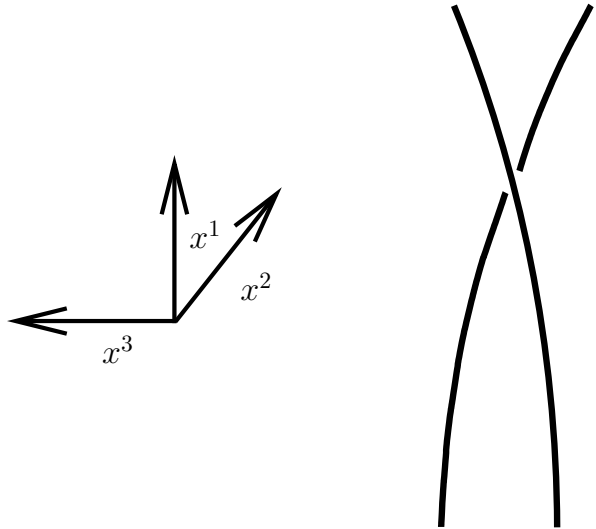}
 \end{center}
\caption{\small  A three-dimensional picture of the process that leads to a 
Morse theory flow. Just two strands are pictured here.
The $x^1$ direction is plotted vertically and the
$x^2$ direction runs into the paper.  We are looking at the picture from along the 
negative $x^2$ axis.  The coordinates $x^2$ and $x^3$ combine to a 
complex variable $z=x^2+ix^3$.  For a particular choice of the direction of 
complex symmetry breaking, a non-trivial Morse theory flow can occur only at 
values of $x^1$ at which the two strands have the same value of $x^3$ and 
thus project to the same point in the
$x^1-x^3$ plane.  In the language of knot theory, we make a two-dimensional 
picture by projecting a knot or link to the $x^1-x^3$ plane.  In this projection, there 
are crossing points, and these are the points at which a non-trivial Morse flow 
may occcur.  \label{crossing}}
\end{figure}

A three-dimensional picture clarifies things (fig. \ref{crossing}). When two strands 
align along the $x^2$
direction, a Morse flow can occur.  The Morse flow occurs on a  
time scale fast compared to the adiabatic evolution, so in the adiabatic picture 
it is essentially instantaneous. The flow involves a Bethe root moving towards 
the positive $x^2$ direction.
The flow can only occur between strands of opposite charge, and will allow positive
charge to move towards positive $x^2$ only.

\begin{figure}
 \begin{center}
   \includegraphics[width=1.2in]{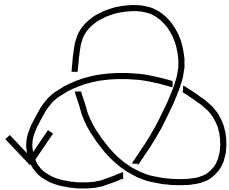}
 \end{center}
\caption{\small  A simple example of a knot projection, with only simple crossings
and simple maxima and minima of the height function.  This figure also
illustrates the fact that the projection to a plane of an oriented knot
allows one to define an integer invariant $p=(1/2\pi)\oint\d s \,\d\theta/\d s$ that equals
the total change in moving around the knot of
 the angle $\theta$ defined by the tangent vector to the knot.  For the example shown, $p=2$.
$p$ is the only invariant of a knot projection that can be written as a local integral
along the knot.
 It depends on the choice of projection and is not an invariant of the knot {\it per se}.}
 \label{winding}
\end{figure}

In general, given any knot, we can usefully project it to the  $x^1-x^3$ plane, and look
at it from the negative $x^2$ direction.    We suppose that the 
embedding of the knot in $\R^3$ is generic
enough so that its tangent vector always has a non-zero projection to the 
$x^1-x^3$ plane, and moreover so 
that the projection of the knot to the plane has only 
simple crossings; finally we will assume that the function $x^1$
has only simple maxima and minima along the knot. 
A simple example of a knot projection is given in fig. \ref{winding}.
  In such a knot projection, the low energy 
  abelian description is valid away from crossings, so away from crossings and local
  maxima and minima, which we discuss in section \ref{creand},
each strand can be labeled by its magnetic charge 1 or $-1$. The charges are unchanged at crossings, unless
a strand of positive charge passes over a 
strand of negative charge, in which case a charge exchange process
(corresponding in Morse theory to a non-trivial gradient flow) is possible.

\begin{figure}
 \begin{center}
   \includegraphics[width=6in]{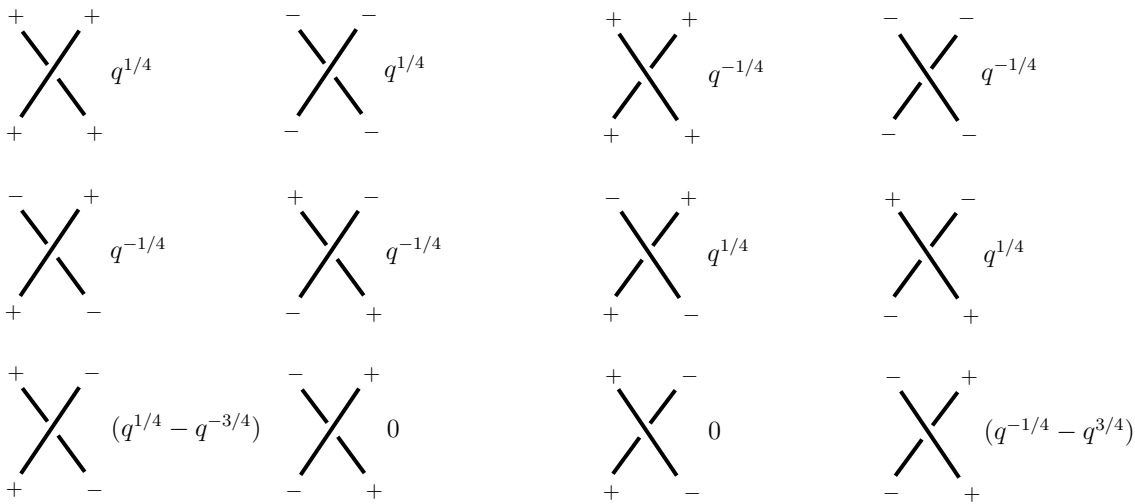}
 \end{center}
\caption{\small  The vertex model assigns the indicated factors to 
every crossing of two strands. The $+$ and $-$ signs labeling the strands express upward flow 
of magnetic charge $+1$ or $-1$; if one turns the
picture upside down, the weights remain unchanged, provided one exchanges all $+$ and $-$
labels.  Charge can be exchanged between strands, but only when a positive charge flows
in from the bottom above a negative charge.  If one reflects the picture from left to right,
while also replacing $q$ with $q^{-1}$, the weights remain invariant; this reflects the behavior
of Chern-Simons theory under reversal of orientation.
 \label{vertexweights}}
\end{figure}
Our previous calculations assign a weight to each possible crossing. These weights are just
$q^{\pm 1/4}$ if the charges are unchanged at the crossing. When the charge jumps,
the weight is $q^{\pm 1/4} - q^{\mp 3/4}$. The weights are summarized in fig. \ref{vertexweights}.
We have arrived at a known vertex model representation of the braid group representations
associated to the Jones polynomials. 
See for example the $R$-matrix on  page $125$ of \cite{Kauffman}
(where $A$ is our $q^{1/4}$) or see \cite{KauffmanTwo}, especially pp. 1777-8,
or fig. 10 of \cite{Vertex}.

\subsubsection{Creation And Annihilation Of Strands}\label{creand}

In order to reproduce the knot invariants, we need to understand 
the loci where the adiabatic approximation is invalid, because two strands are
created or annihilated. Although the adiabatic approximation is invalid near such points, 
the low energy abelian description remains valid. As an immediate consequence, 
conservation of charge in the abelian theory makes it clear that only pairs of strands 
with opposite charge can be created or annihilated.

The map from line operators in a microscopic theory to line operators in an effective low 
energy description is akin to the corresponding map for local operators, but it has a little twist:
the coefficients are not $c$-numbers, but rather quantum mechanical vector spaces 
that have to be transported along the line.  In the present case, the vector spaces are 
one-dimensional (in the $U(1)$ theory, an 't Hooft operator has no structure except its charge), 
but we can still get an overall factor from parallel transport. This factor has to be 
written locally along the loop, and must also be consistent with topological invariance. 
For a knot without any additional structure there is no topological invariant that can be 
written as a local integral along the knot,
but once one is given  a projection of the knot to a plane -- in our case the $x^1-x^3$ plane -- 
there is precisely one such invariant, the total winding number of the tangent
vector to the knot (fig. \ref{winding}). This can be written as $(1/2\pi)\oint \d s\,\d \theta/\d s$, 
where the knot is parametrized by a variable
$s$, and $\theta$ is the angular direction of the tangent vector to the knot in the $x^1-x^3$ 
plane.  To define the sign of this invariant, one needs an orientation of the knot, which
in our case comes from the direction of flow of magnetic charge. 
When the microscopic $SU(2)$ theory
is approximated at long distances as an effective $U(1)$ theory, the effective action for 
an 't Hooft operator may acquire a term
$-i\eta\oint \d s \,\d \theta/\d s$, with a universal coefficient $\eta$.    If the tangent direction 
to a knot changes by an angle $\Delta\theta$,
this will contribute a factor
\begin{equation}
\exp (i\eta \Delta \theta)
\end{equation}
In creation or annihilation of a pair of strands, the change in the tangent angle 
is $\Delta\theta=\pi$ or $-\pi$, depending on whether
the positive charge bends to the left or to the right.  This effect will associate a universal 
factor $\exp(\pm i\pi \eta)$ to each
creation or annihilation event, depending on the direction of flow of charge.  

\begin{figure}
 \begin{center}
   \includegraphics[width=4in]{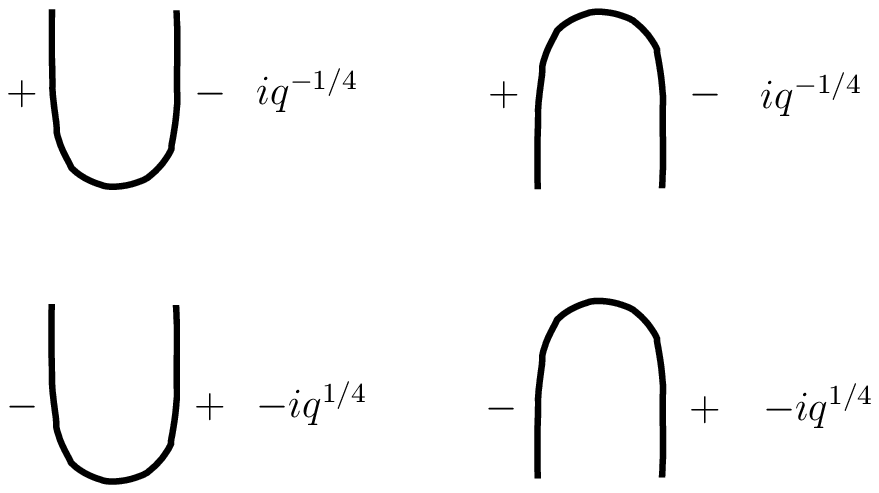}
 \end{center}
\caption{\small  The weights of the vertex model for creation or annihilation of a pair
of strands.  As in fig. \ref{vertexweights}, the weights are invariant under rotating the picture upside down if
one exchanges $+$ and $-$ labels, or under a reflection from left to right if one
replaces $q$ by $q^{-1}$.
 \label{creation}}
\end{figure}

Up to sign, there is a unique choice of $\eta$ that leads to a knot invariant, 
namely  $\exp(i\pi\eta)=\mp iq^{1/4}$.  The choice of sign does not matter, since every knot 
has an even
number of creation and annihilation events; we will take $\exp(i\pi\eta)=-iq^{1/4}$.
The weights for creation and annihilation events with this value of $\eta$ are shown in 
fig. \ref{creation}.
The value of $\eta$
could  possibly be computed directly by 
studying the four-dimensional gauge theory BPS equations near the abelian limit.  
The factor of $q^{1/4}$ should express the difference between instanton number computed
microscopically in the $SO(3)$ theory and instanton number computed in the low energy
$U(1)$ theory.  The factor of $\mp i$ should involve a comparison between fermion determinants
for $SO(3)$ and for $U(1)$.    Globally, the contribution of a given classical solution to the Jones
polynomial is proportional to the sign of the fermion determinant, a subtle invariant that
may receive contributions from charged modes in the microscopic $SO(3)$ theory; to write
this sign as a product of local factors, one apparently must use factors of $i^{\pm 1}$, with
overall signs that depend on how one trivializes the determinant line bundle.

The value of $\eta$  can actually be deduced by combining the information which  is available 
in the abelian description and information available from the conformal block description.
When a pair of strands is created or annihilated, we expect them 
to be fused to the identity.  This means that the two nearby strands,
located at say $z_1$ and $z_2$,
are accompanied by a Bethe root $w$, and that the integration cycle
for this Bethe root is the thimble $\CC_{12}$ connecting $z_1$ and $z_2$.
On the other hand, in section \ref{twop}, we also defined integration cycles
$\tilde \CC_1$ and $\tilde \CC_2$ with the property that the strands at $z_1$ and at
$z_2$ have definite magnetic charges.  (For example, in $\tilde \CC_1$, the Bethe
root accompanies $z_1$, so the charges of the two strands are respectively 1 and $-1$. We order $z_1$ and $z_2$ in order of decreasing $x^3=\mathrm{Im}\,z$.) 
The relation among these cycles turned out to be
\begin{equation}\label{golf}\CC_{12}=\CC_1-\CC_2=q^{-1/4}\tilde\CC_1-q^{1/4}\tilde \CC_2.
\end{equation}
The ratio of the ampitude to create a pair of charges $(1,-1)$ to the amplitude
to create a pair of charges $(-1,1)$ is the ratio of the coefficients on the right
hand side of (\ref{golf}), or $-q^{-1/2}$.  On the other hand, in the abelian
description, this ratio is $\exp(2\pi i\eta)$.  So $\exp(i\pi\eta)=\mp i q^{-1/4}$, as
shown in fig. \ref{creation}.

We have arrived to what is essentially a standard vertex model algorithm for calculating the 
Jones polynomial: given a knot or a link, pick a projection to the $x^1- x^3$ plane such 
that there are
only simple crossings and the function $x^1$ only has simple maxima and minima. 
Divide the link into
segments separated by the maxima, minima and crossings. Label the segments by 
$\pm$ and sum over
all labelings, weighting each labeling with the product of the local  weights at crossings,
maxima, and minima.  
Some simple examples are given momentarily.

\subsubsection{Some Examples And Some Topological Details}\label{details}

\begin{figure}
 \begin{center}
   \includegraphics[width=5.5in]{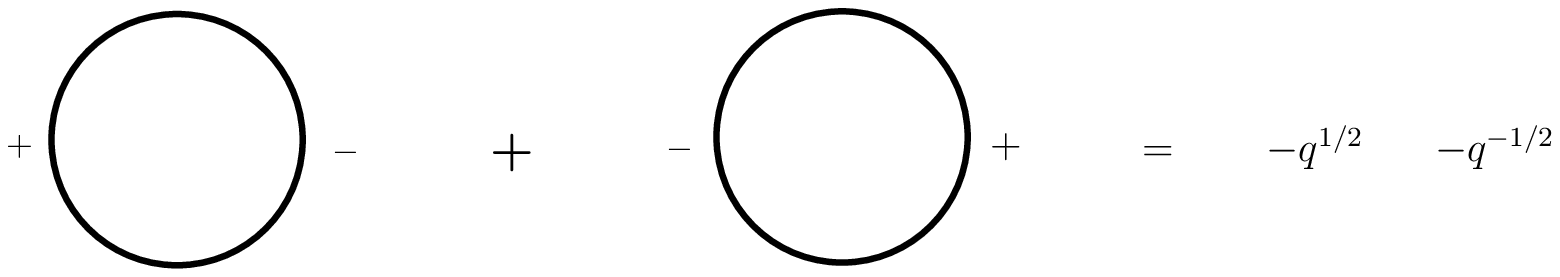}
 \end{center}
\caption{\small  An unknot projected to the plane in the most obvious way.  
There are no crossings, but there is a creation event
and an annihilation event.  In the vertex model, the invariant for the unknot 
is computed by summing over all ways to label the two
sides of the knot (that is, the segments between crossing, creation, and 
annihilation events) by charges $+$ or $-$.  Each labeling is weighted by the product
of the appropriate local factors.  In the present
example, only the two choices shown make nonzero contributions, leading at 
once to the result $-q^{1/2}-q^{-1/2}$.}
 \label{unknot}
\end{figure}
For the simplest example of the use of the vertex model, 
we compute the expectation value of  an unknot, projected to the plane
in an obvious way (fig. \ref{unknot}).  Summing over the two possible labelings
of the diagram, we arrive at the result $-q^{1/2}-q^{-1/2}$.

\begin{figure}
 \begin{center}
   \includegraphics[width=5.5in]{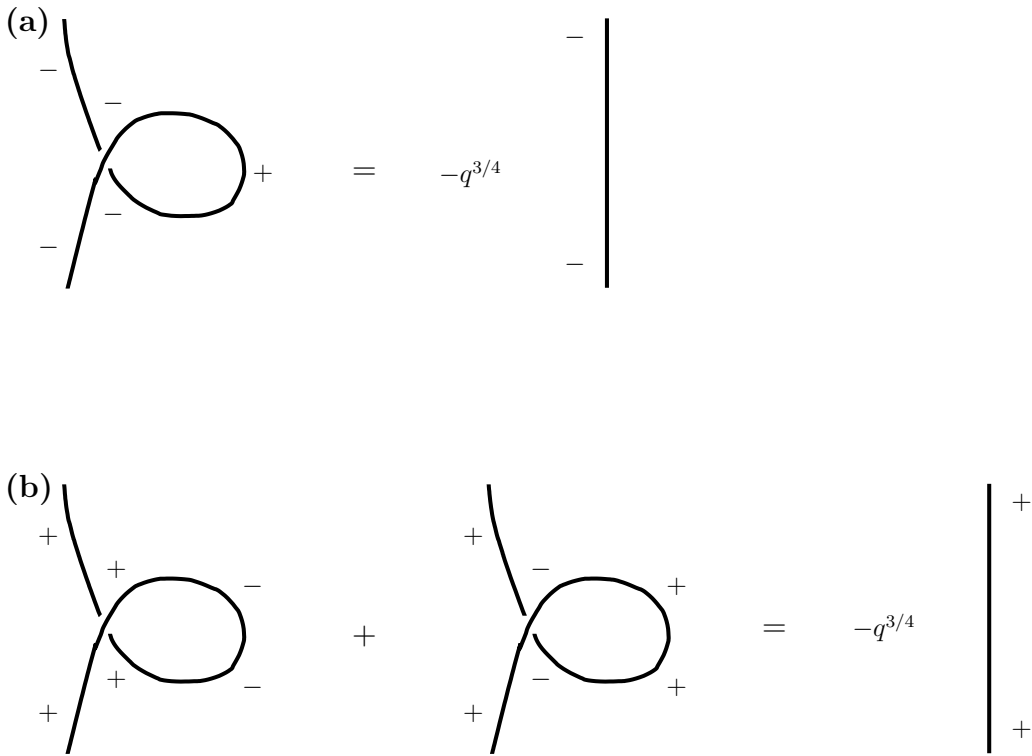}
 \end{center}
\caption{\small  
 \label{twist}  Use of the vertex model to compare two different projections
 of a single strand to the $x^1-x^3$ plane -- with a twist (left) or no twist (right).
 For either sign of the charge carried by the strand,  the twist introduces a factor of $-q^{3/4}$.
 The computation is quite different for the two possible values of the charge carried by the
 strand; for charge $-1$, as shown in (a), the vertex model sum has only one nonzero contribution, 
 corresponding to the indicated labeling, but for
 charge $+1$, there are two possible contributions, shown in (b); they add to the same result.
 In (b), the second contribution involves a charge exchange process in which upper and
 lower strands exchange charge where they cross.  A similar twist with undercrossing
 instead of overcrossing (or with the whole picture replaced by a mirror image) leads
 instead to a factor of $-q^{-3/4}$, as the reader can verify.}
\end{figure}

For a slightly less trivial example, we consider the two projections of a single strand
to the $x^1-x^3$ plane shown in fig. \ref{twist}.    In either (a) or (b), 
the knot projections shown on the left or right  can be deformed into one another,
so one might expect them to be equivalent.
But in the present context,
 this is actually not the case.  The twist of (a) relative to (b) 
 introduces a factor of $-q^{3/4}$, which can
be evaluated by making use of the weights of the vertex model.  It is instructive to
actually do this; the same factor $-q^{3/4}$ arises whether the magnetic charge of
the strand is $+1$ or $ -1$, but the details of the computation are quite different in the two
cases. 

Since the factor $-q^{3/4}$ does 
not depend on the magnetic charge carried by a given strand, 
it is universal: adding a twist of the type shown in the figure 
to any strand in an arbitrary knot or link multiplies
the associated invariant  by $-q^{3/4}$.  A similar twist of the opposite
handedness multiplies the invariant by $-q^{-3/4}$, for similar reasons.  This factor means
that the invariant associated to a knot  (or link) by the quantum field theory depends on a
choice of ``framing.''  A framed knot is a knot that is slightly thickened into a ribbon.  One
keeps track of how the ribbon is twisted and (in the present context) adding a twist multiplies
the invariant by a factor of $-q^{3/4}$.  A knot that is presented with a projection to a plane
comes with a natural framing, given by a slight thickening in the vertical direction, 
normal to the plane.    A little thought (or experimentation with a strip of paper)
shows that although the pictures on the left
and right of fig. \ref{twist}(a) or (b) are topologically equivalent if one ignores the framings,
they do differ by one unit of framing.  

In the context of three-dimensional knot invariants that are associated to two-dimensional
conformal field theory, conformal primary fields in two dimensions are associated to line
operators in three dimensions.  If a conformal primary has dimension $h$, then  in a unit
change in framing,
the corresponding line operator is multiplied by $\exp(2\pi i h)$.  The factor $-q^{3/4}$
is indeed $\exp(2\pi i h)$, where the conformal dimension of the degenerate Virasoro primary
field $V_{-k/2b}$ is
\begin{equation}\label{maxi} h_V(k) =-\frac{k}{2}-\frac{k(k+2)}{4b^2},\end{equation}
and in addition $q=\exp(-2\pi i/b^2)$,
and the vertex weights that we have described are for the minimum charge case $k=1$.

At this stage, an important detail arises.  The vertex weights that we have
described are appropriate for a certain natural normalization of the Jones polynomial,
which has been used in the literature, for instance in \cite{Kauffman}.  However,
a slightly different normalization arises in $SU(2)$ Chern-Simons theory.  In Chern-Simons
theory, the expectation value of an unknot labeled by the two-dimensional representation of
$SU(2)$ is $q^{1/2}+q^{-1/2}$, which differs in sign from what we deduced in fig. \ref{unknot}
using the vertex weights.  (The sign is easily checked in Chern-Simons theory.  Since
$q=\exp(2\pi i/(\k^\vee+2))$ in $SU(2)$ Chern-Simons theory, where $\k^\vee$ is the level, the weak
coupling limit $\k^\vee\to\infty$ corresponds to $q=1$; for $q=1$, the expectation value of a Wilson
loop in any representation is simply the dimension of the representation, or $+2$ for
the two-dimensional representation.)  Similarly, the dimension of a primary field related
to a  representation of $SU(2)$ of spin $j=k/2$ is
\begin{equation}h_{\mathrm{CS}} (k)=\frac{k(k+2)}{4(\k^\vee+2)},\end{equation}
so that the factor acquired in a unit change of framing is $\exp(2\pi ih_{\mathrm{CS}})=q^{3/4}$,
without the minus sign found in fig. \ref{twist}.   Clearly the discrepancy in sign reflects
the fact that $h_V(k)-h_{\mathrm{CS}}(k)=-k/2$, independent of $b$ and $\k^\vee$.

The comparison with Chern-Simons theory is not necessarily a problem for the present paper,
in which we have simply started with the four-dimensional gauge theory equations
(\ref{bpseqns}).  However, one would like to know the best interpretation of these
signs in the context of the duality presented in \cite{fiveknots} between Wilson operators of
Chern-Simons theory and singular monopoles at the boundary.   We believe that the interpretation may be that the dual of a Wilson operator
of spin $k/2$ in Chern-Simons theory is actually a boundary 't Hooft operator that carries angular
momentum $k/2$ and is fermionic when $k$ is odd.  (For 't Hooft operators
defined on the boundary of a four-manifold, the relevant rotation group is $SO(2)$ or rather
its double cover $\mathrm{Spin}(2)$; this group is abelian and 
has one-dimensional representations, labeled by the angular momentum $k/2$.) 
 The fermi statistics for odd $k$ would
give a minus  sign for every crossing (relative to what is presented in fig. \ref{vertexweights}) and
a minus sign for every closed loop; including these signs brings the results
 obtained by braiding of Virasoro
degenerate fields in agreement with the results obtained by braiding in Chern-Simons theory.

Finally, it is instructive to compare the computation in fig. \ref{twist} to an equivalent
computation if the gauge group were simply $G=U(1)$ instead of $SO(3)$.  There would
be two differences.  First, the factors in fig. \ref{creation} associated to creation and
annihilation of a pair of strands would simply be 1.  (Those factors come entirely from
integrating out massive degrees of freedom of the $SO(3)$ theory, in reducing to
an effective abelian description at low  energies.)  Second, charge exchange processes
are absent for $G=U(1)$ (as there are no smooth monopoles), so the second 
contribution in fig. \ref{twist}(b) would be
absent.   A look back to fig. \ref{vertexweights} shows that for $G=U(1)$, evaluation of either
fig. \ref{twist}(a) or (b) gives a simple factor of $q^{1/4}$, instead of $-q^{3/4}$.  
Two comments are in order:\begin{itemize}\item The minus sign of fig. \ref{twist} is absent for $G=U(1)$
(and similarly the minus sign in fig. \ref{unknot} is absent).  \item In Chern-Simons 
theory, the power of $q$ is the quadratic Casimir invariant of the relevant representation 
of $G^\vee$.  The quadratic Casimir of a representation of highest weight $j$ is $j^2$ for 
$G^\vee=U(1)$
and $j(j+1)$ for $G^\vee=SU(2)$.  For $j=1/2$, this gives $q^{1/4}$ or $q^{3/4}$ in the
abelian and nonabelian cases, respectively.\end{itemize}

\subsubsection{Gradient Flow And Strings}\label{morsestrings}

\begin{figure}
 \begin{center}
   \includegraphics[width=3in]{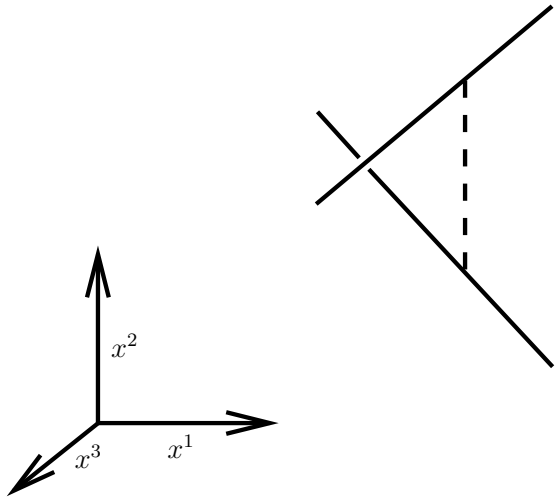}
 \end{center}
\caption{\small  To make the charge exchange process more visible, we have exchanged
the coordinate axes relative to fig. \ref{crossing}.  The $x^2$ axis now runs vertically while
$x^1$ runs horizontally.   Charge exchange occurs at values of $x^1$ at which two strands
differ only in the value of $x^2$, so with the coordinate axes aligned as in this picture,
the charge exchange invoves a flow of charge in the vertical direction, represented by the dotted
line.  We take this to represent the propagation of a BPS soliton.   The soliton propagates
along the axis of symmetry breaking, so it is described by a solution with real symmetry
breaking only -- in fact,  by the bare Miura oper
of equtation (\ref{onl})  with no 't Hooft operator and a single Bethe root.
 \label{soliton}}
\end{figure}
We can give a more concrete physical interpretation to the gradient
flows which occur when strands of appropriate charge cross.   Propagation of magnetic
charge from one strand to another can be described by motion of a magnetic monopole
between the two strands.
If we rotate the coordinates and think of the $x^2$ direction as ``time,'' we should be 
able to see  the relevant object as a time-independent solution in the presence of 
only real symmetry breaking.

\begin{figure}
 \begin{center}
   \includegraphics[width=3in]{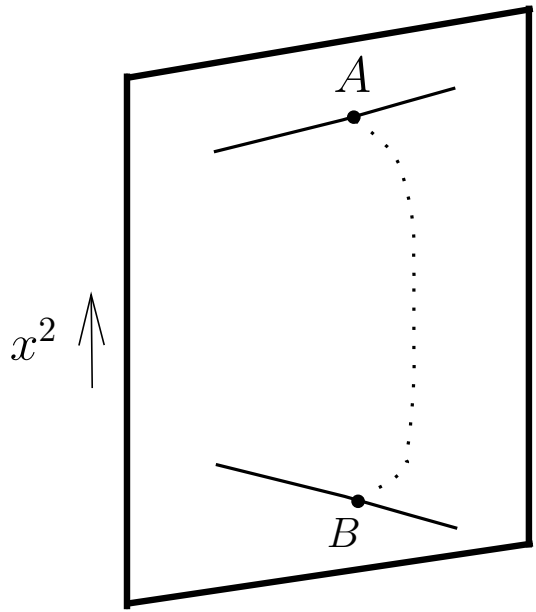}
 \end{center}
\caption{\small  An alternative view of the charge exchange process that was pictured
in fig. \ref{soliton}, to show the role of the $y$ direction (here depicted as the direction normal
to the plane that contains the two strands).  A soliton propagating between two boundary 
points $A$ and
$B$ that are separated by a long distance in the $x^2$ direction will bend away from the
boundary, to reach a value of $y$ corresponding to the solution for the bare Miura oper.
 \label{bending}}
\end{figure}

In fact, 
we have already described precisely the necessary solution: it is associated to 
the bare Miura oper with
a single Bethe root and no singular monopole that was described in eqn. (\ref{onl}).
So we visualize the charge exchange process as propagation of an object in the
$x^2$ direction, as in fig. \ref{soliton}.    This picture is oversimplified, as it ignores
the existence of a fourth dimension, normal to the boundary $\R^3$ that contains
the knots.  An alternative picture  showing the role of the $y$ direction
 is given in fig. \ref{bending}.  When propagating
in the $x^2$ direction, the soliton settles at a value of $y$ that is given by the solution
for the bare Miura oper. (How to compute this value is explained most fully in section
\ref{supermon}.)  Of course, this description is only good if the soliton propagates far
enough in $x^2$ that it has ``time'' to reach the equilibrium value of $y$.  So it is only
good if the strands are sufficiently far separated, or the symmetry breaking is strong enough.

We have gained an intuitive picture of the
solutions of the four-dimensional BPS equations at strong symmetry breaking:
they describe smooth monopole configurations stretched between the singular monopole strands.
The contribution of each configuration to the knot invariant is then the product of two types of factor. One type  arises from the map from the microscopic nonabelian theory to the 
abelian theory,  while the other is computed in the abelian theory. Factors of the first type appear
where pairs of strands are created or annihilated and where a smooth monopole
is emitted or absorbed by a boundary singular monopole.
In the abelian description, the smooth monopoles also look like singular 
monopoles, of charge $2$,
but not attached to the boundary. The whole configuration 
looks like a web of monopole strands, to which the abelian theory 
 associates an overall power of $q$.

\subsubsection{Analog in the Dual Chern-Simons Theory}\label{dualch}

It is entertaining to carry this picture all the way back to three-dimensional Chern-Simons theory,
before all the dualities which brought us to the four-dimensional description studied in the
present paper.
A key step in relating the two pictures is $S$-duality, 
which luckily is very transparent at strong symmetry breaking, as it reduces to 
electric-magnetic duality in the abelian gauge theory. We get immediately
a sum over configurations of massive $W$-bosons stretched between Wilson lines, and only
carrying electric charge towards the positive $x^2$ direction.

This suggests to look for a gauge condition in nonabelian Chern-Simons theory which
would have this effect. It is easy to indentify it: one can pick a ``partial'' axial
 gauge fixing which reduces the $A_2$ component of the gauge field to the Cartan subalgebra.
 We write the nonabelian gauge field as $A=B\frak t_3+W_+\frak t_++W_-\frak t_-$, 
 and impose the $A_2=0$ gauge condition for $W_\pm$;
 it will not be necessary here to make a gauge choice for  the diagonal gauge field $B$.
The action for charged $W$-bosons then reduces to
\begin{equation}
\frac{\k^\vee}{4\pi}\int \d^3x \,\,W_+ D_2 W_-,~~~   D_2=\frac{\partial}{\partial x^2}+[B_2,\,\,\cdot\,\,].
\end{equation}
The equation for the propagator is
\begin{equation}
D_2 G(x,y) = \frac{2\pi}{\k^\vee}\delta^3(x-y)
\end{equation}
and has a solution
\begin{equation}\label{dox}
\frac{2\pi}{\k^\vee} 
\exp\left(\int_{x^2}^{y^2} B_2\right)\, \delta(x^1-y^1) \delta(x^3-y^3) \theta(x^2 - y^2),
\end{equation}
which only describes propagation of charge towards the positive $x^2$ direction.
It is pretty clear that such a partial gauge fixing will lead to a version of the vertex model,
though it may be tricky to compute the precise vertex weights.  The necessary computations
are likely to be somewhat similar to those involved in studying Chern-Simons theory in
an ordinary axial gauge -- for example, see  \cite{AS,Hahn} -- or in a certain almost axial gauge
\cite{Wetering}.  Somewhat analogous is the use of  a complex version
of axial gauge to derive the Knizhnik-Zamolodchikov equations \cite{FK}.

\subsection{Up to Six Dimensions}
With an eye to future categorification, it is useful to lift this abelian picture all 
the way to six dimensions. After all, in the presence of strong symmetry breaking, 
the six-dimensional $(0,2)$ theory is not that mysterious. It is a theory of self-dual 
two-forms coupled to heavy dynamical BPS strings.
The six-dimensional setup which leads to knot invariants involves the six-dimensional 
theory on the product $\R \times M_3 \times D$, where $D$ is a  copy of $\R^2$
with the geometry of a semi-infinite cigar (see eqn. (\ref{cigarl})).
The knot itself is represented by a knotted two-dimensional defect placed at the tip of 
the cigar. The $q$-grading comes from  the conserved angular momentum derived from the rotational symmetry of  $D$. 

In the absence of
symmetry breaking, the fact that the metric on $D$ is cigar-like rather than being the Euclidean
 metric  is important in order to get a well-defined space of 
states for the defect. A defect placed in flat six-dimensional space would be strongly coupled to the
bulk SCFT. On the other hand, in the presence of symmetry breaking the IR physics is free, and
we can hope to recover the space of states from bound states of the dynamical heavy strings
and the defect. The cigar geometry would then not play a significant role, and 
$D$ can be replaced by a flat $\R^2$.

Now, we will specialize to $M_3 = \R^3$, and for brevity 
we will take the knot to be time-independent.
The BPS condition for a time-independent dynamical string is very simple: it must be straight,
and aligned with the symmetry breaking (which we will still take to be the $x^2$ direction). This is literally the six-dimensional lift of the 
condition satisfied by the smooth monopoles, and leads to the same vertex-model 
picture when projected on $x^2$. The defect strands have two possible ground states in the abelian low 
energy description, of opposite two-form charge.  A junction with a dynamical string allow 
the two-form charge to jump.

The interesting question is to reproduce the weights of the vertex model
from six-dimensional considerations. This includes both contributions from the abelian theory
of self-dual two-forms sourced by the strings, and contributions from the 
worldvolume theory of the strings. We will not compute the former here, but we can 
give some insight on the latter.

As summarized in fig. \ref{vertexweights}, the vertex model weight 
for a charge exchange process contributing to the Jones polynomial
is \begin{equation}
\label{twof}q^{\pm 1/4}-q^{\mp 3/4}. \end{equation}
  In the six-dimensional picture,
the Jones polynomial is supposed to come from a sum over BPS states, weighted by
$q^P(-1)^F$, where $P$ is the conserved charge that corresponds to rotation of $D$, while
$F$, which one might loosely call fermion number,  is a certain $R$-symmetry generator.  
We interpret the relative factor $-q^{\pm 1}$ between the two terms in (\ref{twof}) to 
mean
 that a BPS string connecting
two strands in a knot has two physical states, differing by 1 in both angular momentum (to account for the factor of $q^{\pm 1}$)
and fermion number (to account for the minus sign).
To understand this, we can focus on the
two crossing strands, and the string stretched between them. Crucially, a single 
strand is half-BPS,
two non-parallel strands are quarter-BPS, but the configuration of two strands exchanging 
a BPS string breaks symmetry further down to eighth-BPS. So the string breaks two 
supercharges, and hence it must have two ground states, exchanged by the action of the 
broken supercharges. This immediately gives the desired difference in quantum numbers.

Hence the space of approximate ground states for the system can be described as follows.
Let $\mathcal S$ be the set of configurations of the vertex model (labelings of strands
by $+$ or $-$, with smooth strings attached at crossings where a label jumps).  For every smooth string,
introduce  a two-dimensional Hilbert space with quantum numbers
derived from the last paragraph.  To each $s\in \mathcal S$, introduce a Hilbert space $\mathcal H_s$ that is defined as the 
tensor product of the two-dimensional factors associated to the smooth strings.
 Then an approximation $\mathcal H_0$ to the space of BPS
states is $\mathcal H_0=\oplus_{s\in\mathcal S}\mathcal H_s$.
The grading of $\mathcal H_0$ (by angular momentum and fermion number)  is affected by the
abelian field configurations sourced by the system of strings; for example, the self-dual 
abelian tensor fields can carry angular momentum.

In order to compute Khovanov homology, one needs to evaluate the differential
acting in this space of approximate ground states, by searching for instanton
configurations which interpolate between different states in the past and future.
The BPS condition for a time-dependent BPS string is still rather transparent.
The worldsheet should be holomorphic in complex coordinates $x^0 + i x^2$ and $x^1 + i x^3$.
Notice that a time-independent string stretched 
along the $x^2$ direction (and thus parametrized by $x^0$ and $x^2$, with $x^1$ and $x^3$
fixed) indeed has a holomorphic worldvolume.

Thus we expect to be able to build the differential for Khovanov homology from the data of
holomophic curves in $\R \times M_3$ which end on the knot. This avenue seems  promising for future development.

\section{An Effective Superpotential For Monopoles}\label{supermon}

\subsection{Overview Of Results}\label{outline}

Starting in section \ref{symbr}, we interpreted solutions of three-dimensional
supersymmetric equations in terms of configurations containing smooth BPS monopoles.
However, the considerations were purely qualitative.  In this section, we will make the
reasoning quantitative.  We will construct an effective superpotential for smooth BPS
monopoles on $\R^2\times \R_+$ interacting with a Nahm pole and singular monopoles
on the boundary.  This effective superpotential will account in a direct way for all
qualitative results from sections 2 and 3 about what solutions to our equations do or do not
exist for various values of the parameters.  Also, by integrating out some massive
fields, we will be able to recover the Yang-Yang function (\ref{yy}) that has been one of
our main tools.

Let us first consider our underlying supersymmetic equations
\begin{align}\label{bpstwo} (F-\phi\wedge \phi + \t \,\d_A\phi)^+&=0   \cr
                             (F-\phi\wedge\phi-\t^{-1}\d_A\phi  )^- &=0 \cr
                               \d_A\star\phi&=0,                     \end{align}
on $\R\times \R^3$ (where the first factor is the ``time'' direction) and ask
how the solutions of Bogomolny equations for smooth monopoles
can be embedded as solutions of these equations.   This is possible precisely
if $\t=1$ (or $-1$), the only nonzero component of $\phi$ is the time component, which here we will call  $\phi_t$ (rather than $\phi_1$, as before), and
we also set the time component of $A$ to zero.  Then the equations (\ref{bpstwo}) reduce to the Bogomolny equations
\begin{equation}\label{bog} F=\star \d_A\phi_t.\end{equation}

For the Bogomolny equations to have smooth monopole solutions, the field $\phi_t$
must have an expectation value at infinity, which means that the real symmetry breaking
parameter  $\ca_1$ of section \ref{symbr} must be nonzero.  On the other hand, the
complex symmetry breaking parameter $\ca$ of section \ref{simvar} must vanish (or
$\phi_t$ would not be the only nonzero component of $\phi$).

The basic solution of the Bogomolny equations on $\R^3$ is the one-monopole solution
for $G=SO(3)$ or $SU(2)$. It has
has magnetic charge $\m=2$ (in units in which the basic singular monopole has charge
$\m=1$).  The moduli space of the one-monopole solution is $\P=\R^3\times S^1$,
where $\R^3$ measures the center of mass position of the monopole on the spatial
manifold $\R^3$, and $S^1$ is parametrized by a collective coordinate $\vartheta$
for the $U(1)$ gauge symmetry
that is unbroken at infinity.   Of course, $\P$ is a hyper-Kahler manifold; this follows from
the unbroken supersymmetry of the Bogomolny equations on $\R^3$.  However, since
we will be considering perturbations that break some of the supersymmetry, for our purposes
it is more useful to merely look at $\P$ as a complex manifold in one of its complex
structures.  We decompose $\R^3$ as $\R^2\times \R$ (which we will eventually replace with
$\R^2\times \R_+$).   The motion of the smooth monopole along $\R^2$ is parametrized
by a chiral superfield $\sW$.  And the position $y$ of the smooth monopole in the $\R$
direction combines with the collective coordinate $\vartheta$ to a second chiral
superfield
\begin{equation}\label{work}\sY=\ca_1 y+i\vartheta.   \end{equation}

Now we want to construct an effective superpotential $\W(\sW,\sY)$ that describes this
situation.  If $\ca=\zeta=0$, there is a smooth monopole solution for every value of $\sW$ and
$\sY$, which means that every value of $\sW$ and $\sY$ is a critical point of $\W$, so
$\W$ must vanish (modulo an irrelevant constant).  On the other hand, if either
$\ca$ or $\zeta$ is nonzero, then there is no supersymmetric monopole solution, meaning
that $\W$ has no critical point.  But turning on $\ca$ and $\zeta$ preserves the symmetries
of adding a constant to $\sW$ or $\sY$.  So $\W$ must be invariant modulo an additive
constant under constant shifts of $\sW$ or $\sY$; in other words, $\W$ must be a linear
function of $\sW$ and $\sY$.  In fact, the form of $\W$ is
\begin{equation}\label{thelf}\W=\ca \sW+\zeta \sY.    \end{equation}
This follows from the following considerations.  Invariance under rotations of $\R^2$
implies that $\ca$ and $\sW$ can only appear as the product $\ca \sW$.  On the other
hand, although at $\zeta=0$, the superpotential $\W$ is given microscopically by
the single-valued function (\ref{superone}), as soon as $\zeta$ becomes nonzero
it is given by the Chern-Simons function (\ref{supertwo}), which is only single-valued
mod $2\pi i\Z$.  Since $\sY$ is similarly single-valued mod $2\pi i\Z$ (because of the
angular nature of $\vartheta$), the effective superpotential (\ref{thelf}) has a multivaluedness
that just matches that of the microscopic description.

Now let us turn off $\ca$ and $\zeta$ but replace $\R^2\times \R$ by $\R^2\times \R_+$,
where $\R_+$ is the half-line $y\geq 0$ and we assume the usual Nahm pole at $y=0$.
We assume symmetry breaking for $y\to\infty$ with $\phi_t\to {\mathrm{diag}}(\ca_1,-\ca_1)$
while of course $A$ and (at $\ca=0$) $\vec\phi$ vanish for $y\to\infty$.
In any solution, the fields approach these asymptotic values exponentially fast for $y\to\infty$.
This reflects the fact that near $y=\infty$, the gauge symmetry
is reduced from $SU(2)$ to $U(1)$ by the expectation
value of $\phi_1$, and all charged fields have masses proportional to $|\ca_1|$.
In particular, in the absence of singular or smooth monopoles, the solution for the full
system (\ref{bpstwo}) is given
by a solution  of Nahm's equations (the relevant solution is described in \cite{Krontwo})
in which the fields approach their vacuum values
exponentially fast for $y\to\infty$.

Next, still with $\ca=\zeta=0$,  let us add a smooth monopole with positions $\sW$, $\sY$.  
There is not an exact
solution for the smooth monopole, because the Nahm pole forces charged components
of $\vec\phi$ to have nonzero values that ``repel''  the monopole to $y=\infty$.  However,
these charged fields are exponentially small for large $y$, so a smooth monopole located
at large $y$ is exponentially close to being a solution.   This being so, a smooth monopole 
that is located at large $y$ must be governed by an effective
superpotential.   This superpotential  must be a single-valued function of $\sY$ that
vanishes exponentially for $y\to\infty$.  These conditions are satisfied by a linear combination
of exponentials $\exp(-n\sY)$, $n=1,2,3\dots$.  However, it will soon become clear
that the expected qualitative picture emerges if and only if at $\ca=\zeta=0$, $\W$ is linear
in $\exp(-\sY)$:
\begin{equation}\label{forg} \W=\Lambda \exp(-\sY),\end{equation}
for some constant $\Lambda$.

In principle, it should be possible to compute this result by evaluating the microscopic
superpotential (\ref{superone}) for an approximate solution consisting of a smooth
monopole at large $y$ in the presence of a Nahm pole.   The exponentially small 
term should come from $W$ boson exchange between the boundary and the monopole.
Instead of attempting
such a computation, we will take a shortcut in this paper: we will consider
a representation of the smooth monopole and the Nahm pole by a configuration of branes,
and in that context the exponential superpotential (\ref{forg}) will emerge from a simple
brane instanton.

Postponing that analysis to section \ref{realm}, let us discuss the implications of (\ref{forg}).
First we assume that $\ca=\zeta=0$, so that (\ref{forg}) is the full superpotential.   We see
at once that $\W$ has no critical point so there is no supersymmetric solution in the presence
of the smooth monopole.  This is in full accord with the analysis in section \ref{symbr}.
With real symmetry breaking only and $\zeta=0$, a supersymmetric solution is expected
to be unique (this remains true even in the presence of singular monopoles at $y=0$, which
we have not yet included in $\W$) and does not require smooth monopoles.  Now let us see
what happens when we turn on $\ca$ and $\zeta$.  We construct the full superpotential
by simply adding the various terms that we have found so far:
\begin{equation}\label{morg}\W=\ca \sW+\zeta\sY +\Lambda\exp(-\sY).\end{equation}
The justification for including $\ca$ and $\zeta$ in this way is the same as
 before ($\Lambda$ may now depend on $\zeta$).
Let us look for critical points.  We see at once that there is no critical point unless
$\ca=0$ and $\zeta\not=0$.  If these conditions are satisfied, there is a one-parameter family
of critical points, parametrized by $\sW$, with
\begin{equation}\label{norg} \exp(-\sY)=\frac{\zeta}{\Lambda}.\end{equation}
(The uniqueness of the solution for $\exp(-\sY)$ holds precisely because in (\ref{forg})
we took $\W$ to be linear in $\exp(-\sY)$.)    We found a similar result in section \ref{realsym}
where we described in eqn. (\ref{onl}) a family of solutions that exist precisely if
$\ca=0$ and $\zeta\not=0$.  The solution in question corresponded to a Miura oper
with only real symmetry breaking, a single Bethe root at an arbitrary point $w\in \R^2$,
and no singular monopoles.  We interpret the Bethe root $w$ as the value of the superfield
$\sW$ -- in other words, the position of the smooth monopole in $\R^2$ -- while the position
of the smooth monopole in the $\R_+$ direction is determined in (\ref{norg}).  The reason
that this description makes sense is that for small $\zeta$, the smooth monopole is located
at large $y$, where the effective superpotential $\W$ is valid.

The next step is to include singular monopoles.  As in section \ref{addsing}, in  the presence
of singular monopoles of charge $k_a$ located
at positions $z_a,\,a=1,\dots,d$ in the complex plane,
it is convenient to introduce the polynomial $K(z)=\prod_{a=1}^d(z-z_a)^{k_a}$.  In
section \ref{scattering}, we will argue that the only effect of the singular monopoles is to
multiply the exponential term by $K(\sW)$, so that the superpotential becomes
\begin{equation}\label{bigsuper}\W=\ca \sW+\zeta\sY +\Lambda K(\sW)\exp(-\sY).\end{equation}
Now let us examine the implications of this formula.    Suppose first that $\zeta=0$.
Then the conditions for a critical point are
\begin{equation}\label{igsuper}K(\sW)=0=\ca+\Lambda K'(\sW)\exp(-\sY).\end{equation}
The first condition says that $\sW$ must equal one of the $z_a$.
For $\ca=0$ and all $k_a=1$, the second condition then has no solutions.   If $k_a>1$ for some $a$, 
still with $\ca=0$,
the second condition is satisfied for  arbitrary $Y$.  
All this is in keeping with what we found in
sections \ref{solutions} and \ref{symbr}.  
 Now suppose that $\ca\not=0$.     If $k_a=1$, the second condition
 in (\ref{igsuper}) determines $Y$ uniquely, and if $\ca$ is small, then $Y$ is large so that
 the analysis is valid.  For $k_a>1$, the second condition in (\ref{igsuper}) cannot be satisfied.
All these statements match what was
found in section \ref{simvar} from a quite different point of view.\footnote{In section \ref{simvar},
we found, for general $k_a$ and $\ca\not=0$, a solution with no smooth monopoles at $W=z_a$ and
a solution with $k_a$ of them.  (These correspond to the two ways of solving 
$PQ=(z-z_a)^{k_a}$ such that $P$ and $Q$ have no common zero at $z=z_a$.)
Since (\ref{bigsuper}) is the superpotential for just one
smooth monopole, it describes a solution with $k_a$ smooth monopoles only
if $k_a=1$.   The general analysis for arbitrary $k_a$ can be 
made and matched to section \ref{simvar} using the superpotential 
(\ref{giantsuper}) for an arbitrary number of smooth
monopoles.}

Now let us consider the case that $\zeta\not=0$.  The most illuminating way to proceed
is to integrate out the massive field $\sY$ to generate an effective superpotential
for $\sW$.  For fixed $\sW$, the condition $\partial\W/\partial \sY=0$ has the unique
solution $\exp(-\sY)=\zeta/\Lambda K(\sW)$.  Setting $\sY$ to this value and evaluating
$\W$, we find (modulo an irrelevant constant)
\begin{equation}\label{twig} \W=\ca \sW + \zeta\log K(\sW)=\ca\sW+\zeta\sum_a k_a
\log(\sW-z_a).  \end{equation}
But this is the Yang-Yang function (\ref{yy}) for the case of a single Bethe root $\sW=w$,
modulo terms that depend only on the $z_a$ and not on $\sW$; the present derivation is
not sensitive to those terms.

At this point, the reader hopefully would like to see a similar derivation leading to the general
Yang-Yang function with any number of Bethe roots.  For a general case with $q$ smooth monopoles,
we describe the positions of the $i^{th}$ smooth monopole by chiral superfields
$\sW_i$, $\sY_i$, $i=1,\dots,q$.  The definition of these fields is somewhat subtle and is discussed in section
\ref{scattering}.  The expectation values of the $\sW_i$ will turn out to be the Bethe roots $w_i$
of section \ref{opbethe}.  As in that discussion, it is convenient to introduce the polynomial
$Q(z)=\prod_{i=1}^q(z-\sW_i)$.  It turns out that the generalization of the superpotential
(\ref{bigsuper}) to
an arbitrary number of smooth monopoles is
\begin{equation}\label{giantsuper}\W=\ca\sum_i \sW_i +\zeta\sum_i \sY_i+\Lambda
\sum_i \frac{K(W_i)}{Q'(W_i)}\exp(-Y_i).   \end{equation}

To recover the qualitative results of section \ref{teqone},
 we first set $\zeta=0$, $\ca\not=0$. To find a critical point,  the $\sW_i$ must
each equal zeroes $z_a$ of $K$.  Assuming that the charges $k_a$ are all 1, no more than one
of the $\sW_i$ may equal the same $z_a$ (otherwise a zero of $Q'(\sW_i)$ cancels a
zero of $K(\sW_i)$ and the condition $\partial\W/\partial \sY_i=0$ is not obeyed).  Summing
over all values of $q$, there are a total of $2^d$ solutions -- each $z_a$ may or may not
be equal to one of the $\sW_i$.  If we take $\ca\to 0$, then all but one of these solutions (the
one with no smooth monopoles at all) disappear, with $\sY_i\sim \log(1/\ca)$.

For $\zeta\not=0$, just as in the derivation of (\ref{twig}), it is convenient to integrate
out the massive fields $\sY_i$.  Modulo terms that do not depend on the $\sW_i$,
the superpotential that we arrive at is precisely the Yang-Yang function:
\begin{equation}\label{longyang} \W=\sum_i\ca\sW_i+\zeta\sum_i\log(K(\sW_i)/Q'(\sW_i))
=\sum_i\ca\sW_i+\zeta\sum_{i,a}\log(\sW_i-z_a)-2\zeta\sum_{i<j}\log(\sW_i-\sW_j).\end{equation}

The attentive reader may notice one gap in what we have said.  In the case $\zeta=0$,
we have not analyzed the problem with $k_a>1$ for some $a$.  To do this, it is important to
consider the case that $W_i=W_j=z_a$ for some $i,j$, but the coordinates that we have used to
describe the monopole moduli space are actually not adequate when $W_i=W_j$.  
We explain a better
description momentarily.

\subsection{Coordinates for Monopoles}\label{scattering}

The moduli space of several smooth BPS monopoles on $\R^3$ is the subject of
a rich mathematical theory \cite{AH}.  From this theory we only need a small part:
when the monopoles are widely separated in space (or equivalently when the symmetry
breaking is strong), an effective abelian description of the moduli space is possible.
In this description, the monopoles are regarded as ``point'' Dirac monopoles that interact
with each other via the abelian gauge multiplet.  Each monopole has a position in $\R^3$
and an angular coordinate $\vartheta$ that is a collective coordinate for charge rotations.
As in section \ref{outline}, once we pick a particular complex structure on the moduli space,
the position and angular coordinate of each monopole  combine to a pair of chiral superfields 
$\sW$, $\nY$.   However, there is  a subtlety in the definition of $\nY$, which is the reason
that we have changed our notation from section \ref{outline}.

 The angular part of the coordinate $\nY$ 
parametrizes the freedom to do a $U(1)$ gauge transformation on the smooth monopole 
solution before ``gluing'' it to the abelian solution. Hence $e^{\nY}$ is an holomorphic 
section of the $U(1)$ gauge bundle over the $\sW$-plane. This is the reason  that 
in the presence of boundary singular monopoles, the exponential superpotential $\exp(-\nY)$ 
for a single smooth 
monopole needs a prefactor $K(\sW)$: the superpotential should be a function, 
but the singular monopoles make $e^{-\nY}$ into the section of a bundle 
$\otimes_a {\cal O}(z_a)^{-k_a}$, where the exponents are
the charges of the singular monopoles in the abelian effective field theory. So 
we compensate for this by multiplying by $K(\sW)$. A more intuitive explanation 
is that the superpotential
encodes the interaction of the BPS monopole with the off-diagonal part of the complex Higgs field
$\varphi$, which has a zero of order $k_a$ at $z_a$.  So the superpotential acquires a factor
$(z-z_a)^{k_a}$.

On the other hand, in the abelian theory, the BPS monopole behaves like a singular monopole.  At any
given value of $y$, one can restrict the $U(1)$ gauge bundle of the low energy description to the $\sW$ plane.  As one increases $y$
so that one passes the location of a BPS monopole, 
 the $U(1)$ gauge bundle on the $\sW$ plane jumps.  
In other words, a second BPS monopole to the right of the first will feel the presence of a singular monopole of charge $+2$ at the location $\sW_1$ of the first monopole. Hence we expect a superpotential
\begin{equation}
\W=K(\sW_1)\exp(-\nY_1) + \frac{K(\sW_2)}{(\sW_2 - \sW_1)^2}\exp(-\nY_2)
\end{equation}
and similarly for several monopoles with increasing values of the real parts of $\nY_i$:
\begin{equation}
\W=\sum_i \frac{K(\sW_i)}{\prod_{j<i} (\sW_i - \sW_j)^2}\exp(-\nY_i)
\end{equation}

This superpotential is equivalent to the relevant part of (\ref{giantsuper})  as long as the $\sW_i$
are distinct.
The two are related by the change of variables
\begin{equation} \label{rela}
\exp(-\nY_i) = \frac{\prod_{j<i} (\sW_i - \sW_j)}{\prod_{j>i} (\sW_i - \sW_j)} \exp(-\sY_i)
\end{equation}
This re-definition does not affect the  part of the superpotential linear in $\ca$ and $\zeta$, since
 $\sum_i \nY_i = \sum_i \sY_i$.  However, it is the $\nY_i$, not the $\sY_i$ whose real
 parts are the actual positions of the monopoles in the $y$ direction; moreover, the
 difference between the $\nY_i$ and the $\sY_i$ is divergent when two or more $\sW_i$ coincide.
The superpotential expressed in terms of the 
$\nY_i$ reproduces correctly the counting of solutions at $\zeta =0$ for 
arbitrary values of the $k_a$. For $\ca=0$, if $K(z) = z^{k_a}$ the superpotential is
\begin{equation}
\W=\sum_{i=1}^q \frac{\sW_i^{k_a}}{\prod_{j<i} (\sW_i - \sW_j)^2}\exp(-\nY_i)
\end{equation}
This function is extremized for arbitrary $\nY_i$ if the $\sW_i$ are all zero and the number
$q$ of smooth monopoles is no greater than $k/2$,
because the prefactors have a zero of order at least $2$ when the $\sW_i$ are all zero.
This reproduces what we found in section \ref{solutions}.
If we do the same computation in terms of the $\sY_i$, we would seem to get solutions
even when the number of monopoles at the origin is greater then $k_a/2$, but the $\sY_i$
are not good variables when the $\sW_i$ coincide.    We will leave the case $\ca\not=0,\,\zeta=0$
to the reader.  For $\zeta\not=0$, the difference between the $\sY_i$ and the $\nY_i$ is 
not important.

It is interesting to match the coordinates in the low energy description to the
exact nonabelian description of the monopole moduli space.
The exact monopole moduli space is parametrized  \cite{Hurtubise} 
by a scattering matrix for the
operator $\D_y=D_y+i[\phi_t,\,\cdot\,]$. The scattering matrix takes the form
\begin{equation}
S = \begin{pmatrix} Q(z) & P(z) \cr \tilde P(z) & R(z)\end{pmatrix} \qquad QR - P \tilde P =1
\end{equation}
where $Q$ is a monic polynomial of order $q$, $P$ and $\tilde P$ are 
polynomials of degree up to $q-1$, and $R$ is a polynomial of degree up to $q-2$. 
$P$ and $Q$ are necessarily coprime (this follows from the condition $QR-P\tilde P=1$), and
both $\tilde P$ and $R$ are uniquely determined
by $Q$ and $P$.   One can define coordinates $\sW_i$, $\sY_i$ on the monopole
moduli space by 
$Q(\sW_i)=0$ and $P(\sW_i) = \exp \sY_i$. However, this definition does 
not work well if the $\sW_i$
are not distinct.

To find a parametrization that works better as long as the symmetry breaking is strong,
we can proceed as follows.
For a single monopole, the scattering matrix takes the form.
\begin{equation}
S_1 = \begin{pmatrix} (z - \sW_1) &~ e^{\nY_1} \cr -e^{- \nY_1} &0\end{pmatrix}
\end{equation}
This expression can be matched naturally to the fact that at low energies, the smooth
monopole can be approximated as a singular Dirac monopole.
It tells us that there are two solutions $\psi_\pm$ of the equation $\D_y\psi=0$ that behave as
\begin{align}
\psi_+ &\sim e^{\ca_1 y/2}\begin{pmatrix}1 \cr 0 \end{pmatrix} 
\qquad & y<<0 \cr  \psi_+ &\sim (z-\sW_1) e^{\ca_1 y/2}
\begin{pmatrix}1 \cr 0 \end{pmatrix} + e^{\nY_1 -\ca_1 y/2}
\begin{pmatrix}0 \cr 1 \end{pmatrix}\qquad &  y>>0
\end{align}
\begin{align}
\psi_- &\sim e^{-\ca_1 y/2}\begin{pmatrix}0 \cr 1 \end{pmatrix}\qquad  & y>>0
 \cr  \psi_- &\sim (z-\sW_1) e^{-\ca_1 y/2}\begin{pmatrix}0 \cr 1 \end{pmatrix} 
 + e^{-\nY_1 +\ca_1 y/2}\begin{pmatrix}1 \cr 0 \end{pmatrix} \qquad & y<<0
\end{align}
Here $\psi_+$ ($\psi_-$) is the unique solution which is small for $y<<0$ ($y>>0$).
A singular monopole solution in the abelian theory would have given the same exponential
growth for $y>>0$ ($y<<0$), and the subexponential correction is due to the exponentially decreasing  corrections for the smooth monopole solution.

For a configuration of many well-separated smooth monopoles,
the scattering matrix is a product
\begin{equation}\label{wellsep}
S = S_q S_{q-1} \cdots S_1.
\end{equation}
where $S_a$ is the scattering matrix due to the $a^{th}$ monopole and the monopoles
are taken to be ordered in the $y$ direction.
It is natural to parametrize the moduli space by 
\begin{equation}
S_a = \begin{pmatrix} (z - \sW_a) & ~e^{\nY_a} \cr -e^{- \nY_a} &0\end{pmatrix}.
\end{equation}
This enables us to write $P$ and $Q$ in terms of the $\sW_i$ and $\nY_i$, and finally,
using $P(\sW_i)=\exp(\sY_i)$, to  express the $\sY_i$ in terms of the $\sW_i$ and
the $\nY_i$.  

\subsection{Realization Via $M$-Theory And Branes}\label{realm}

Here we will explain an $M$-theory approach to understanding the exponential
superpotential (\ref{forg}).  We begin with some preliminaries.

\subsubsection{$M$-Theory Preliminaries}\label{prelims}

The six-dimensional $(0,2)$ model of type $\sf A_1$ can be realized on a pair of parallel
M5-branes.  Thus, we begin on $\R^{11}$ with coordinates $x^0,\dots,x^{10}$, and
we consider two M5-branes parametrized by $x^0,\dots,x^5$ and located at
 $x^6=\dots=x^{10}=0$.   This system preserves 16 global supersymmetries.  
 Their generators can be understood as eleven-dimensional spinors $\varepsilon$ that
 obey
 \begin{equation}\label{zog}\Gamma_0\Gamma_1\cdots\Gamma_5\varepsilon=\Gamma_6\Gamma_7\cdots
 \Gamma_{10}\varepsilon=\varepsilon. \end{equation}
 Here the gamma matrices obey $\{\Gamma_\mu,\Gamma_\nu\}=2g_{\mu\nu}$.
 
 To simplify the picture, we introduce symmetry breaking, separating the two M5-branes
 in, say, the $x^6$ direction.  So now we place one at $x^6=0$ and the other at $x^6=L$, for
 some $L$.  At low energies, this system is described by a pair of abelian tensor multiplets,
 coupled to BPS strings.  The strings arise from M2-branes stretched between the two
 M5-branes.  The string tension is $T=T_{\mathrm{M2}}L$, where $T_{\mathrm{M2}}$ is
 the M2-brane tension.
 
 We will consider a string whose world-volume is parametrized by $x^0$ and $x^1$,
 and that is located at specified values of  $x^2,\dots,x^5$.  The string is of course represented
 by an M2-brane that stretches from $x^6=0$ to $x^6=L$, though this direction will just
 factor out of the following analysis.
 
 The string described in the last paragraph preserves those supersymmetries 
 whose generator obeys
 \begin{equation}\label{zorf}\Gamma_0\Gamma_1\Gamma_6\varepsilon=
 \varepsilon\end{equation}
as well as (\ref{zog}).  Altogether, there are eight unbroken supersymmetries, corresponding
to $\N=4$ supersymmetry in the two-dimensional sense.  There is an $SO(4)$ symmetry
group rotating the $x^2,\dots, x^5$ coordinates, and a second $SO(4)$ symmetry, which
we will call $SO(4)_R$, that rotates $x^7,\dots,x^{10}$.

However, we will soon modify the construction in a way that will break half of the supersymmetry
and also reduce $SO(4)\times SO(4)_R$ to a maximal torus.  So it will help to focus
on the relevant symmetries to begin with.  We consider the
$\N=2$ subalgebra consisting of supersymmetries that (in addition to the conditions
already given) are invariant under a combined rotation of (say) the $x^4-x^5$ plane
together with an $R$-symmetry rotation of the $x^9-x^{10}$ plane:
\begin{equation}\label{dorz}
\left(\Gamma_4\Gamma_5+\Gamma_9\Gamma_{10}\right)\varepsilon=0.
\end{equation}
Given (\ref{zorf}), this is equivalent to
\begin{equation}\label{orz}\left(\Gamma_2\Gamma_3+\Gamma_7\Gamma_8\right)\varepsilon=0.
\end{equation}
We note that the equations (\ref{dorz}) and (\ref{orz}) are exchanged if we exchange
$W=x^2+ix^3$ with $Z=x^4+ix^5$, and similarly exchange $x^7,x^8$ with $x^9,x^{10}$.  
So in particular, our conditions on $\varepsilon$ are symmetrical between $W$ and $Z$.

The $\N=2$ supersymmetry algebra singled out by the above conditions has a $U(1)\times
U(1)$ group of $R$-symmetries generated by $J=\Gamma_7\Gamma_8$ and $J'=\Gamma_9
\Gamma_{10}$.    We want to compare $J$ and $J'$ to $R$-symmetry generators that we will
call $J_+$ and $J_-$ that only act, respectively, on supersymmetry generators that have
positive or negative two-dimensional chirality, in other words, that obey $\chi\varepsilon=\pm 
\varepsilon$, where $\chi=\Gamma_0\Gamma_1$ is the two-dimensional chirality.
The conditions  given above can be combined to give
\begin{equation}\label{omly}\chi\varepsilon=JJ'\varepsilon\end{equation}
and this implies that (with a suitable choice of sign for $J_+$ and $J_-$, and normalizing
them so that they square to 1 on supersymmetry generators of the appropriate chirality)
$J$ and $J'$ can be expressed as  $J=J_++J_-$, $J'=J_+-J_-$.

In particular, an exchange $J\leftrightarrow J'$ amounts to $J_\pm\to \pm J_\pm$, an operation
known as the mirror symmetry automorphism of the $\N=2$ algebra.  The automorphism
of the above-described  $\N=2$ algebra  that exchanges $W$ and $Z$ also exchanges
$J$ and $J'$, so it is a mirror symmetry.  Hence, if we view $W$ as a chiral superfield in the
two-dimensional worldsheet theory of the string,
we must view $Z$ as a twisted chiral superfield, in the sense of \cite{GHR}.

\subsubsection{Reduction To Gauge Theory}\label{gaugered}

So far we have half-BPS strings, but no gauge theory description of them.

To get a gauge theory description, we compactify one direction, say the $x^5$
direction, on a circle of radius $R$. 
$M$-theory on a circle reduces at long distances
to Type IIA superstring theory.  The M5-branes become D4-branes and the theory
on the D4-branes is at long distances a $U(2)$ gauge
theory, broken to $U(1)\times U(1)$ by the separation between the D4-branes.  What
is relevant to us  is the $SU(2)$ subgroup, broken at low energies to $U(1)$.  
The string that was originally described via a stretched M2-brane
is now represented by a D2-brane stretched between the two D4-branes.

This situation has been studied in \cite{diac}.  The D2-brane stretched between two
D4-branes carries magnetic charge and corresponds to a smooth BPS monopole in the low
energy $SU(2)$ gauge theory.  

Before compactifying the $x^5$ direction, the low energy theory along the string was
described by the chiral superfield $W=x^2+ix^3$ and the twisted chiral superfield
$Z=x^4+ix^5$.  After the compactification, we can replace $Z$ by the single-valued
field $\exp(Z/R)$, which is still a twisted chiral superfield.  
However, for matching to the theory of BPS monopoles, another variable 
is more useful.   The moduli of the BPS
monopole corresponding to the D2-brane are the positions of the underlying M2-brane
in $x^2,x^3,$ and $x^4$, and the {\it dual} of its position in $x^5$.  The duality in question
is a $T$-duality in the two-dimensional effective field theory governing the string.  
This $T$-duality is a mirror symmetry in the two-dimensional sense. 
It replaces $x^5/R$ with a new angular coordinate $\vartheta$.  
 So while
$Z/R=x^4/R+ix^5/R$ is a twisted chiral superfield, $Y=x^4/R+i\vartheta$ is an ordinary
chiral superfield, just like $W$.  Of course, the single-valued chiral superfield is not
$Y$ but $e^Y$.

We conclude with two comments:
\begin{itemize}
\item The fact that the angular coordinate $\vartheta$ of the BPS monopole is $T$-dual
to the angular position $x^5$ is part of the relation between M5-branes on a circle and
D4-brane gauge theory.  The symmetry that rotates $x^5$ becomes instanton number in
the $4+1$-dimensional gauge theory, while the symmetry that rotates $\vartheta$ is electric
charge.
\item The single-valued field $\Omega=\exp(Z/R)$ can be understood as  a map to $\C^*$; it can be neither
zero nor infinity.  In section \ref{mod}, we modify the problem to make it possible to have
$\Omega=0$.
\end{itemize}

\subsubsection{Reducing On A Half Space}\label{mod}
To get a non-trivial superpotential, we will have to break some of the translation
symmetries of the problem.  In fact, we are interested in gauge theory on a half-space,
so we want to restrict $y=x^4$ to be non-negative.  

The gauge theory problem studied in the present  paper
arises if $x^4$ and $x^5$ parametrize not $\R\times S^1$, as is the case in our presentation
so far, but rather a copy of $\R^2$ with a cigar-like metric
\begin{equation}\label{cigarl}\d s^2=\d y^2+f(y)\left(\d x^5\right)^2. \end{equation}
Here $f(y)\sim y^2/R^2$ for $y\to 0$ and $f(y)\to 1$ for $y\to\infty$.  In fact, this
was the starting point in the derivation in \cite{fiveknots}.

As a complex manifold, $\R^2$ is the same as $\C$, so we can now parametrize
the $x^4$ and $x^5$ directions by a $\C$-valued chiral superfield $\Omega$, which asymptotically
at large $y$ (but only there) can be written 
\begin{equation}\label{exo}\Omega=\exp(Z/R), ~~Z=x^4+ix^5.\end{equation}
In contrast to the concluding remark of section \ref{gaugered}, $\Omega$ is now $\C$-valued
rather than $\C^*$-valued, and in particular there is no problem in having $\Omega=0$.  

We will make use of this shortly in generating a superpotential.

\subsubsection{The Instanton}\label{thinst}

We now want to describe an M2-brane instanton that will generate the superpotential
that we are looking for.

The instanton is supposed to correct the physics of a string that is parametrized
by a worldsheet coordinate $X=x^0+ix^1$.  The string is located at definite values of $W$ and $\Omega$, say $W=W_0$,
$\Omega=\Omega_0$.

We can understand qualitatively what sort of instanton can generate a superpotential.
We consider a two-dimensional model with $\N=2$ supersymmetry  whose 
chiral ring is generated at the classical level 
by the chiral superfield $W$ and  whose twisted chiral ring 
is generated by the
twisted chiral superfield
$\Omega$.  The chiral ring is the ring of observables of a twisted $B$-model, and
the superpotential that we want to generate will give a deformation of this chiral ring.
A hypothetical superpotential  must be generated by configurations that preserve the $B$-model
supersymmetry.  These are configuations in which
the chiral superfield $W$ is constant, while the twisted chiral superfield $\Omega$ is holomorphic 
as a function of the worldsheet coordinate.
(It is a familiar fact that $B$-model supersymmetry requires a chiral superfield to be constant.
The fact that $B$-model supersymmetry allows a twisted chiral superfield to be holomorphic
is mirror to the perhaps more familiar fact that $A$-model supersymmetry allows a chiral
superfield to be holomorphic.)

The instanton that generates the superpotential is accordingly given by $W=W_0$
while $\Omega$ is a nontrivial but simple holomorphic function of  $X$:
\begin{equation}\label{yelf}\frac{\Omega}{\Omega_0}=X-X_0,\end{equation}
(Note that this formula only makes sense because $\Omega$ is allowed to vanish.)
Here $X_0$ is a constant, which we interpret as the instanton position;
as always in instanton physics, to calculate physical amplitudes, 
it is necessary to integrate over the instanton moduli, which here mean $X_0$ as well
as some fermionic moduli associated to the supercharges under which the instanton
solution is not invariant.

The worldvolume of an M2-brane instanton is supposed to be a three-manifold.  
The three-manifold we want is just the product of the two-manifold $S$ that was defined in
(\ref{yelf}) with the one-manifold $0\leq x^6\leq L$ (all at $x^7=\dots=x^{10}=0$).  

Since it is invariant under $B$-model supersymmetry, and has no moduli except what
follows from translation invariance and supersymmetry (the parameter $\Omega_0$ corresponds
roughly to a constant value that $\Omega$ would have in the absence of the instanton), 
this sort of instanton will
generate a superpotential.  To understand just what superpotential will be generated,
we use the asymptotic formula (\ref{exo}) and look at the disturbance in the string
that is generated by the instanton, at great distances.  For large values of $y=x^4$,
we can write
\begin{equation}\label{omo}\exp((y+ix^5)/R)=\Omega_0(X-X_0).\end{equation}
We see that as $X$ circles once around $X_0$ in the clockwise direction (at large
values of $|X-X_0|$ so that the formula (\ref{omo}) is valid), $x^5$ increases by $2\pi R$.
To produce this effect, the operator inserted at $X=X_0$ must be a twist field.  
As the $T$-dual of $x^5$ is the angular variable $\vartheta$,  a twist field is 
$\exp(-i\vartheta)$, and this must be the $\vartheta$-dependence of a superpotential
that captures the effects of the instanton.  The holomorphic expression must therefore
be $\W=\Lambda \exp(-Y)$, where $Y=y+i\vartheta$, for some constant $\Lambda$.
This is precisely the result claimed in (\ref{forg}).

For further confirmation, and also to check the sign in the exponent of $\W$, let us
consider the behavior of the field $y$ at large distances, far from $X=X_0$.
At long distances, the fluctuations in $y$ are described by a free-field path integral
\begin{equation}\label{donel}\int Dy\,\exp\left(-\frac{1}{4\pi R^2}\int \d x^0\,\d x^1\,|\nabla y|^2\right).
\end{equation}
When the operator $\exp(-y/R)$ is inserted in such a path integral at a point $X=X_0$, the result 
is that at large distances, $y/R$ grows as $|\log(X-X_0)|$.  But this is exactly what we see
in (\ref{omo}).  

A final comment is that if the  worldvolume dimension of the string were bigger than 2, we would have considered the instanton
as a fluctuation around a vacuum defined by a limiting value of $y$ (and all the other
worldvolume fields) for $X\to\infty$, and we would have asked for the instanton to approach
this limiting value at infinity.  In two dimensions, because of the usual infrared divergences
-- which appear, for instance, in the logarithmic growth mentioned in the last paragraph -- such
a formulation is not valid.

\section{Opers And Branes}\label{opbranes}

The purpose of the present section is to place some of the ingredients that have appeared
in the present paper in a wider context.

We  continue to study $\N=4$ super Yang-Mills theory, with a twist
that preserves half the supersymmetry, on the four-manifold $M_4=\R\times C\times \R_+$, and 
with  the usual Nahm pole boundary condition at the finite end of $\R_+$.  The
novelty, compared to what has been said so far, is that we will view the problem from
the point of view of compactification on $C$ from four to two dimensions.  
In general
\cite{BVJ,HM}, assuming for simplicity that $C$ has genus at least 2 (we relax this 
condition in section \ref{mondr}), 
compactification on $C$ gives at low energies  
a two-dimensional sigma-model in which the target is $\M_H$, the moduli space of
solutions of Hitchin's equations \cite{Hitchin}.   
The Nahm pole boundary condition must reduce at low energies to a brane in this
sigma-model, and this brane must be half-BPS because the Nahm pole boundary
condition is half-BPS in four dimensions.

\subsection{Back to $\t=1$}\label{tone}

We begin by analyzing the case $\t=1$.
 For simplicity, we take $G$ to be $SU(2)$ or 
$SO(3)$.    In section \ref{bc}, we found that at $\t=1$, the Nahm pole boundary condition
(in the absence of singular monopoles)  describes  a Higgs bundle $(E,\varphi)\to C$ 
that is endowed with 
a line sub-bundle $\L\subset E$ that is nowhere $\varphi$-invariant.  Viewing $E$ as
a rank 2 complex bundle of trivial determinant, the inclusion $\L\subset E$ is part of
an exact sequence:
\begin{equation}\label{exacto} 0 \to \L\to E\to \L^{-1}\to 0.\end{equation}
Here we use the fact that, as $E$ has trivial determinant, the quotient $E/\L$ must be 
isomorphic to $\L^{-1}$.

We view $\varphi$ as a holomorphic map $E\to E\otimes K$, where $K$ is the canonical
bundle of $C$.  We can restrict
$\varphi$ to $\L$, to get a holomorphic map $\L\to E\otimes K$, and then using
the projection $E\to \L^{-1}$, we get a map $\varphi:\L\to \L^{-1}\otimes K$.  The condition
that  $\L$ is nowhere $\varphi$-invariant means precisely that the map
$\varphi:\L\to \L^{-1}\otimes K$ is everywhere nonzero.  In other words, this
map is an isomorphism.

Tensoring with $\L$, we learn that $\L^2$ is isomorphic to $K$, so that $\L$ is
a square root $K^{1/2}$ of $K$.  If $G=SU(2)$, a solution of the Nahm pole
boundary condition involves a choice of $K^{1/2}$, while if $G=SO(3)$, since
we really should be working with the adjoint bundle $\mathrm{ad}(E)$ rather than $E$,
the choice of $K^{1/2}$ does not matter.  In what follows, we assume that either
$G=SO(3)$ or we have picked a particular square root of $K$.

It is possible to make a non-trivial extension $0\to K^{1/2}\to E\to K^{-1/2}\to 0$,
and we will exploit this fact in section \ref{general}.  However, for Higgs bundles, we 
want $E$ to be a direct sum $K^{1/2}\oplus K^{-1/2}$, since in the case of a non-trivial
extension, the Higgs fields that we are about to write would not exist.  If we write
$E$ in column form
\begin{equation}\label{equad} E=\begin{pmatrix}K^{-1/2}\cr K^{1/2}\end{pmatrix}.\end{equation}
then up to an automorphism of $E$, a possible Higgs field
$\varphi$ takes the form
\begin{equation}\label{varph}\varphi=\begin{pmatrix}0 & 1\cr q & 0 \end{pmatrix},\end{equation}
where $q$ is a quadratic differential.
To be more exact, we assume the upper right matrix element of $\varphi$ to be nonzero
as otherwise $\L$ would be $\varphi$-invariant (and the Higgs bundle $(E,\varphi)$ would
be unstable, as explained in \cite{Hitchin}).  Given this, by a bundle automorphism
$\mathrm{diag}(\lambda,\lambda^{-1})$, we can take the upper right matrix element to be 1,
and by a lower triangular bundle automorphism, we can make the diagonal matrix elements of
$\varphi$ vanish.  Finally, for $E$ as in (\ref{equad}), the lower left matrix element  of $\varphi$
is a quadratic differential (an element of $H^0(C,K^2)$), which we call $q$.

Let $\T\subset \M_H$ be the submanifold parametrizing the Higgs bundles $(E,\varphi)$
described in the last paragraph.  At $\t=1$, the brane in $\M_H$  
defined by the Nahm pole is supported
on $\T$.  What sort of subvariety is $\T$?    As in \cite{Hitchin}, let 
$I$ be the complex structure on $\M_H$
in which it parametrizes Higgs bundles, $J$ the complex structure in which $\M_H$ parametrizes
flat bundles with connection $\A=A+i\phi$, and $K=IJ$.  The Hitchin fibration is the map from $\M_H$ to the space of
quadratic differentials that maps $(E,\varphi)$ to $\mathrm{Tr}\,\varphi^2$. This map
is holomorphic in complex structure $I$.  For the Higgs field in (\ref{equad}), we have
$\Tr\,\varphi^2=2q$, so there is a unique such $\varphi$ for every desired value of
$\Tr\,\varphi^2$.  Accordingly, $\T$ is a holomorphic section of the Hitchin fibration;
in fact it is the holomorphic section constructed in \cite{Hitchin}.  Actually, $\T$ is complex
Lagrangian from the point of view of complex structure $I$.  That assertion
means that the complex symplectic form
\begin{equation}\label{Om}\Omega_I=\frac{1}{4\pi}\int_C\d\bar z\,\d z\,\Tr\,\delta A_{\bar z}\delta\phi_z
\end{equation}
vanishes when restricted to $\T$.  This is the case since, as the holomorphic type of $E$
is fixed for all Higgs bundles that represent points in $\T$,  $\delta A_{\bar z}$ is zero (up to a gauge transformation) when restricted to $\T$. 

Since $\T$ 
is complex Lagrangian in complex structure $I$, we can identify as follows the supersymmetry
of the half-BPS brane produced by the Nahm pole.  This is a brane of type $(B,A,A)$,
that is, it is a $B$-brane in complex structure $I$, but an $A$-brane from the point of view of
$J$ or $K$.

\subsection{General $\t$}\label{general}

At general $\t$, we are dealing with a flat bundle rather than a Higgs bundle.
The Nahm pole still gives a line sub-bundle $L\subset E$, so we still have an exact sequence  
\begin{equation}\label{texaco} 0\to \\L\to E\to \L^{-1}\to 0,\end{equation}
 as in (\ref{exacto}).   The covariant derivative 
$\D/\D z$ now gives a holomorphic map $E\to E\otimes K$.  We can still restrict
this map to $\L$ and project the image to $\L^{-1}\otimes K$, to get a linear map
$\D/\D z:\L\to \L^{-1}\otimes K$.  The condition that $\L$ is nowhere invariant under
$\D/\D z$ implies, just as in section \ref{tone}, that this map is an isomorphism from $\L$
to $\L^{-1}\otimes K$, and again we conclude that $\L=K^{1/2}$.

The difference from section \ref{tone} is that now the bundle $E$ is not a direct
sum $K^{1/2}\oplus K^{-1/2}$ but a non-trivial extension.  Indeed, as we assume that the genus of 
$C$ exceeds 1,  a bundle
that holomorphically is a direct sum  $K^{1/2}\oplus K^{-1/2}$ would not admit a flat connection.

Non-trivial extensions of $K^{-1/2}$ by $K^{1/2}$ are all isomorphic; this is so because such
an extension is determined by an element of $H^1(C,K)\cong \C$, and the choice of a nonzero
element does not matter, up to a bundle automorphism.

The simplest example of a flat bundle that from a holomorphic point of view is the extension
described in the last paragraph can be found by placing on $C$ a Kahler metric of 
scalar curvature
$R=-1$.  Let $\omega$ be the spin connection of such a metric and $e$ the vierbein.
The flat connection is
\begin{equation}\label{omurf}\A=\omega\ct_3+e_{\bar z}\ct_- +e_z\ct_+.\end{equation}
In differential geometry, since $\omega$ is the spin connection, the flat bundle  $E$ is the
spin bundle of $C$, or more precisely the direct sum $K^{1/2} \oplus K^{-1/2}$ of the
two spin bundles of opposite chirality.  But in this basis, the complex structure of
$E$ is defined by the $(0,1)$ part of $\A$, which is $\A_{\bar z}=\omega_{\bar z}\ct_3
+e_{\bar z}\ct_-$; this is lower triangular, but not diagonal, so $E$ is an extension rather
than a direct sum.    

Having found a single flat connection $\A$ on the bundle $E$, it is straightforward
to find them all.  We do not want to change $\A_{\bar z}$ (since the holomorphic structure of
$E$ is supposed to be unchanged), but we can change $\A_z$ by $\A_z\to \A_z+\lambda_z$,
where (to preserve flatness) $\lambda_z$ is annihilated by $\D_{\bar z}$.  For $E$ as described in the last
paragraph, the relevant choice is $\lambda_z=q\ct_-$, where $q$ is a quadratic differential.

Mathematically, a flat bundle $E\to C$ that from a holomorphic point of view fits in a non-split
exact sequence (\ref{texaco}) is called an oper; see \cite{Frenkel} for a detailed explanation.
We have learned that, at general $\t$, the brane defined by the Nahm pole boundary
condition is supported on the variety  of opers.   Actually, we should be
more precise, because the Nahm pole boundary
condition depends in general on a parameter $\zeta$ that was introduced in section
\ref{motivation}, and the complex connection $\A$ that obeys the oper condition is in general
not $A+i\phi$ but the more general connection $\A^\zeta$ defined in eqn. (\ref{generic}).
We write $\V_\zeta$ for the subvariety of $\M_H$ defined by requiring that $\A^\zeta$
obeys the oper condition.

 When restricted to $\V_\zeta$, $\A^\zeta_{\bar z}$ is fixed, up to a gauge
transformation, so the complex symplectic form
\begin{equation}\label{tufo}\Omega_{I_\zeta}=\frac{1}{4\pi}\int_C\d\bar z\,\d z\,\Tr\,\delta\A^\zeta_{\bar z}\delta\A_z
^\zeta
\end{equation}
vanishes.  Accordingly, the brane $\V_\zeta$ is a complex Lagrangian brane, just as in section
\ref{tone}, but now in a rotated complex structure.  In the context of the present paper,
the rotated complex structure is $I_\zeta$, defined in section \ref{motivation}.   $\V_\zeta$
might be called a brane of type $(B,A,A)_\zeta$, being related to complex structure $I_\zeta$
as a $(B,A,A)$ brane is to complex structure $I$.

As long as
$\zeta\not=0,\infty$, the complex structures $I_\zeta$ are all equivalent.    
If we simply set $\zeta=i$, we get the usual variety of opers for complex structure $J$;
alternatively, in the limit $\zeta\to 0$, $\V_\zeta$ reduces to the holomorphic section 
$\V$ of the Hitchin fibration, described in section \ref{tone}. 

\subsection{$S$-Duality}\label{sduality}

A particularly simple boundary condition in $\N=4$ super Yang-Mills theory is the Neumann
boundary condition for gauge fields, 
extended to the whole supermultiplet in a half-BPS fashion.  In terms
of branes, this is the boundary condition for a family of D3-branes ending on a single NS5-brane
in the absence of a gauge theory $\theta$-angle.

Upon compactification on $C$ and reduction to two dimensions, this boundary condition gives a brane $\B_{\NS5}$ on $\M_H$ 
corresponding to a trivial
flat line bundle over $\M_H$.  In other words, the support of the brane $\B_{\NS5}$ is all of
$\M_H$, and its Chan-Paton connection is trivial.  The brane $\B_{\NS5}$ is of type $(B,B,B)$,
meaning that it is a $B$-brane in every complex structure.  This reflects the fact that the
trivial bundle over $\M_H$ is holomorphic in every complex structure.

Under  $S$-duality or electric-magnetic duality, 
the D3-NS5 boundary condition is converted to a D3-D5
boundary condition, still with $\theta=0$.  On the other hand, $S$-duality acts in the
dimensionally reduced theory as $T$-duality on the fibers of the Hitchin fibration \cite{BVJ,HM}.
Hence the brane $\B_{\NS5}$ must be mapped by $S$-duality to a brane ${\B}_{\uD5}$ supported
on a section of the Hitchin fibration.  Moreover, the 
$S$-dual of a brane of type $(B,B,B)$ is  of type $(B,A,A)$ 
(this is shown in \cite{KW}). For a middle-dimensional brane to be an $A$-brane, its support must be a Lagrangian
submanifold, and its Chan-Paton bundle must be flat.  
So the section of the Hitchin fibration on which $\B_{\uD5}$
is supported must be complex Lagrangian from the point of view of complex structure $I$.
On the other hand, concretely the D3-D5 boundary condition, for the case of a single D5-brane,
is described by the Nahm pole \cite{gw} at $\t=1$. (This value of $\t$ corresponds to
unbroken supersymmetry of type $(B,A,A)$.)  Our analysis above determines the section of the
Hitchin fibration that corresponds to the Nahm pole; it corresponds to the family of Higgs
bundles described in (\ref{equad}) and (\ref{varph}).

The D3-NS5 boundary condition can be deformed by turning on $\theta$ and more
generally by turning on a $U(1)$ gauge field on the NS5-brane.  These deformations, which are
described in \cite{gw}, preserve the half-BPS nature of the boundary condition but 
rotate the unbroken supersymmetry.  A particular deformation in this family, described in 
section 12 of \cite{KW}, gives a brane -- the canonical coisotropic brane $\Bcc$ -- that is 
important in the gauge theory approach to the geometric Langlands correspondence.  
$\Bcc$ is  a rank one brane supported on all of $\M_H$, with a Chan-Paton bundle whose
curvature is a linear combination of the Kahler forms of $\M_H$.  The precise combination
depends on a parameter analogous to our $\zeta$.  
With a choice that is convenient for geometric Langlands, the curvature of the Chan-Paton
bundle is a multiple of $\omega_J$ (the Kahler form for complex structure $J$) and
then $\Bcc$ is of type $(A,B,A)$. 

 The $S$-dual of the deformation of $\B_\NS5$ that gives
$\Bcc$ is a deformation of $\B_\uD5$ that is obtained by rotating the Nahm pole in the
space of fields $\vec A$ and $\vec \phi$.  This
is analyzed in \cite{gw}, and the appropriate type of ``rotation'' was briefly described in
section \ref{motivation}.    As we have seen, the rotated Nahm pole boundary condition 
leads to a brane
$\Boper$ that is supported on the variety of opers, that is on $\V_\zeta$ for some $\zeta$.
If $\Bcc$ is defined in the standard fashion as a brane of type $(A,B,A)$, then its $S$-dual
must have the same supersymmetry.  In that case, $\Boper$ is supported on the ordinary 
variety of opers, with $\zeta=i$.  (In the present paper, it is more natural for $\zeta$ to be real.)

That the $S$-dual of the brane $\Bcc$ is the brane $\Boper$ supported on the
variety of opers is important in mathematical treatments of the geometric Langlands
correspondence.  See for example \cite{Frenkeltwo} for an explanation of the role of opers
in the geometric Langlands correspondence.  
The facts that we have just described give a gauge theory way to
understand the $S$-duality between $\Bcc$ and $\Boper$.  It has  been argued \cite{NW} that the $S$-duality between
these two branes is important in understanding  the AGT correspondence \cite{AGT} as well as recent developments relating
supersymmetric gauge theory and integrable systems \cite{NS}.  The role of the $S$-duality between $\Bcc$ and $\Boper$ is
more explicit in  \cite{NStwo}.

\subsection{Monodromy Defects}\label{mondr}

The concept of a Higgs bundle can be generalized by allowing singularities at isolated
points $p_i\in C$.  For what follows, the case of interest will be a regular singularity.
To introduce a regular singularity near a point $p$ in $C$, we pick a local complex coordinate
$z$ near $C$ and then we introduce polar coordinates $r,\theta$ with $z=re^{i\theta}$.  
We select elements $\alpha,\beta,\gamma$ in the Lie algebra $\frak t$ of a maximal torus
$T\subset G$, and look for solutions of Hitchin's equations with a singularity at $p$ of the form
\begin{align}\label{justo}A&=\alpha\,\d\theta+\dots \cr 
\phi&=\beta\frac{\d r}{r}-\gamma\,\d\theta+\dots \end{align}
where the ellipses refer to additional terms that are less singular than $1/r$.  
We call this sort of codimension two singularity a monodromy defect.  
The general theory of Hitchin's equations adapts well to this situation \cite{Simpson}
and one can define a moduli space $\M_H(C;p,\alpha,\beta,\gamma)$ of solutions that
is a hyper-Kahler manifold with properties rather similar to what one has in the absence of
the monodromy defect.  Everything we will say generalizes in an obvious way to the case
of any number of monodromy defects.

Once we introduce monodromy defects, the limitation of some of the above statements
to the case that the genus of $C$ is at least 2 can be dropped.  All above statements hold
for $C$ of any genus in the presence of a sufficient number of monodromy defects
(for $G=SU(2)$, the required number is 3 if $C$ has genus 0, and is 1 if $C$ has genus 1).

In the context of $\N=4$ super Yang-Mills on $\Sigma\times C$, where $\Sigma$ is another
two-manifold, one can consider a monodromy defect supported on $\Sigma\times p$,
with $p\in C$.  The singular solution (\ref{justo}) of 
Hitchin's equations embeds naturally as a solution of the four-dimensional equations
(\ref{bpseqns}).  In the limit that $C$ is small compared to $\Sigma$, the $\N=4$ theory
on $\Sigma\times C$ reduces to a sigma-model on $\Sigma$ with target $\M_H(C;p,\alpha,\beta,
\gamma)$.  In this description, there is  an additional parameter $\eta$ that arises
\cite{GuW} as a theta-angle for the abelian subgroup of $G$
that is unbroken along $\Sigma\times p$.  So quantum mechanically, a monodromy defect
really has four parameters $\alpha,\beta,\gamma,
\eta$.   Under $S$-duality, a monodromy defect of the
above-described type in $G$ gauge theory is mapped to a similar monodromy defect
in $G^\vee$ gauge theory.  The transformation of the parameters under $S$-duality
is $(\alpha,\eta)\to (\eta,-\alpha)$ while $\beta$ and $\gamma$ are rescaled (for more
detail see section 2.4 of \cite{GuW}).  

If we drop the subleading terms represented by the ellipses in 
(\ref{justo}), we find that the monodromy of the complex flat
connection $\A=A+i\phi$ is $U=\exp(-2\pi(\alpha-i\gamma))$.  The subleading terms
do not modify the monodromy as long as $U$ is regular -- meaning that the subgroup of
$G$ that commutes with $U$ has dimension equal to $r$, the rank of $G$.  If $U$ is not regular,
there is an important subtlety, explained in detail in \cite{GuW}, section 3.3.  For brevity,
we will here consider only the case that $G=SU(2)$, so that the nonregular values of $U$
are only $\pm 1$.  If $U=\pm 1$, the monodromy $V$ of a connection of the form (\ref{justo})
is not necessarily conjugate to $U$; on the contrary, generically it is in the ``unipotent'' conjugacy
class containing the element
\begin{equation}\label{zonk}U'=\pm \begin{pmatrix}1 & 1 \cr 0 & 1 \end{pmatrix}.\end{equation}
A general element $V$ of this conjugacy class is $\pm 1$ plus an arbitrary 
 nilpotent matrix:
\begin{equation}\label{onk}V=\pm 1 + \begin{pmatrix} x & y \cr z & -x \end{pmatrix},~~x^2+yz=0.
\end{equation}
The equation $x^2+yz=0$ describes an $\sf A_1$ singularity $\C^2/\Z_2$.  The singular point
is located at $x=y=z=0$ where the monodromy  $V$ is precisely $\pm 1$; in other words,
this is the case that the subleading terms in (\ref{justo}) do not correct the monodromy.   In
setting $\alpha$ and $\gamma$ to special values at which $U=\pm 1$, we will assume that
$\beta$ remains generic.  In this case, even if $U=\pm 1$, the solution has a ``symmetry breaking
direction'' built in, given by the singular term in the connection
proportional to $\beta$.  The effect of this is to
blow up the $\sf A_1$ singularity, replacing $\C^2/\Z_2$ with $T^*\CP^1$.  This important
fact is established in \cite{Simpson}.

The precise meaning of this $T^*\CP^1$ is that if $U=\pm 1$ and $\beta=0$, 
then $\M_H(C;\alpha,\beta,\gamma)$ has a locus of $\sf A_1 $ singularities, which 
parametrizes Higgs bundles
for which the monodromy around $p$ is precisely $\pm 1$.  But
if $\beta\not=0$  with $U$ still equal to $\pm 1$, then this singular locus is blown up, replacing
the singularities by a family of  $\CP^1$'s.  

The moduli space $\M_H(C;p,\alpha,\beta,\gamma)$ is invariant under shifting
$\alpha$ by a cocharacter -- for $G=SU(2)$, this means that it is invariant under 
$\alpha\to \alpha+\mathrm{diag}(i,-i)$, 
which is a shift that can be induced by a gauge transformation that has a singularity at $p$.
However, we will be interested in brane constructions that are not invariant under such
shifts of $\alpha$,  and for this reason, it will be best for our purposes 
not to view $\alpha$ as a periodic
variable.

\begin{figure}
 \begin{center}
   \includegraphics[width=3.5in]{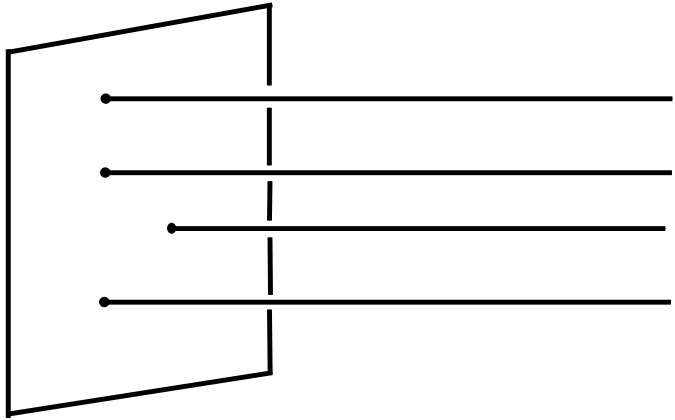}
 \end{center}
\caption{\small  Monodromy defects in $C\times \R_+$. supported on $p_i\times \R_+$ where
the $p_i$ are points in $C$.  $C$ is represented by the rectangle.  We assume a Nahm
pole boundary condition at $y=0$.}
 \label{defects}
\end{figure}
The first brane that we want to consider is the oper brane.  It is defined as usual by the
Nahm pole boundary condition.  For this, we take $\Sigma$ to be $\R\times \R_+$, where
$\R_+$ is the usual half-line $y\geq 0$,
and we impose the Nahm pole boundary condition at $y=0$.  Suppressing the $\R$ or time
direction, the picture on $C\times \R_+$ is sketched in fig. \ref{defects}: there are 
monodromy defects supported on $p_i\times \R_+$,  with respective  parameters
$(\alpha_i,\beta_i,\gamma_i,\eta_i)$, and a Nahm pole boundary condition
at $y=0$.  Of course, we need  to explain what
sort of singularity we want where the monodromy defect ends on a boundary with the Nahm pole.
As usual, this kind of question is answered by finding a model solution with the desired singularity.
For the present case, 
this has been done in section 3.6 of \cite{fiveknots}, and in greater generality in 
\cite{Mikhaylov}.

Let us set $\alpha-i\gamma=\lambda\,\mathrm{diag}(i,-i)$, for a complex parameter $\lambda$, and consider a Higgs bundle
$E$ with a singularity of this type at, say, $z=0$.  If $E$ (viewed in complex structure $I_\zeta$)
 is also an oper, then it
can be described by the classical stress tensor
\begin{equation}\label{lork}t=-\frac{\lambda(\lambda+1)}{z^2}+\dots, \end{equation}
where we have  omitted less singular terms.  This
formula is 
just like (\ref{roog}), with $j_a=k_a/2$ replaced by $\lambda$.  Flat sections of $E$ correspond
to holomorphic solutions of the differential equation
\begin{equation}\label{stork}\left(\frac{\partial^2}{\partial z^2}+t\right)\psi=0.\end{equation}
For generic $\lambda$, one 
can find two linearly independent solutions with $\psi_1=z^{-\lambda}
\left(1+\sum_{i=1}^\infty
c_iz^i\right)$, $\psi_2=z^{\lambda+1}\left(1+\sum_{i=1}^\infty \tilde c_i z^i\right)$.    
This means that, as expected, the monodromy is
\begin{equation}\label{dusc}U=\mathrm{diag}(\exp(-2\pi i\lambda),\exp(2\pi i\lambda)).\end{equation}

What happens if instead $\lambda=k/2$ with $k\in \Z$?  There is always a solution
$\psi_2=z^{k/2+1}(1+\sum_{i=1}^\infty \tilde c_iz^i)$, but if we look for a solution with
$\psi_1=z^{-k/2}(1+ c_1z+\dots)$, we find that generically when we carry this
expansion to order $z^{k/2+1}$, we need logarithmic terms of order
$ z^{k/2+1}\log z+\dots$. The logarithmic terms are simply a multiple of $(\log z)\psi_2$.
Accordingly, the monodromy around
$z=0$ is actually generically of the unipotent form
\begin{equation}\label{omogg}
\begin{pmatrix} \psi_1\cr \psi_2\end{pmatrix}\to (-1)^k\begin{pmatrix} 1 &  s\cr
    0 & 1\end{pmatrix}\begin{pmatrix}\psi_1 \cr \psi_2\end{pmatrix}\end{equation}
    for some complex constant $s$.

So if we want the monodromy around $z=0$ to be trivial, we need to impose
one condition on the subleading coefficients in the stress tensor
(\ref{lork}), so as to get $s=0$.  This means that having trivial monodromy around $z=0$ is a middle-dimensional
condition.
 Indeed,  without this condition, a monodromy defect for $G=SU(2)$ increases the complex
dimension of $\M_H$ by 2, but the trivial monodromy condition fixes 1 of the 2 parameters.

As this point is important, we will dwell on it a bit. Generically,
 the monodromy
around the defect is an element of $SL(2,\C)$ (complex dimension 3) that obeys
1 constraint specifying its conjugacy class, leaving 2 complex parameters.  For example,
when $\alpha=\gamma=0$, the conjugacy class is two-dimensional, as exhibited explicitly
in (\ref{onk}).  The condition of trivial monodromy (which is defined only when $U=\exp(-2\pi(\alpha-i\gamma))$ equals $\pm 1$, and has no analog for other values) fixes 1 of the 2 parameters
 associated to the defect, so it leaves 1 parameter.  One can think of this 1 parameter
 as the direction of symmetry breaking associated to the term $\beta \,\d r/r$ in eqn. (\ref{justo}).
 The choice of a symmetry-breaking direction determines a point in a copy of $\CP^1$;
 this $\CP^1$ is the projectivization of the fiber of $E$ at the point $p\in C$ where the
 monodromy defect lives.  A more detailed explanation of the origin of this $\CP^1$ is as follows.
 First of all, because of the equation $x^2+yz=0$, the unipotent conjugacy class described
 in eqn. (\ref{onk}) is explicitly isomorphic as a complex manifold
 to $\C^2/\Z_2$, with an $\sf A_1$ singularity at $x=y=z=0$.
The singularity is precisely the point at which the group element $V$ in (\ref{onk}) equals
 $\pm 1$.  In the context of the
 construction of $\M_H$ as a hyper-Kahler manifold, the $\beta$ parameter is a Kahler
 parameter that blows up the $\sf A_1$ singularity, replacing the conjugacy class $\C^2/\Z_2$
 by its resolution, the Eguchi-Hansen manifold $T^*\CP^1$.  In the blowup, the singular
 point at the origin is replaced by a copy of $\CP^1$.    See \cite{Simpson} for the interpretation
 of $\beta$ as a blowup parameter, and \cite{GuW} for a leisurely explanation of some of these
 matters.

We have essentially already run into the fact that in this situation, vanishing monodromy
is a middle-dimensional condition.  Let us specialize to the case that $C=\CP^1$ (we could
similarly treat the case that $C=\C$ with an irregular singularity at infinity).  We know
from section \ref{opbethe} that for a given set of singular points $z_a$ and charges $k_a$,
$a=1,\dots,d$,
there are finitely many opers with monodromy-free singularities.  The condition that
a flat $G_\C$ bundle  should
be an oper is a middle-dimensional condition.   To reduce to a finite set of opers
with monodromy-free singularity,
the condition of vanishing monodromy must also be middle-dimensional.  (This assertion
tacitly assumes that the two conditions intersect in a transverse fashion, which is in fact the
case.)

In fact, dropping the oper condition, we can explicitly describe  the moduli space of solutions of 
Hitchin's equations
on $C$, with monodromy defects  characterized by $\lambda_a=k_a/2$,  for which
the complex connection $\A$ has trivial monodromy around those points. As $C$ is simply
connected,  a flat bundle on $C$ with no monodromy around the points $p_a$ is completely
trivial as a flat bundle.  The only possible moduli arise because the symmetry breaking
associated to the parameters $\beta_a$ (which we assume to be all nonzero)
generates a copy of  $\CP^1$ at each singular point
$p_a$.  
To get the moduli space, 
we must divide the product of these $\CP^1$'s by the automorphism group of the
trivial flat bundle $E$; this is a copy of $SL(2,\C)$.  So finally the locus $\mathcal U$ of 
solutions of Hitchin's equations corresponding to flat  bundles with
trivial monodromy at each singular point is isomorphic to $(\CP^1)^d/SL(2,\C)$.
This is a complex submanifold of $\M_H$ in complex structure $J$ (it is defined by a condition
on the monodromies, which are holomorphic in that complex structure). Its dimension is $d-3$,
which is one-half the dimension of $\M_H$.   In fact, $\mathcal U$ is
complex Lagrangian from the point of view of complex structure $J$; this is true roughly
because each $\CP^1 $ is complex Lagrangian in $T^*\CP^1$.  So the brane $\Btriv$
supported on $\mathcal U$ with trivial Chan-Paton bundle is a half-BPS brane of type
$(A,B,A)$.

This gives us a new way to think about opers of trivial monodromy.  They are intersection
points of two Lagrangian submanifolds of type $(A,B,A)$ -- one is the variety of opers and
one parametrizes bundles with   trivial monodromy.  So the
opers of trivial monodromy  give a basis
for the space of supersymmetric open strings stretching between the brane $\Boper$ and
the brane $\Btriv$.  We call this the space of $(\Boper,\Btriv)$ strings.  Technically
here we want the space of $(\Boper,\Btriv)$ strings in the $B$-model of type $J$.

We can study this space of supersymmetric string states using $S$-duality, which converts the $B$-model of type $J$ to the
$A$-model of type $\omega_K$.  $S$-duality converts the brane $\Boper$ to the canonical
coisotropic brane $\Bcc$, as we learned in section \ref{sduality}.  It turns out that, as we describe shortly, $\Btriv$ is mapped to
itself by $S$-duality (with the usual transformation of the monodromy defect parameters
$(\alpha_a,\beta_a,\gamma_a,\eta_a)$).  So the $S$-dual of the space of $(\Boper,\Btriv)$ strings
is the space of $(\Bcc,\Btriv)$ strings, now viewed in the $A$-model of type $\omega_K$.
The key aspect of this problem is that although the support of $\Btriv$ is 
Lagrangian for $\omega_K$, it is actually symplectic for $\omega_J$ -- indeed, 
the support of $\Btriv$ is
a complex submanifold in complex structure $J$, and accordingly has $\omega_J$ as a 
Kahler form.  This being the case, the problem of describing the space of $(\Bcc,\Btriv)$ strings is
governed by the analysis of quantization and branes in   \cite{GuWtwo}.  The space of
$(\Bcc,\Btriv)$ strings 
is  obtained by quantizing the support  $\mathcal U$ of $\Btriv$; here $\mathcal U$
is viewed as a symplectic manifold with symplectic structure $\omega_J$.

 \begin{figure}
 \begin{center}
   \includegraphics[width=3.5in]{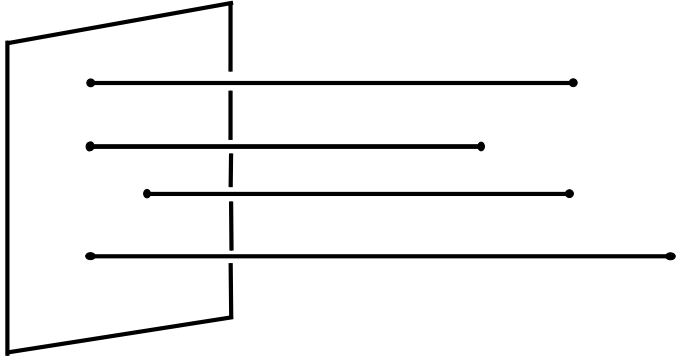}
 \end{center}
\caption{\small  This figure differs from fig. \ref{defects} only in that each monodromy defect
line ends on a singular monopole, indicated by a black dot on the right.   Since the defect
lines themselves are selfdual (with a suitable transformation of their parameters), the brane
defined by ending of the defect lines  is also selfdual. }
 \label{ends}
\end{figure}

\subsection{A Selfdual Brane}\label{simplex}
There is a simple gauge theory explanation of why $\Btriv$ is selfdual. 
Forgetting about supersymmetry for a moment, we can think of a monodromy defect
line as the Dirac string associated to a magnetic monopole that may have been improperly
quantized.  Hence a monodromy defect line can end on a singular magnetic monopole
(fig. \ref{ends}).  Since monodromy defects are mapped to themselves by $S$-duality
(with some transformation of the parameters), pictures in which the  monodromy defects
end on singular monopoles are similarly mapped to themselves by duality. 

Supersymmetry imposes some constraints on the values of the parameters at which
such pictures exist.
In the context of the $B$-model of type $J$, the monodromy around a given
defect line must be trivial if the defect line is going to end.  This means that, in this
$B$-model, the picture of fig. \ref{ends} only exists if $\alpha=\gamma=0$ (here we will
view $\alpha$ and $\eta$ as periodic variables).
Of course, that is anyway the only case that the brane $\Btriv$ can be defined.
Dually, in the $A$-model of type $\omega_K$, a picture like that of fig. \ref{ends}
only exists if $\gamma=\eta=0$.  (For example, $\eta$ must vanish because the worldsheet
theta-angle $\eta$ fails to preserve the topological supersymmetry of the $A$-model
if the support of the monodromy defect ends at a place where the $U(1)$ bundle along
the monodromy defect is not trivialized.)

In the context of the present paper, opers with trivial monodromy arise most directly 
from singular
monopoles at $y=0$.  However, without changing anything essential, 
we can move the singular monopoles away from the boundary as long as we 
connect them to the boundary
via monodromy defects,
as in fig. \ref{ends}.  This has the advantage of making it obvious that  
opers with trivial monodromy
are intersection points of two branes, and also making clear the selfduality of one of these
branes.  

In the general context of a defect line ending on a singular monopole, the monopole may
be incorrectly quantized.  However, for $\lambda=k/2$, which is equivalent to
$\gamma=0$, $\alpha=(k/2)\mathrm{diag}(i,-i)$, the monopole at the end of the string
obeys Dirac quantization, but the string is observable because we assume $\beta\not=0$.

\subsection{Application To The Gaudin Model}\label{applic}

The selfduality of the brane $\Btriv$ provides a gauge theory explanation of 
the main result of \cite{FFR,Frenkel}: opers on $\CP^1$ with trivial monodromy correspond to
simultaneous
eigenvectors of the commuting Hamiltonians of the Gaudin model.  Let us consider
the duality between the space of $(\Bcc,\Btriv)$ strings and the space of $(\Boper,\Btriv)$
strings.    The following discussion assumes familiarity with the framework of \cite{GuWtwo}.

To construct the space of $(\Bcc,\Btriv)$ strings, we have to quantize a moduli space
$(\prod_{a=1}^d\CP^1_a)/SL(2,\C)$, where $\CP^1_a$ is a copy of $\CP^1$ attached
to the monodromy defect at $z=z_a$.  Quantization of $\CP^1_a$ gives an irreducible
representation $R_a$ of $SU(2)$ of spin $j_a=k_a/2$, and quantization of
$(\prod_{a=1}^d\CP^1_a)/SL(2,\C)$ gives a quantum Hilbert space $\frak H$ that 
is the $SU(2)$-invariant part of $\otimes_a R_a$, 
\begin{equation}\label{ifo} \frak H=(\otimes_aR_a)^{SU(2)}. \end{equation}
The classical commuting Hamiltonians of Hitchin's integrable systems can be interpreted
(in the $A$-model of type $\omega_K$)  as $(\Bcc,\Bcc)$ strings.  So they act
on the space $\frak H$ of $(\Bcc,\Btriv)$ strings.  In fact, the Hitchin Hamiltonians
become the commuting Hamiltonians of the Gaudin model.  To demonstrate the last
statement, one interprets the generators of the $SU(2)$ action on $R_a$ as arising
from first order differential operators on $\CP^1_a$, whence the Gaudin Hamiltonians
(\ref{gaudham}) become second order differential operators.  The ``symbols'' (or coefficients
of the leading terms) of these operators are functions on the base of the Hitchin fibration
that are precisely the Hitchin Hamiltonians.  So, reading this in reverse, the Gaudin Hamiltonians
represent a quantization of the Hitchin Hamiltonians (and this quantization is unique, given 
the commutativity of the Hitchin Hamiltonians, modulo 
the possibility of adding $c$-numbers).  

To understand the eigenvectors and eigenvalues of the commuting Hamiltonians,
we use the equivalence of $(\Bcc,\Btriv)$ strings in the $A$-model of type $\omega_K$
to $(\Boper,\Btriv)$ strings in the $B$-model of type $J$.   The latter strings simply
correspond to intersection points of the classical branes $\Bcc$ and $\Boper$.  So
opers with trivial monodromy give a basis for the quantum Hilbert space $\frak H$ of the
Gaudin model.  In the $B$-model description, the commuting Hamiltonians simply become
functions on the variety $\V$ of opers, which is the support of the brane $\Boper$.  Hitchin's classical
Hamiltonians are holomorphic functions on the space of quadratic differentials on $C$
(with poles of prescribed type at the positions $z_a$ of the monodromy defects).  
The support of $\V$ is the space of stress tensors on $C$ (with prescribed poles at the $z_a$). 
The space of stress tensors differs from the space of quadratic differentials only because
of the $c$-number conformal anomaly.  This matches the additive $c$-number ambiguity
in the quantization of the Hitchin Hamiltonians.  The eigenvalues of the quantized Hamiltonians
corresponding to a given oper are simply given by the stress tensor associated to that oper.

\begin{figure}
 \begin{center}
   \includegraphics[width=3.5in]{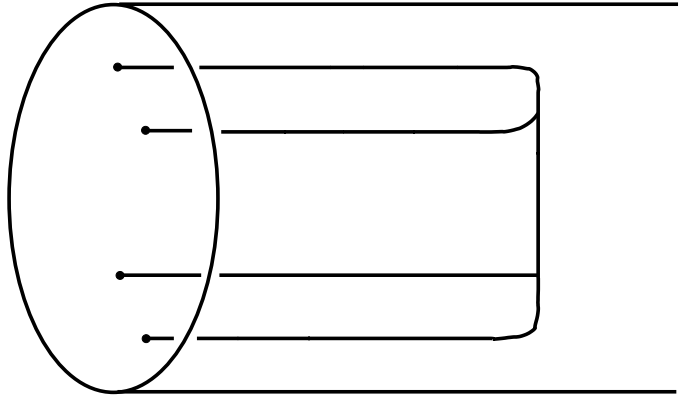}
 \end{center}
\caption{\small  Another selfdual brane can be constructed by replacing
the ends of monodromy defects, which we used in fig. \ref{ends}, with junctions of
monodromy defects, as depicted here. }
 \label{junctions}
\end{figure}

Recently \cite{NStwo}, a ``noncompact'' version of the Gaudin model has been described
in which the finite-dimensional representations $R_a$ are replaced by infinite-dimensional
ones.  The eigenvectors of the commuting Hamiltonians are again expressed as opers,
now with certain conditions on their monodromies.  It is natural to suspect that this
construction again reflects the existence of a selfdual brane.  There actually is a good candidate
-- a selfdual brane that is constructed by replacing ends of monodromy defects,  as in
fig. \ref{ends}, by junctions of such defects, as in fig. \ref{junctions}.  Such a junction is defined
by a solution of Hitchin's equations on a small two-sphere $S$ linking the junction with
singularities (of a type depending on the parameters $\alpha_a,\beta_a,\gamma_a,\eta_a$)
at the intersection points of $S$ with the monodromy defects.

\vskip .5cm
{\bf Acknowledgments} We thank D. Bar-Natan, L. Kauffman, S. Lewallen, P. Li,  R. Mazzeo,
G. Moore,  R. Schoen, P. Seidel, L.-F. Tam, and V. Toledano-Laredo for discussions and comments.  We also thank M. Turansick for assistance with the figures.

\appendix

\section{Three-Dimensional BPS Equations From Six Dimensions}\label{fromsix}
The time-independent configurations we consider in section \ref{teqone} are solutions of the 
3d BPS equations
\begin{equation} \label{sixbps}
[\D_i, \D_j]=0 \qquad  \sum_{i=1}^3[\D_i,\D_i{}^\dagger]
\end{equation}
for a 3d connection together with three adjoint scalar fields, packaged together in the operators $\D_i$ as 
\begin{align} \D_1 & = \frac{D}{Dx^2}+i\frac{D}{Dx^3}    \cr
                                   \D_2 & = [\phi_2-i\phi_3,\,\cdot\,] \cr
                             \D_3 & = \frac{D}{Dy}-i[\phi_1,\,\cdot\,]  \end{align}
Or, in a complex notation,
\begin{align}\label{oggi} \D_1 & =2\frac{D}{D\bar z} \cr
                                                              \D_2 & = 2[\varphi,\,\cdot\,]\cr
                                                               \D_3& = \frac{\D}{\D y}. \end{align}
                                                                                                                           
This system of equations can be generalized to  a one-real-parameter family of 3d BPS 
equations, which can be written as in  \ref{sixbps}, 
but with a different choice of operators $\D_i$: 
\begin{align}\label{bpszeta}\D_1 & =2\frac{D}{D\bar z}  +2 \zeta [\bar \varphi,\,\cdot\,] \cr
                                                              \D_2 & = - 2 \zeta \frac{D}{D z} + 2[\varphi,\,\cdot\,]\cr
                                                               \D_3& = \frac{\D}{\D y}. \end{align}
The generalization was studied in section \ref{analog}.

This family of 3d equations can be usefully derived from six dimensions.  We start
in $\R^6$  with coordinates $x^a,$ $x^{a+3}$, $a=1,2,3$.  Then we constrain
a gauge field by requiring that the field strength, seen as an element of the $SO(6)$ Lie
algebra, lies in a specified $SU(3)$ subgroup.  As long as one is in six dimensions, the choice
of a subgroup does not matter; it just amounts to the choice of an identification of $\R^6$ with
$\C^3$.  But if we require that the fields actually only depend on the first three coordinates
$x^a$, and are invariant under constant shifts of $x^{a+3}$, then the choice of an $SU(3)$
subgroup does matter.  So after dimensional reduction to three dimensions, one can
obtain a family of inequivalent three-dimensional equations depending on a parameter.

A simple way to show that the family is of real dimension one, modulo equivalences, 
is as follows.  First, a choice of embedding of $SU(3)$ in $SO(6)$, parametrized by  
$SO(6)/U(3)\sim SU(4)/U(3) \sim \CP^3$, is equivalent to the choice of a complex line 
in the space of 6d spinors of positive chirality. After dimensional reduction to 3d, 
the inequivalent sets of 3d equations are parametrized by such a choice modulo the 
$SO(3) \times SO(3)$ group of space rotations and rotations of the three scalars 
$\phi_i$. Although this group is six-dimensional,  just like $\CP^3$, it does not act freely
on $\CP^3$; rather, a generic point in $\CP^3$ preserves 
an $SO(2)$ subgroup of $SO(3)\times SO(3)$. 
For example, all the 3d equations parametrized by $\zeta$ are invariant under a 
simultaneous phase rotation of $D/Dz$ and $\phi_z$. 
In general,  $SO(3) \times SO(3)$ acts as $SO(4) \subset SU(4)$ on the space of 6d 
spinors.   So a complex spinor of $SU(4)$ is a complex vector of $SO(4)$, and
its real and imaginary parts break $SO(4)$ to $SO(2)$. 
Hence the family of 3d equations obtained as dimensional reduction of the 6d 
equations is of real dimension one, and 
(\ref{bpszeta}) is a generic representative. 

A consequence of this picture is that we can change $\zeta$ by an $SO(6)$ rotation.
Indeed, we can change $\zeta$ as desired by acting with an appropriate element of a group
that we will call  $SO(2)_\zeta$, which rotates 
$D/Dx^2$ and $[\phi_2,\,\cdot\,] $ into each other, and also rotates 
$D/Dx^3$ and $[\phi_3,\,\cdot\,]$ 
into each other. 
To be precise, the $SO(2)_\zeta$ rotation acting on the $\D_i$ defined at  $\zeta=0$  will give a 
slightly rescaled 
version of the $\D_i$, with a prefactor $(1+\zeta^2)^{-1/2}$. This prefactor can be absorbed 
by a simple rescaling 
of the $z$ coordinate. 
 
 Now let us discuss the Nahm pole boundary condition that has been so important in the
 present paper.   
If we assume a dependence on $y$ only, and further assume that $A_{\bar z}=0$
(so that we can disregard $\D_1$), the equations (\ref{sixbps}) with the $\D_i$ defined as in
(\ref{oggi}) reduce to Nahm's equations.  The Nahm pole boundary condition is defined
by requiring that for $y\to 0$, the fields can be approximated by a certain singular solution
of Nahm's equations.

There is a similar boundary condition 
for the 3d BPS equations at generic $\zeta$. Indeed,  $SO(2)_\zeta$ maps a 
solution of (\ref{sixbps}) which only depends on $y$
to a solution of (\ref{bpszeta}) which only depends on $y$. 
More explicitly, taking the general form of the $\D_i$ in (\ref{bpszeta}), 
we can look for solutions which depend on $y$ only, and such that 
$\D_1$ reduces to $2\partial/\bar\partial\bar z$,  i.e. $A_{\bar z} =- \zeta \bar \varphi$.
Then $\D_2 =  2(1 + \zeta^2) [\varphi,\,\cdot\,]$ and hence we can embed solutions 
of the Nahm equations as solutions of 
the general 3d BPS equations, at the price of a rescaling of the complex scalar 
$\varphi$ by $1+ \zeta^2$. 
This leads to the rotated Nahm pole boundary condition 
which we found useful in this paper. 

Of course, 
what we have just described is not the only embedding of the Nahm pole 
which would be compatible with the general 3d BPS equations. For example, at 
$\zeta \neq 0$, we could have chosen
to look for an embedding in which $\D_2$ rather than $\D_1$ is trivial; this would
lead to what we might call anti-opers -- flat bundles with an oper-like constraint on their
antiholomorphic structure, rather than on their holomorphic structure.
Any rotation of our choice of Nahm pole by the $U(2)$ subgroup of $SU(3)$ which 
preserves $\D_3$ would produce a possible boundary 
condition, but we will generally stick to the ``oper'' Nahm pole.

We will conclude with an alternative explanation of the meaning 
of the parameter $\zeta$.  For finite, non-zero $\zeta$, 
the $\D_i$ can be rescaled and interpreted as a generic 
complex 3d connection. In Cartesian coordinates, we can denote the 
components of the connection as 
$\tilde \D_a$. The complex equations tell us that the connection is flat.
Then we have a moment map constraint, which set to zero a certain constant linear combination 
of the commutators $[\tilde \D_a, \tilde \D^\dagger_b]$. 

From this point of view,
$\zeta$ only appears in the choice of moment map equation. A generic linear combination of the
 commutators  is described by a $3 \times 3$ matrix of coefficients $\omega^{ab}$, 
 \begin{equation}\label{mct}  \sum_{a,b} \omega^{ab} [\tilde \D_a,\tilde \D_b{}^\dagger]=0.
\end{equation}
From eqn. (\ref{bpszeta}), we have a useful relation: $\omega^{z \bar z}/\omega^{\bar z z}=\zeta^2$. 
If $\zeta^2=1$, $\omega^{ab}$ is symmetric, but in general that is not so.

Generically, under linear coordinate redefinitions, there is a one-dimensional parameter 
space of possible 
$\omega^{ab}$. For example, if the symmetric part of $\omega^{ab}$ is positive definite, as 
it is for (\ref{bpszeta}),
we can make it into the identity matrix $\delta^{ab}$. Then the antisymmetric part 
$B^{ab}$ can be rotated to live in the $z, \bar z$ plane, 
and its magnitude is controlled by a single real parameter, which we can identify with $\zeta$.

\section{Three-Dimensional BPS Equations From Four And Eight Dimensions}\label{foureight}

In this appendix, we will discuss how the 3d BPS equations of parameter $\zeta$ can arise from
time-independent solutions of the four-dimensional  BPS equations (\ref{bpseqns}).
We will generalize the statement that 
the $\zeta=0$  equations in three dimensions  arise from the 4d   equations at $\t=1$
if we drop the dependence on one coordinate, say $x^1$, and also set $A_1=\phi_y=0$. 

First  we will show that this is not a feature of a specific choice. We can 
start with any choice of $\t$, set $\d/\d x^1=0$ and set $A_1$ and  $\phi_y$ 
to any two linear combinations 
of the other three components of  $\phi$, and the resulting 3d equations 
will be equivalent to 
the 3d BPS equations discussed in the last appendix for some value of the parameter $\zeta$ in (\ref{bpszeta}). 
 
For that purpose, it is rather convenient to rewrite the 4d BPS equations in a compact form, 
as a dimensional reduction of BPS equations in eight-dimensional Yang-Mills theory.
A succinct way to describe the desired eight-dimensional equations is to pick a $\mathrm{Spin}(8)$ spinor $\epsilon$ of definite
chirality and require
\begin{equation}
F_{IJ} \Gamma^{IJ} \epsilon =0.
\end{equation}
 If the curvature $F_{IJ}$ is understood as an element
of the Lie algebra of $SO(8)$, then 
the equations restrict the curvature to a $\mathrm{Spin}(7)$ subalgebra of $SO(8)$. 
These are really 7 equations, because of the obvious relation
\begin{equation}
\epsilon^T F_{IJ} \Gamma^{IJ} \epsilon =0,
\end{equation}
as $\Gamma^{IJ}$ are antisymmetric. 

Dimensional reduction to four dimensions breaks $SO(8)$ to a subgroup that we will
call $SO(4)_s \times SO(4)_R$, acting respectively on the first four and last four
coordinates.  The spinor $\epsilon$ decomposes into 
a piece $\epsilon_L$ which is left chiral under both $SO(4)_s$ and $SO(4)_R$, and a piece 
$\epsilon_R$ which is right chiral under both $SO(4)_s$ and $SO(4)_R$. 
If both $\epsilon_L$ and $\epsilon_R$ are non-zero, they fix a choice of a 
twisted $SO(4)'_s$ diagonally embedded in $SO(4)_s \times SO(4)_R$,
such that $\epsilon_L$ and $\epsilon_R$ are $SO(4)'_s$ scalars.
Then the $7$ equations decompose under $SO(4)'_s$ into a triplet of self-dual two-forms, a 
triplet of anti-self-dual forms 
and a scalar equation.  This is the form familiar from (\ref{bpseqns}).

\def\a{{\underline a}}
\def\aa{{\overline a}}
\def\b{{\underline b}}
\def\bb{{\overline b}}
We write $a'$ as an abbreviation for $a+4$ and adopt a complex notation
with $\a$ as an abbreviation for $a+ia'$ and $\aa$ as an abbreviation for $a-ia'$.
In order to bring the 8d equations explicitly to the form (\ref{bpseqns}),
it is useful to combine the  $\Gamma$ matrices to raising  operators
\begin{equation}
\gamma_\a = \Gamma_a + i \Gamma_{a+4}
\end{equation}
and lowering operators
\begin{equation}
\gamma_{\aa} = \Gamma_a - i \Gamma_{a+4}
\end{equation}
with $a=1,\dots,4$.
We write $|\Omega\rangle$ for a state annihilated by the lowering operators,
and $|\mho\rangle$ for its complex conjugate, a state annihilated by the raising operators.
Being invariant under $SO(4)'_s$, $\epsilon$ is a linear combination of $|\Omega\rangle$
and $|\mho\rangle$; being real, it is actually 
 $\epsilon =e^{-i \alpha} |\Omega \rangle+e^{i\alpha}|\mho\rangle$, for some real $\alpha$.

Then the 8d equations can be written in terms of the 
$(2,0)$, $(1,1)$ and $(0,2)$ components of the curvature $F_{\a\b}$, $F_{\a \bb}$ and 
$F_{\aa\bb}$:
\begin{align}
e^{-i \alpha}  F_{\a\b} + e^{i \alpha} \frac{1}{2} \epsilon_{ab}\,^{cd} 
F_{\overline c \overline d} &=0 \cr \sum_a F_{\a \aa}&=0.
\end{align} 
When we reduce to 4d, the first equation tells us that the selfdual part of 
$ {\mathrm {Re}} \, (e^{-i \alpha} F_{\a \b}) $ vanishes, as does
the anti-selfdual part of $ {\mathrm {Im}} \,( e^{-i \alpha} F_{\a\b}) $. 
With $\phi=\sum_a A_{a+4}\,\d x^a$, 
we recover the familiar 4d equations
\begin{align} \label{dolk}(F-\phi\wedge \phi + \t \,\d_A\phi)^+&=0   \cr
                             (F-\phi\wedge\phi-\t^{-1}\d_A\phi  )^- &=0 \cr
                               \d_A\star\phi&=0,                     \end{align}                                                                    
with $\t = \tan \alpha$. 

If we start from the 8d form of the equations, it is clear that 
solutions which are independent of some of the eight directions 
preserve additional supersymmetry. 
For example, any solution such that $F_{I8}=0$ for some $I$ also satisfies  
\begin{equation}
F_{IJ} \Gamma^{IJ} \Gamma^8 \epsilon =0
\end{equation}
and hence preserves the supersymmetry generated by the real anti-chiral spinor $\Gamma^8 \epsilon$ of $SO(8)$.
The 7 equations remain independent, and describe a reduction of $SO(7)$ to  
$G_2$ preserving a 7d spinor $\epsilon_7$.

Solutions  that satisfy $F_{I8}=0$ and $F_{I7}=0$ preserve generically four spinors: 
$\epsilon$, $\Gamma^7 \epsilon$, $\Gamma^8 \epsilon$, $\Gamma^{78} \epsilon$.
The 7 equations then describe the reduction of $SO(6)$ to $SU(3)$ 
preserving the supersymmetries
 generated by a 6d complex spinor $\epsilon_6$ and its complex conjugate. They 
decompose into $3$ complex equations and a real moment map condition
\begin{equation}
[\D_i, \D_j]=0 \qquad \sum_i [\D_i, \D^\dagger_{\bar i}] =0,
\end{equation}
as discussed in Appendix \ref{fromsix}.

This is exactly the situation we are in whenever in the four-dimensional equations (\ref{dolk}),
for any value of $\t$,
we set $\d/\d x^1=0$ and set $A_1$, $\phi_y$ 
to any two linear combinations of the remaining three scalar fields $\vec \phi$ in $\phi$.
Any such choices will produce a 3d reduction of the 6d BPS equations, and 
hence, according to the analysis in Appendix \ref{fromsix}, will be equivalent to 
the standard 3d BPS equations  for some $\zeta$. The 3d BPS equations admit the 
oper-Nahm pole boundary condition. 
This will induce a boundary condition in the original 4d BPS equations, which will be 
some deformation of the standard Nahm pole boundary condition.
Vice-versa, with this boundary condition, the usual vanishing theorems will guarantee that
 time-independent solutions 
arise from solutions of the corresponding 3d BPS equations.

Finally, we will describe a simple explicit choice of reduction from 4d to 3d 
which gives whatever $\zeta$ we wish.   Starting from $\alpha=0$ and the standard
reduction with $A_1=\phi_y=0$, 
we make simultaneous $SO(2)$ rotations in the 
$(a, a+4)$ planes for $a=1,2,3$, i.e. rotations of $D/Dx^a$ and $[\phi_a,\,\cdot\,] $ into each 
other by 
angles $\theta_a$.  (We do not make such a rotation for $x^4=y$, as this would not behave
well when we introduce a boundary at $y=0$.)
The rotation  multiplies the creation and destruction operators by phases
$e^{\pm i \theta_a/2}$, 
and hence the vacuum $|\Omega\rangle$ by the phase $e^{- i \sum_a \theta_a/2}$. Hence 
it shifts the angle $\alpha$ 
by $\sum_a \theta_a/2$, and acts correspondingly  on the $\t$ parameter. 

In order to preserve the $SO(2)$ symmetry that rotates $x^2$ and $x^3$,
it is natural to keep $\theta_2 = \theta_3$. Given how the rotation transforms $\D_2$
and $\D_3$ in (\ref{oggi}), we will then have clearly $\zeta = \tan \theta_2$. 

Concerning the relation between $\theta_1$ and $\theta_2$, there are two
particularly  natural choices.
If we want to keep three-dimensional topological symmetry along the boundary, 
we should keep $\theta_1 = \theta_2 = \theta_3=\theta$. 
A rotation by these angles will 
change $\t$ to $\tan( 3 \theta/2 + \pi/4)$, and set $\zeta$ to $\tan \theta$. On 
the other hand, if we 
content ourselves with two-dimensional  symmetry, we can keep $\t=1$, 
by setting $\theta_1 = - 2 \theta_2 = - 2 \theta_3 = - 2\theta$. Again, $\zeta$ will be $\tan \theta$.
With this second choice, we deform only the Nahm pole 
boundary condition, and not the four-dimensional
equations.

\section{On Boundary Conditions And A Special Solution Of The BPS Equations}\label{bco}

Here we will describe the Nahm pole boundary condition for the 3d BPS equations with generic
$\zeta$, allowing for singular monopoles on the boundary, and describe
explicitly the model solution for the case of just one singular monopole.  We work
throughout on $\R^2\times \R$ (the generalization to $C\times \R_+$ is straightforward).

 We will write the BPS equations simply as a
flatness condition
\begin{equation} \label{sixbpst}
[\tilde \D_i, \tilde \D_j]=0 \end{equation}
for a complex 3d connection $\tilde \D_i=\d_i+[\A_i,\,\cdot\,]$
together with a moment map constraint.  Just as in eqn. (\ref{bpszeta}),
the indices $i=1,2,3$ refer to $\bar z$, $z$, and $y$.
The definition of the $\tilde \D_i$ differs  
 from eqn. (\ref{bpszeta}) by a rescaling of $\D_2$.  
 
 This affects the relative normalization of the $ [\tilde \D_i,\tilde \D_i{}^\dagger]$ terms in the moment map constraint.
 Of course, we can always rescale the $y$ coordinate with respect to $z$, $\bar z$. If we write the moment map constraint as  
\begin{equation}\label{mct}  \sum_{i,j} \omega^{ij} [\tilde \D_i,\tilde \D_j{}^\dagger]=0,
\end{equation}
 for a constant diagonal matrix $\omega^{ij}$, the statement invariant under
 scaling is that $\omega^{22} = \zeta^2 \omega^{11}$. 
 We will find it convenient to set $\omega^{11} = \zeta^{-2}$, $\omega^{22} = 1$, $\omega^{33}=1$. 
 If $\zeta^2 =1$, then 
 \begin{equation}
\omega^{ij}\partial_i \partial^\dagger_j = \partial_y^2 + 2 \partial_z \partial_{\bar z}
\end{equation}
 is the Laplace operator for a Euclidean metric on the half-space $\R^2\times\R_+$
 that is normalized in a slightly unconventional way 
 \begin{equation} \label{euc}
\d s^2 = \d y^2 + 2 |\d z|^2.
 \end{equation}
 This normalization will be useful later. 

The flatness condition (\ref{sixbpst}) tells us that  $\tilde \D_i = g \partial_i g^{-1}$ for a complex 
gauge transformation $g$, that is, a map from $\R^2\times\R_+$ 
to $G_\C$. The moment map condition (\ref{mct}) is invariant under
unitary ($G$-valued)  gauge transformations $g\to Ug$. We can eliminate the gauge-invariance
by introducing the gauge-invariant hermitean matrix $h = g^\dagger g$. Then the 
moment map equation 
can be conjugated to 
\begin{equation}\label{conj}
 \omega^{ij}\partial_i (h^{-1} \partial^\dagger_j h) = 0
\end{equation}
or 
\begin{equation}\label{conjtwo}
 \omega^{ij}\partial_i \partial^\dagger_j h = \omega^{ij}(\partial_i h) h^{-1}( \partial^\dagger_j h)
\end{equation}
When $\zeta^2 =1$, this equation says that the map $h$ from the half-space
$\R^2\times\R_+$ to the quotient space $G\backslash G_\C$ (endowed with its 
natural $G_C$-invariant
metric $ \mathrm{Tr}\, ( h^{-1} \d h)^2/2$) is harmonic.  Problems of this type are much-studied,  but
usually   (for
example, see \cite{LiTam}) in the context of a hyperbolic metric on  $\R^2\times\R_+$, rather than
a Euclidean metric, as in our case.

For simplicity, we will specialize to the case $G=SU(2)$, so that $G\backslash G_\C$
is a copy of hyperbolic threespace $H^3$ or $\mathrm{AdS}_3$.  
We can write 
\begin{equation}\label{firstparam} g= \begin{pmatrix} 
Y^{-1/2} & 0 \cr 0 & Y^{1/2}\end{pmatrix} 
\begin{pmatrix} 1 & -\Sigma \cr 0 & 1\end{pmatrix}   \end{equation}
for a real function $Y$ and a complex function $\Sigma$. 
This is a general parametrization, in the sense that every $g\in SL(2,\C)$ 
can be uniquely written in this
form, modulo a unitary gauge transformation $g\to Ug$.  With this parametrization, we have 
\begin{equation}\label{param}
h = \begin{pmatrix} Y^{-1} & -\Sigma Y^{-1}  \cr - \bar \Sigma Y^{-1} &~
 |\Sigma|^2 Y^{-1} + Y \end{pmatrix} 
\end{equation}
In these coordinates, the natural metric on $H^3$ takes a familiar form
\begin{equation}\label{hypmetric}
\frac{1}{2} \mathrm{Tr} \,\left( h^{-1} \d h\right)^2 = \frac{\d Y^2 +  \d\Sigma \,\d\bar \Sigma}{Y^2}.
\end{equation}

In general, in terms of the variables $Y$ and $\Sigma$, the equations for $h$ become 
\begin{align}\label{twoq}
\omega^{ij}\left(\partial_i\left( Y^{-1} \partial^\dagger_j Y \right)+ 
Y^{-2} \partial_i\Sigma \partial^\dagger_j \bar \Sigma \right)=0 \cr \omega^{ij}
\partial_i  \left(Y^{-2} \partial^\dagger_j \bar \Sigma \right)=0.
\end{align}

In the framework
of section \ref{flatpole},
we want a boundary condition that is determined by the properties of the ``small section.''
If we write $s$ for the small section in the complex gauge $\A_i=0$,
then in the unitary gauge with $\D_i=g\partial_i g^{-1}$, the small section becomes $gs$.
We must require $g s$ to go as $y^{1/2}$ as $y \to 0$, while $g$ itself diverges as $y^{-1/2}$.
This means that $h$ diverges as $y^{-1}$ while $hs$ and $s^\dagger h$ are finite and 
$s^\dagger h s$ goes as $y$.   
The standard Nahm pole solution corresponds to $	s=\begin{pmatrix}z \cr 1\end{pmatrix}$ and
\begin{equation}\label{zelm}
g = \begin{pmatrix} 
y^{-1/2} & 0 \cr 0 & y^{1/2}\end{pmatrix} \begin{pmatrix} 1 & -z \cr 0 & 1\end{pmatrix}.
\end{equation}
This formula, which is familiar from eqn. (\ref{inggauge}), is equivalent to
\begin{equation}
h = \begin{pmatrix} y^{-1} & -z y^{-1}  \cr - \bar z y^{-1} &~ |z|^2 y^{-1} + y \end{pmatrix} .
\end{equation} Comparing to the general parametrization  (\ref{param}), we see that the standard Nahm pole
solution is $Y=y$, $\Sigma=z$. 
(The normalization $\omega^{11} = \zeta^{-2}$, $\omega^{22} = 1$, $\omega^{33}=1$ was
chosen to ensure that this is a solution for all $\zeta$.) 
In other words, this solution is 
the ``identity'' map from the half-space $\R^2\times\R_+$ endowed with the
Euclidean metric (\ref{euc}) to the half-space endowed with the hyperbolic metric
(\ref{hypmetric}).  For $\zeta^2=1$, the assertion that this gives a solution is simply
the statement that the ``identity'' map between half-spaces endowed with these
two metrics is harmonic.

In general, if  $s = \begin{pmatrix}P\cr Q\end{pmatrix}$, we want to require that $Y\sim y$ and 
$P - Q \Sigma \sim y$ as $y \to 0$.   The last statement means that if we set $\sigma(z)=P/Q$,
then $\Sigma=\sigma$ at $y=0$.  The fact
that $Y\to 0$ for $y\to 0$ means that the boundary $y=0$ of the half-space $\R^2\times \R_+$
is mapped to the conformal boundary at infinity of the hyperbolic space $H^3$.
That conformal boundary is a copy of $\CP^1$.  
By adjoing $\CP^1$ to $H^3$, one makes the usual conformal compactification $\overline H^3$
of $H^3$.
The choice of an oper without monodromy determines a holomorphic map $\sigma(z)$
from $\R^2\cong\C$ to $\CP^1$, and the condition $\Sigma|_{y=0}=\sigma$ means that,
as a map 
of the boundary  of the half-space to $\CP^1$, $h$ coincides with $\sigma$.  So our problem is 
this:  given a holomorphic map $\sigma$ from the boundary of the half-space to the
conformal boundary of the hyperbolic space, we want
to extend $\sigma$ to
 a map $h:\R^2\times \R_+\to \overline H^3$
that obeys (\ref{conj}) when restricted to $y>0$. For $\zeta^2=1$, we are simply trying to extend the
given map $\sigma$ to a harmonic map from the half-space $\R^2\times \R_+$  to $\overline H^3$.
 (Technically, we assume that the map $\sigma$ has only
polynomial growth so that it extends to a holomorphic map from the one-point compactification
of $\C$ to $\CP^1$, and we similarly require that $h$ extends to a continuous map from the
one-point compactification of the half-space to $\overline H^3$.) 

For any $\sigma(z)$, at least away from the branch points of the map $\sigma$ -- in other
words, the zeroes of $\d\sigma/\d z$ -- it is not difficult to write 
a systematic expansion of $Y$ and $Z$ for
$y \to 0$, involving powers of $y$ and powers of $\log y$. 
The expansion roughly starts with 
$Y =  y |\sigma'(z)| + \cdots$ and 
$\Sigma = \sigma(z) + \cdots$, and the coefficients are rational functions in 
derivatives of $\sigma(z)$, and of three 
undetermined real functions of $z$ and $\bar z$. The denominators of these rational functions are 
powers of $\sigma'(z)$ and its complex conjugate.
So, away from the zeroes of 
$\sigma'(z)$, 
boundary condition behaves well, and cuts in half the degrees of freedom of a solution. 
The branch points are precisely the points with $PQ'-QP'=0$ -- in other words, the points
at which there are singular monopoles.

We still need to show that it is possible for a solution 
to be smooth away from the boundary in the presence of branch points or in other
words singular 
monopoles on the boundary.  The basic problem is to find a model solution in the presence
of just one singular monopole; we then ask for the behavior  near
every singular monopole to match the model solution.
 In order to describe a singular monopole of charge $k$, we 
 consider the special case  $s = \begin{pmatrix}z^{k+1}/(k+1)\cr 1\end{pmatrix}$, or in
 other words $\sigma(z)=z^{k+1}/(k+1)$.  We also
 make use of the invariances of the BPS  equations.
 The equations (\ref{twoq}) are invariant under
 scale transformations $y\to \lambda y, \,z\to \lambda z$ with real $\lambda$,
 and under rotations $z\to e^{i\theta }z$. They are also invariant under reflections $z\to \bar z$ of $\R^2$, accompanied, if $\zeta^2\not=1$, by  $\Sigma \to \bar \Sigma$.

 The boundary conditions $Y\sim
1/y$ for $y\to 0$, $\Sigma|_{y=0}=z^{k+1}/(k+1)$ are invariant
 under all  these symmetries, accompanied by obvious rescalings
of $Y$ and $\Sigma$ (which correspond to $SL(2,\C)$ transformations of the hyperbolic
space).  We expect the solution
of the moment map condition that obeys the boundary condition to be unique, so
it must be invariant under all these symmetries.
Hence we require 
$Y$ to be of the form $y |z|^k e^{u(\rho)}$ and 
$\Sigma$ to be of the form $z^{k+1}e^{u(\rho)}v(\rho)/(k+1)$,  
for  real functions $u$ and  $v$ of $\rho=y/|z|$.  

Then the equations for $h$ turn into two unfortunately rather complicated-looking
 non-linear PDEs:
\begin{align}
v(\rho ) \left(\zeta^2 \left((4 k+3) \rho ^2-8\right)-\rho ^2\right) u'(\rho )-4 k (k+1)
   \rho  \zeta^2 v(\rho ) &\cr +\rho  \left(\rho ^2+\left(\rho ^2+4\right) \zeta^2\right) v(\rho )
   u''(\rho )-\left(\rho ^3+\left(\rho ^2+4\right) \rho  \zeta^2\right) v(\rho ) u'(\rho
   )^2 &\cr +\rho  \left(\rho ^2+\left(\rho ^2+4\right) \zeta^2\right) v''(\rho )+\left(-\rho
   ^2-\left(\rho ^2+8\right) \zeta^2\right) v'(\rho ) &=0 \cr 
(k+1)^2 \rho ^2 \left(\rho ^2+\left(\rho ^2+4\right) \zeta^2\right) u''(\rho )& \cr + v'(\rho )
   \left(2 \left(\rho ^2+\left(\rho ^2+4\right) \zeta^2\right) v(\rho ) u'(\rho )-4 (k+1)
   \rho  \zeta^2 v(\rho )\right)& \cr +(k+1) \rho  u'(\rho ) \left((k+1) \rho ^2
   \left(\zeta^2+1\right)-4 \zeta^2 v(\rho )^2\right)+4 (k+1)^2 \zeta^2 \left(v(\rho
   )^2-1\right)&\cr +\left(\rho ^2+\left(\rho ^2+4\right) \zeta^2\right) v(\rho )^2 u'(\rho
   )^2+\left(\rho ^2+\left(\rho ^2+4\right) \zeta^2\right) v'(\rho )^2 &=0. \end{align}
These equations involve $v$ and the first two derivatives of $u$ and $v$, but not $u$ itself.
Indeed, a constant shift of $u$ is a symmetry of the equations, though not of the desired 
boundary conditions for $y \to 0$. 

As PDEs for $v$ and the derivative $u'$, these equations have a space of solutions which 
is locally three-dimensional. 
The requirement that the solution should be smooth as $z \to 0$ poses two constraints. 
It turns out that at large $\rho$, $u$ behaves as $k \log \rho$, so that $Y \sim y^{k+1}$, 
while $v(\rho)$ scales as $\rho^{-k}$, so that $\Sigma \sim \delta z^{k+1}$ for some constant $\delta$.
The solution admits for large $\rho$ a convergent power series expansion in  $1/\rho$, which depends on $\delta$. 

On the other hand, the boundary condition at $\rho\to 0$ is more forgiving, and only 
imposes a single further constraint on the solution, which basically reduces to the 
requirement that $v \to 1$ as $\rho \to 0$. 
It is not difficult to check numerically that $\delta$ can be tuned so that the solution 
satisfies the constraint, and it is hopefully possible to 
prove this rigorously for any non-zero finite $\zeta$. As $\delta$ is tuned, given the behavior for large $\rho$ imposed in the last
paragraph, there are two possible behaviors for $v(\rho)$ as $\rho$ becomes small. If $\delta$ 
is small, $v(\rho)$ does not reach $1$, and goes to zero as $\rho \to 0$. If $\delta$ is large, 
it crosses $1$ at some finite $\rho$, and then blows up before reaching $\rho=0$. The solution we 
are after corresponds to the critical value of $\delta$ which separates these two behaviors. 

\begin{figure}
 \begin{center}
   \includegraphics[width=3.5in]{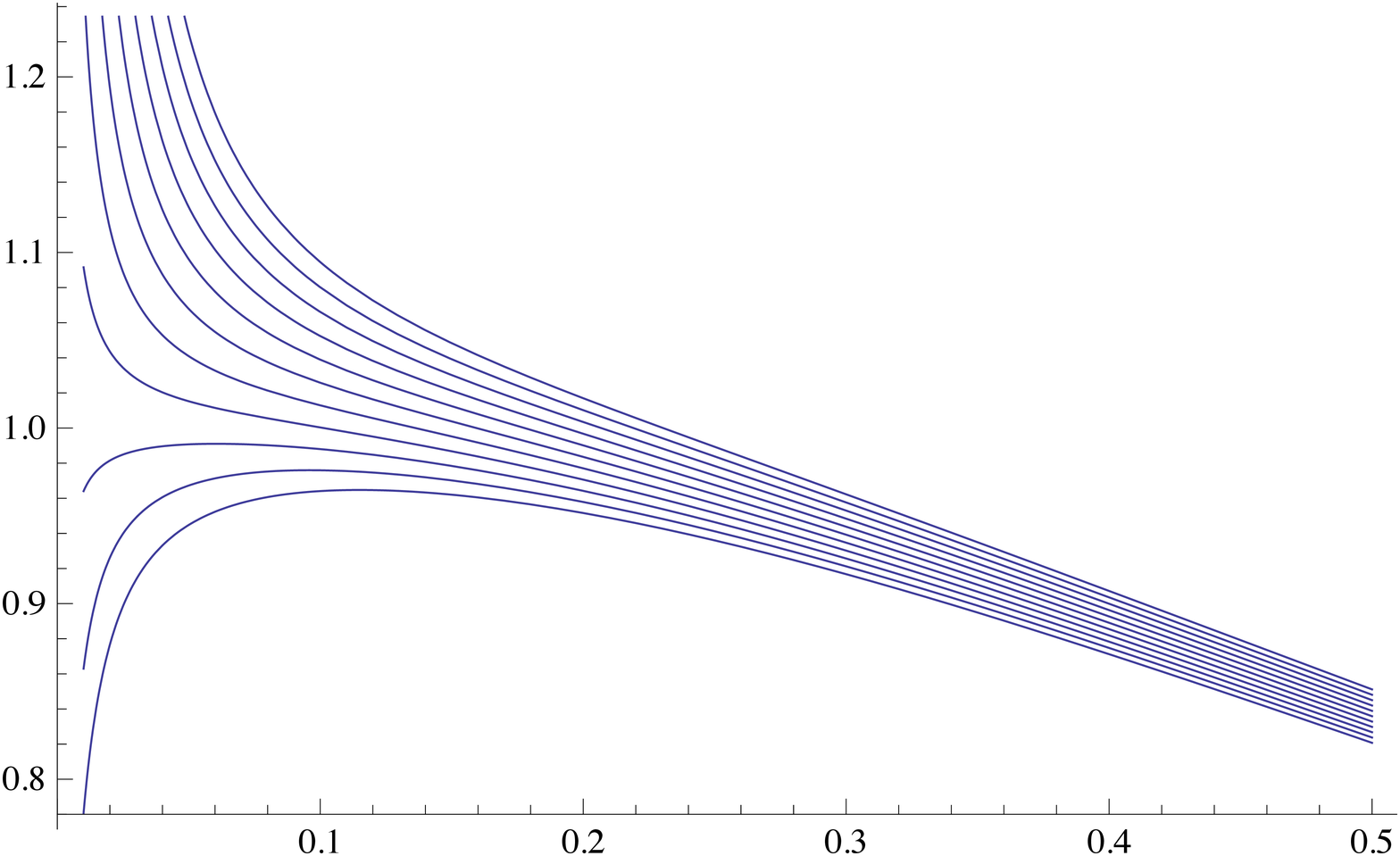}
 \end{center}
\caption{\small  The numerical solutions as $\delta$ is varied across the critical value, for $k=1$.  }
 \label{braidbo}
\end{figure}

\bibliographystyle{unsrt}

\end{document}